\documentclass[preprint2]{aastex}
\usepackage{epsf}
\usepackage{epsfig, subfigure, amsmath, times}

\def\pac{Paczy\'{n}ski}

\def\lsim{{{}_{{}_<}^{~}\atop {}^{{}^\sim}_{~}}}
\def\that{{\hat t}}
\def\amax{{A_{\rm max}}}
\def\VEV#1{\left\langle #1\right\rangle}
\def\ten#1{\times 10^{#1}}
\def\umin{u_{\rm min}}
\newcommand{\be}{\begin{equation}}
\newcommand{\ee}{\end{equation}}
\newcommand{\chisq}{\chi^2}
\newcommand{\lab}[1]{\label{#1}}

\begin{document}

\title{Microlensing Optical Depth towards the Galactic Bulge Using
Clump Giants from the MACHO Survey}

\author{      
    P.~Popowski\altaffilmark{1},
    K.~Griest\altaffilmark{2},
    C.L.~Thomas\altaffilmark{2},
    K.H.~Cook\altaffilmark{3},
    D.P.~Bennett\altaffilmark{4},
    A.C.~Becker\altaffilmark{5},
    D.R.~Alves\altaffilmark{6},
    D.~Minniti\altaffilmark{7},
    A.J.~Drake\altaffilmark{7},
    C.~Alcock\altaffilmark{8},
    R.A.~Allsman\altaffilmark{9},
    T.S.~Axelrod\altaffilmark{10},
    K.C.~Freeman\altaffilmark{11},
    M.~Geha\altaffilmark{12},
    M.J.~Lehner\altaffilmark{13},
    S.L.~Marshall\altaffilmark{14},
    C.A.~Nelson\altaffilmark{3},
    B.A.~Peterson\altaffilmark{11},
    P.J.~Quinn\altaffilmark{15},
    C.W.~Stubbs\altaffilmark{8},
    W.~Sutherland\altaffilmark{16},
    T.~Vandehei\altaffilmark{2},
    D.~Welch\altaffilmark{17} \\
    (The~MACHO~Collaboration) \\
	}

\altaffiltext{1}{Max-Planck-Institute for Astrophysics,
    Karl-Schwarzschild-Str.\ 1, Postfach 1317, 85741 Garching bei M\"{u}nchen, Germany\\
    Email: {\tt popowski@mpa-garching.mpg.de}}

\altaffiltext{2}{Department of Physics, University of California,
    San Diego, CA 92093, USA\\
    Email: {\tt griest@astrophys.ucsd.edu, clt@ucsd.edu, vandehei@astrophys.ucsd.edu }}

\altaffiltext{3}{Lawrence Livermore National Laboratory, Livermore, CA
    94550, USA\\
    Email: {\tt kcook, cnelson@igpp.ucllnl.org}}

\altaffiltext{4}{Department of Physics, University of Notre Dame, IN 46556, USA\\
    Email: {\tt bennett@emu.phys.nd.edu}}

\altaffiltext{5}{Astronomy Department,
    University of Washington, Seattle, WA 98195, USA\\
    Email: {\tt becker@astro.washington.edu}}

\altaffiltext{6}{Laboratory for Astronomy \& Solar Physics, Goddard Space Flight Center, Code 680, Greenbelt, MD 20781, USA\\
    Email: {\tt alves@lasp680.gsfc.nasa.gov}}

\altaffiltext{7}{Departmento de Astronomia, Pontifica Universidad Catolica, Casilla 104, Santiago 22, Chile\\
Email: {\tt dante, ajd@astro.puc.cl}}

\altaffiltext{8}{Harvard-Smithsonian Center for Astrophysics, 60 Garden St., Cambridge, MA 02138, USA\\
Email: {\tt calcock, cstubbs@cfa.harvard.edu}}

\altaffiltext{9}{NOAO, 950 North Cherry Ave., Tucson, AZ  85719, USA\\
  Email: {\tt robyn@noao.edu}}

\altaffiltext{10}{Steward Observatory, University of Arizona, Tucson,
AZ  85721, USA\\
Email: {\tt taxelrod@as.arizona.edu}}

\altaffiltext{11}{Research School of Astronomy and Astrophysics,
        Canberra, Weston Creek, ACT 2611, Australia\\
 Email: {\tt kcf, peterson@mso.anu.edu.au}}

\altaffiltext{12}{Carnegie Observatories,  813 Santa Barbara Street,
     Pasadena, CA 91101, USA\\
     Email: {\tt mgeha@ociw.edu}}

\altaffiltext{13}{Department of Physics and Astronomy, University of
Pennsylvania, PA 19104, USA\\ 
Email: {\tt mlehner@hep.upenn.edu}}

\altaffiltext{14}{SLAC/KIPAC, 2575 Sand Hill Rd., MS 29, Menlo Park,
CA 94025, USA\\
Email: {\tt marshall@slac.stanford.edu}}

\altaffiltext{15}{European Southern Observatory, Karl-Schwarzschild-Str.\ 2,
        85748 Garching bei M\"{u}nchen, Germany\\	
	Email: {\tt pjq@eso.org}}

\altaffiltext{16}{Institute of Astronomy, University of Cambridge,
Madingley Road, Cambridge. CB3 0HA, U.K.\\
    Email: {\tt wjs@ast.cam.ac.uk}}

\altaffiltext{17}{McMaster University, Hamilton, Ontario Canada L8S 4M1\\
    Email: {\tt welch@physics.mcmaster.ca}}

\begin{abstract}
Using 7 years of MACHO survey data, we present a new determination of
the optical depth to microlensing towards the Galactic bulge.  We select the
sample of 62 microlensing events (60 unique) on clump giant sources
and perform a detailed efficiency
analysis. We use only the clump giant sources because these are bright 
bulge stars and are not as strongly affected by blending as other
events.  Using a
subsample of 42 clump events concentrated in an area of 4.5 deg$^2$ 
with 739000 clump giant stars, we find 
$\tau =  2.17^{+0.47}_{-0.38} \ten{-6}$ at $(l,b) = 
(1 \hbox{$.\!\!^\circ$} 50, -2 \hbox{$.\!\!^\circ$} 68)$, somewhat
smaller than found in most previous MACHO studies, but in excellent 
agreement with recent theoretical predictions.  We also present the 
optical depth in each of the 19 fields in which we detected events,
and find limits on optical depth for fields with no events.
The errors in optical depth in individual fields are dominated by 
Poisson noise.
We measure optical depth gradients 
of $(1.06 \pm 0.71) \times 10^{-6} {\rm deg}^{-1}$ and
$(0.29 \pm 0.43) \times 10^{-6} {\rm deg}^{-1}$
in the galactic latitude 
$b$ and longitude $l$ directions, respectively.  
Finally, we discuss the possibility of anomalous duration distribution 
of events in
the field 104 centered on $(l,b) 
= (3 \hbox{$.\!\!^\circ$} 11, -3 \hbox{$.\!\!^\circ$} 01)$ as well as
investigate spatial clustering of events in all fields.
\end{abstract}

\keywords{Galaxy: center, stellar content, structure --- gravitational
lensing --- surveys}

\section{Introduction}
\lab{sec:intro}
The structure and composition of our Galaxy is one of the 
outstanding problems in contemporary astrophysics. 
Microlensing is a powerful tool
to learn about massive objects in the Galaxy.
The amount of matter between the source and observer
is typically described in terms
of the microlensing optical depth, which is defined as the probability that a 
source flux will be gravitationally magnified by more than a factor of 1.34.
Early predictions (Griest et al.\ 1991; \pac\ 1991) of the optical depth 
towards the Galactic center 
included only disk lenses and found values near $\tau = 0.5\ten{-6}$.
The early detection rate (Udalski et al.\ 1993, 1994a) seemed higher, 
and further calculations (Kiraga \& Paczy\'{n}ski 1994) added bulge stars to bring the prediction up to $0.85 \ten{-6}$.  The first measurements
were substantially higher than this: $\tau \geq 3.3 \pm 1.2 \ten{-6}$
at $(l,b) \approx (1 \hbox{$.\!\!^\circ$} 0, -3 \hbox{$.\!\!^\circ$} 9)$
(Udalski et al.\ 1994b) based upon 9 events and 
$\tau = 3.9^{+1.8}_{-1.2} \ten{-6}$ at $(l,b) = 
(2 \hbox{$.\!\!^\circ$} 52, -3 \hbox{$.\!\!^\circ$} 64)$ 
(Alcock et al.\ 1997a) based upon 13 clump-giant events and an 
efficiency calculation.  Many additional theoretical studies ensued, including
additional effects, especially non-axisymmetric components such
as a bar (e.g., Zhao, Spergel \& Rich 1995; Metcalf 1995; 
Zhao \& Mao 1996; Bissantz et al.\ 1997; Gyuk 1999; Nair \&
Miralda-Escud\'e 1999; Binney, Bissantz \& Gerhard 2000;
Sevenster \& Kalnajs 2001; Evans \& Belokurov 2002; Han \& Gould 2003).  
Values in the range $0.8 \ten{-6}$ to 
$2 \ten{-6}$ were predicted for various models, and values as large
$4\ten{-6}$ were found to be inconsistent with almost any model.

More recent measurements using efficiency calculations found values of
$\tau = 2.43^{+0.39}_{-0.38} \ten{-6}$ at 
$(l,b) = (2 \hbox{$.\!\!^\circ$} 68, -3 \hbox{$.\!\!^\circ$} 35)$ 
from 99 events in 8 fields using
difference image analysis (Alcock et al.\ 2000a; MACHO),
$\tau = 2.59^{+0.84}_{-0.64} \ten{-6}$ at 
$(l,b) \approx (1 \hbox{$.\!\!^\circ$} 0, -3 \hbox{$.\!\!^\circ$} 9)$ 
from 28 events using difference image analysis
(Sumi et al.\ 2003; MOA)\footnote{We would like
to notice here that the results from the difference image analyses
cannot be easily compared to the ones from clump analyses. The
analyses that use clump giant sources have reasonably good control
over the location of sources they are sensitive to: the bulge clump
giants dominate over any other possible locations. Difference
image determinations are sensitive to all sources along the line of
sight, and thus the determined optical depth is characteristic for
the direction only and cannot be claimed to be the optical depth
toward the bulge. A typically employed remedy is to assume that the
only two populations of sources that matter in star counts are: the
bulge one and the foreground disk one, and that only bulge sources
contribute to the optical depth. Following on this assumption,
one can correct a measured optical depth by a fudge factor that is
supposed to account for the number of inefficient disk sources. 
It is not clear whether
such correction is needed and the size of it is very uncertain.
The corrections applied by both Alcock et al.\ (2000a)
and Sumi et al.\ (2003) increase the optical depth by about 25\%.
The results we quote here do not include these controversial 
corrections.}, 
$\tau = 2.0 \pm 0.4 \ten{-6}$ at 
$(l,b) = (3 \hbox{$.\!\!^\circ$} 9, -3 \hbox{$.\!\!^\circ$} 8)$
from around 50 clump-giant events
in a preliminary version of this paper
(Popowski et al.\ 2001a; MACHO), and $\tau = 0.94 \pm 0.29 \ten{-6}$ 
at $(l,b) = (2 \hbox{$.\!\!^\circ$} 5, -4 \hbox{$.\!\!^\circ$} 0)$
from
16 clump-giant events (Afonso et al.\ 2003; EROS). 

Blending is a major problem in any analysis of the 
microlensing data involving point spread function photometry.
The bulge fields are crowded, so that
the objects observed at a certain atmospheric seeing are blends of several 
stars.
At the same time, typically only one star is lensed.
In this general case, a determination of the events' parameters
and the analysis of the detection efficiency of microlensing
events is very involved and vulnerable to a number of possible systematic 
errors.
If the sources are bright, one can avoid these problems.  First, 
parameters of the actual microlensing events are typically better constrained. 
Second, it is sufficient to estimate detection efficiency
based on the sampling of the light curve alone.  This eliminates the need
of obtaining deep luminosity functions across the bulge fields. 
Red clump giants are among the brightest and most
numerous stars in the bulge, so we focus on these stars here.
An optical depth determination using all observed microlensing events would be
desirable, but would require an accurate calculation of the blending efficiency
for non-clump stars, demanding
much additional input including
HST quality images and luminosity functions over much of the bulge.

The structure of this paper is the following.
In \S~\ref{sec:data} we briefly describe the MACHO experiment as the source of the
data used here. Section \ref{sec:selection} is devoted to the selection of microlensing 
events. In particular, we discuss how we select our sample of 62 clump
giant events (60 unique) and its relation to the catalog of over
500 events constructed by Thomas et al. (2004,
companion paper). In \S~\ref{sec:blending} we test our sample for signatures of
blending. Optical depth toward the Galactic bulge is derived in \S~\ref{sec:optical_depth}.
In \S~\ref{sec:clustering} we consider spatial distribution of events and discuss the
significance of some apparent clusters seen in the data.
We summarize our results in \S~\ref{sec:discussion}.

\section{Data}
\label{sec:data}
\begin{figure}[t]
\epsfig{figure=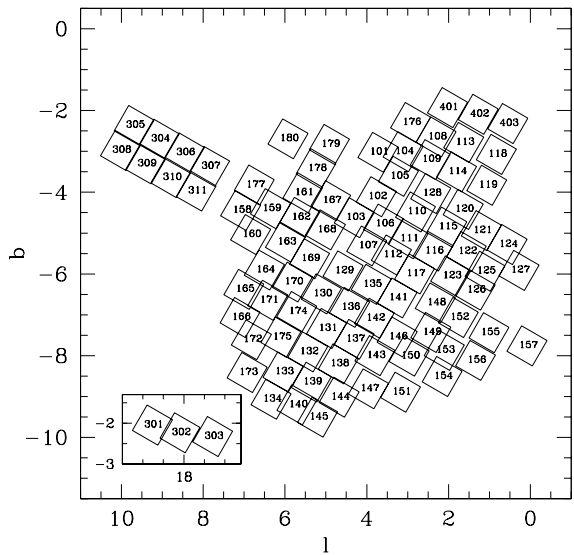, width=0.48\textwidth}
\caption{Location of all 94 MACHO fields [including 3 high-longitude fields
at $(l,b) \sim (18, -2)$].  The clump regions in fields 301 through 311
are expected to be contaminated by disk stars, and thus these fields are not included in the optical depth determination. \label{fig:field_locs}}
\end{figure}

The MACHO Project had full-time use of the 1.27 meter telescope at
Mount Stromlo Observatory, Australia from July 1992 until December
1999.  Details of the telescope system are given by Hart et al.\
(1996), and details of the camera system by Stubbs et al.\ (1993) and
Marshall et al.\ (1994).  Briefly, corrective optics and a dichroic
were used to give simultaneous imaging of a 43'$\times$ 43' fields in
two bands, using eight $2048 \times 2048$ pixel CCD's.  
A total of 32700 exposures were taken in 94 fields (Figure~\ref{fig:field_locs}) towards the Milky Way bulge 
resulting in around 3 Tbytes of raw image data and light curves
on 50.2 million stars. 
The coverage of fields varies greatly from 12 observations
of field 106 to 1815 observations of field 119. Note that the 
observing strategy changed several times during the project, so even
in a given field the frequency of observations changes from year to year.
In addition, the bulge was not observed at all during the prime LMC observing times, so all bulge lightcurves have gaps during November-February.

In this paper, we analyze 7 seasons of Galactic bulge data.
We select a subsample of 62 clump-giant events (60 unique 
events and 2 duplicate events) from 337 selection criteria c (to be
defined in \S 3.1) events
listed in our catalog of more than 500
bulge microlensing events constructed by Thomas et
al.\ (2004, companion paper).
For the optical depth determination, we exclude eleven fields in 
300 series (301, 302, 303, 304, 305, 306, 307, 308, 309, 310, 311). 
The excluded fields are close to the Galactic plane 
but are relatively distant from the Galactic center. As a result, 
they are dominated by the disk stars, which complicates the 
morphology of the color-magnitude diagram and renders the selection 
of clump giants much more difficult. 

For photometric calibrations, we used global relations that express 
Johnson's $V$ and Kron-Cousins' $R$ in terms
of the MACHO intrinsic magnitudes $b_M$ and $r_M$ as:
\be
V = b_M - 0.18(b_M-r_M) + 23.70,
\ee
\be
R = r_M + 0.18(b_M-r_M) + 23.41.
\ee
For more details see Alcock et al.\ (1999).
The limiting magnitude in $V$-band is about 21.5, and a typical
seeing is 2.2'' and almost identical for the blue and red filters.

\begin{figure}[t]
\epsfig{figure=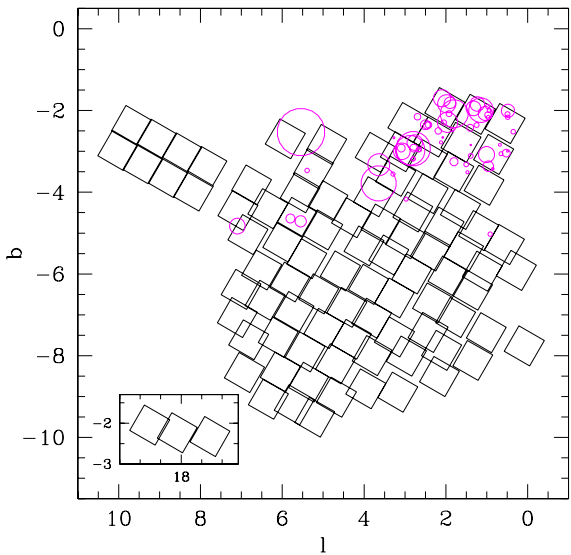, width=0.48\textwidth}
\caption{Spatial distribution of events with clump giants as sources.  
Radii of circles are scaled as $ t/\epsilon(t, A_{max}) $ (See \S 5).
The clump region in fields 301 through 311 (Fig. 1) are expected to be
contaminated by disk stars, and thus clump giant events are not
selected in those fields.
\label{fig:event_locs}}
\end{figure}

\section{Selection of Events for Optical Depth Determination}
\label{sec:selection}
\subsection{Event Selection}
\label{ssec:event_sel}
The microlensing events are selected based on several levels of cuts using 
statistics calculated for the lightcurves.
We denote as level-0, all 50.2 million lightcurves in our database. 
Using a small set of variability and ``bump-finding" statistics, a few percent
of the lightcurves are advanced to level-1.  The complete set of over 150
statistics, including non-linear fits to microlensing lightcurve shape,
is calculated for all these lightcurves and around 90000 are 
advanced to level 1.5.
Final selection (level-2) is made after fine tuning the selection criteria
to maximize inclusion of genuine microlensing events and minimize inclusion
of variable stars, supernovae, noise, etc.
A more thorough description of the most useful statistics is given in 
Alcock et al.\ (2000b).  Table~\ref{tab:cuts} gives a brief summary
of the selection criteria used here, many of which measure signal/noise
and goodness-of-fit to a microlensing lightcurve shape.

To differentiate the current set of 
level-2 cuts from the cuts used previously and from the cuts used
to select LMC microlensing events, we denote
the current set of cuts as ``selection criteria c".  It is crucial to
use exactly this set of selection criteria on the artificial microlensing
events we create to calculate our detection efficiency.

A total of 337 stars passed selection criteria c, 62 of which are
clump giants (a total of 531 events are reported in the companion
paper; these were found with
a less restrictive selection).  Figure~\ref{fig:event_locs} shows the
position of the clump events on the sky. 
Clump giant stars are found by final additional cuts on the color-magnitude diagram (CMD).  This selection differs from the selection used in Alcock et al.\ (1997a). A detailed description of the clump giant selection is given in the next subsection.
The full lightcurves of the microlensing events are
shown in the companion paper (Thomas et al.\ 2004).

\subsection{Clump Giant Selection} 
\label{ssec:clump_sel}

We pay very close attention to the selection of an appropriate clump
region.
Selection of bulge clump giant sources is important since any
non-giant or non-bulge source stars mistakenly included will distort
the optical depth result due to being heavily blended or being at the
wrong distance. 
Therefore, our goal is to select a clean sample of unblended 
{\em bulge} sources.
It is very hard to distinguish bulge red giants from disk clump giants
based on the CMD location alone. Therefore, we decided to 
not include bright red stars in our clump sample. As a result, we
limit clump giants to the relatively narrow strip associated with the 
extinction-based selection described below.
Also, we introduce specific cut to eliminate bright main sequence
stars that are in the foreground Galactic disk.
While clump giants form a 
well defined population the patchy extinction towards the Galactic 
center makes selection more complicated.  We have attempted to correct 
for this effect in a more sophisticated way than in the past, 
but the possibility for contamination still exists.

\begin{figure}[t]
\epsfig{figure=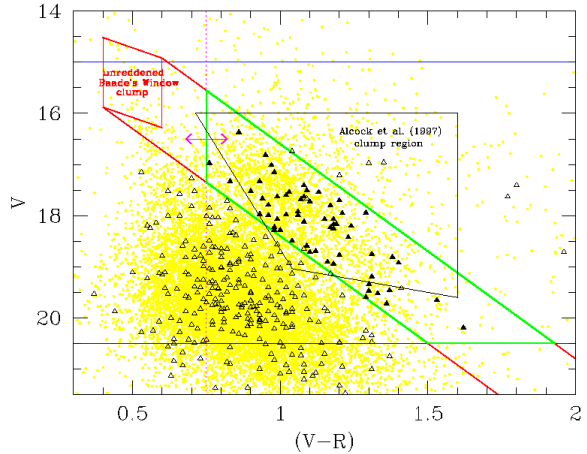, width = 0.48\textwidth}
\caption{Color-magnitude diagram of MACHO objects.  Triangles are
events and filled triangles are clump events. The underlying
population is shown in yellow. \label{fig:cmd}}
\end{figure}

The question, which of our sources may be clump giants, is first investigated
through the analysis of the global properties of the color-magnitude
diagram in the Galactic bulge.
Using the accurately measured extinction towards Baade's Window 
(Stanek 1996 with zero point correction according to Gould, Popowski,
\& Terndrup 1998 and Alcock et al.\ 1998) allows us to locate {\it bulge} 
clump giants on the dereddened color-magnitude diagram. 
Such diagram can be then used to predict the positions of clump
giants on the color-apparent magnitude diagram for fields with different
extinction.
Based on Baade's Window data we conclude that unreddened clump giants
are present in the color range $(V-R)_0 \in (0.40,0.60)$, and
they concentrate along a line $V_0 = 14.35 + 2.0 \, (V-R)_0$, where
the zero point of this line is likely uncertain at the level of at
least 0.1 mags\footnote{The $(V-R)_0$ color range given here is somewhat bluer
than derived by Popowski et al.\ (2001a,b) and reported 
by Popowski et al.\ (2003a). The reason for this change
is the detection of anomalous extinction law toward the bulge fields
probed by microlensing surveys, which in turn affects the dereddened
colors. A possibility of smaller than normal $R_V$ coefficient in the
bulge direction was suggested by Popowski (2000)
and spectacularly confirmed by Udalski (2003) and Sumi (2004) based on
the OGLE data. Very similar conclusions can also be drawn from the
extinction calibrations derived by Popowski, Cook, \& Becker (2003b)
based on the MACHO data.}.
We assume that the actual clump giants scatter in $V_0$ magnitude around these 
central values, but
by not more than 0.6-0.7 mag toward both fainter and brighter $V_0$.
This range of magnitudes is designed to include both the population
scatter and the distance dispersion of clump giants along the line of sight.
The parallelogram-shaped box in the upper left corner
of Figure \ref{fig:cmd} is approximately defined by the above assumptions. 
To be exact, we define this parallelogram to have the following vertex
points in $[(V-R)_0, V_0]$ space: (0.40, 14.52), (0.40, 15.88), (0.6,
16.28), and (0.6, 14.92). 
With the assumption that the clump populations in the whole bulge
have the same properties as the one in the Baade's Window,
the parallelogram described above
can be shifted by the reddening vector
to mark the expected locations of clump giants in different fields.
The solid green lines are the boundaries of the region where one could find
the clump giants in fields with different extinctions.

A universal selection of clump giants outlined above (see also Popowski et al.\
2001a, 2003a) allows some contamination by main sequence stars in high 
extinction areas of the sky. Popowski et al.\ (2001a,b) tried to remedy
this problem introducing a universal color cut $(V-R) > 0.7$. However,
the stars in high-extinction regions may become so red that the
main sequence enters so-defined clump region from the blue.
On the other hand, moving this generally fixed limit of $(V-R)=0.7$ to the red
would remove legitimate
clump giants from low-extinction regions. Therefore, one is driven toward
introducing a non-universal clump selection that makes use of the properties
of the CMD at a given sky location.

In our non-universal selection,
we decided to keep our previous general procedure of selecting clump giants
intact except for 1) the adjustments coming from the different form of
the reddening-law adopted here, 2) addition of a blue color cut tied
to the characteristics of the CMD.  We checked that the region that is
small enough to be sensitive to the local changes of extinction and
large enough to form a reliable CMD approximately coincides with the,
so-called, tile in the MACHO database. Each tile is a 4 by 4 arc-minutes
patch and contains a few thousand identified stars.  As our bulge fields
contain more than 10000 tiles, it would be very impractical to define
blue edges of clump regions on an individual basis. Therefore, we
formed a training set of 240 tiles (in places corresponding to
suspected microlensing candidates from the first 5 years of the
experiment, but rejecting all tiles in 300 series fields and keeping
only one copy of repeated tiles).  This set of tiles is fairly
representative and contains regions with a range of stellar densities
and stellar extinctions.  We decided to set a blue clump limit
at $(V-R)$ half the way between the central clump and main sequence
over-density at the $V$-magnitude of the clump.  The 240 CMDs were
visually inspected and the most likely clump boundary was selected.
The next task was to relate these boundaries to the global properties
of individual CMDs. We computed first four moments of the
$(V-R)$ color of each CMD (no luminosity weighting).  We used the mean
and dispersion as the only parameters in the linear
fit. Mathematically, we requested:
\begin{multline}
(V-R)_{\rm boundary} = \alpha + \beta \, \left< V-R \right>_{CMD} \\
+\gamma \, f\left(\sigma(V-R)_{CMD}\right), \label{eqn:boundary1}
\end{multline}
where we tested three types of f(x), namely $f(x)=x$, $f(x)=x^2$, and $f(x)= \ln(x)$.
All forms of equation (\ref{eqn:boundary1}) produced very similar results,
and we adopted $f(x) = \ln(x)$, requesting that the clump giants be redder than:\begin{multline}
(V-R)_{\rm boundary} = (-0.001 \pm 0.033) \\
+ (0.872 \pm 0.014) \, \left< V-R \right>_{CMD} \\
+ (-0.042 \pm 0.012) \, \ln\left(\sigma(V-R)_{CMD}\right).\label{eqn:boundary2}
\end{multline} 

The errors reported in equation (\ref{eqn:boundary2}) were normalized 
to produce 
$\chi^2/d.o.f. = 1$.\footnote{The correlation matrix is:\\
$\left(
\begin{array}{rrr}
1.0000 & -0.9084 & 0.9720 \\
-0.9084 & 1.0000 & -0.7872 \\
0.9720 & -0.7872 & 1.0000 \\
\end{array}
\right)
$
}
The clump blue boundary determined by visual inspection that deviates
most from the 
fit reported in equation (\ref{eqn:boundary2}) is at $3.28 \sigma$ away. 
In general, the number of $3\sigma$ points is not
very different than expected from a normal distribution of errors,
and therefore we reject no points as outliers.
Moreover, in the cases with largest deviations, the 
blue limits 
suggested by equation (\ref{eqn:boundary2}) are equally reasonable as originally
selected boundaries, and, therefore, do not hint at any small subset of CMDs 
that do not obey a general relation.
The average scatter of the visually-selected $(V-R)_{\rm boundary}$
around relation (\ref{eqn:boundary2}) is below 0.02 mag, which, 
being an order of magnitude smaller than a typical range of 
clump colors, is more than satisfactory.

Our final set of clump selection criteria are described by the
following equations:
\be
V_{\rm base} \geq 15.0 \;\;\;\;\; {\rm and} \;\;\;\;\; V_{\rm base} \leq 20.5, \label{eqn:clumpcut1}
\ee
\be
V_{\rm base} \geq 4.2 \, (V-R)_{\rm base} + 12.4, \label{eqn:clumpcut2}
\ee
\be
V_{\rm base} \leq 4.2 \, (V-R)_{\rm base} + 14.2, \label{eqn:clumpcut3}
\ee
\be
(V-R)_{\rm base} > (V-R)_{\rm boundary}, \label{eqn:clumpcut4}
\ee
\be
{\rm field} < 301 \;\;\;\;\; {\rm or} \;\;\;\;\; {\rm field} > 311, \label{eqn:clumpcut5}
\ee
where subscript ``base'' referring to $V$-magnitude and $(V-R)$-color
indicates baseline magnitudes, i.e. the ones in the limit of no
microlensing-induced amplification.
The cut on bright sources in (\ref{eqn:clumpcut1}) is intended to eliminate
events that are in the foreground or may have saturated photometry,
the cut on faint sources is designed to avoid stars with uncertain
photometry or highly affected by blending.
In practice, the cut on bright sources from (\ref{eqn:clumpcut1}) is
always weaker than the combination of cuts (\ref{eqn:clumpcut2}) and 
(\ref{eqn:clumpcut4}).
Cuts (\ref{eqn:clumpcut2}) and (\ref{eqn:clumpcut3}) determine the reddening
channel that marks the possible locations of clump giants.
Cut (\ref{eqn:clumpcut4}) admits only stars with red enough color, which
removes contamination by foreground main sequence stars. Finally, cut
(\ref{eqn:clumpcut5}) excludes 11 disk dominated/contaminated fields,
namely the ones in the 300 series.

The slope of relations (\ref{eqn:clumpcut2}) and (\ref{eqn:clumpcut3}) has the
interpretation of the coefficient of selective extinction $R_{V,VR}
\equiv A_V/E(V-R)$. The coefficient we use here $R_{V,VR} = 4.2$
was found by Popowski et al.\ (2003b) in Baade's Window. We checked
that clump selection based on $R_{V,VR} = 4.5$ produces very similar
results. On the other hand, $R_{V,VR} = 5.0$ seems to be too steep.
Despite this, Popowski et al.\ (2001a) selection was quite successful,
because their intrinsic clump color was assumed to be redder and
they had very few events in high-extinction regions, which are
primarily affected by this change.
Finally, let us note that the properties of extinction law and the 
zero-point needed to
determine absolute extinction in the bulge are not known precisely,
which could have negatively affected our selection. This was not the case.
We visually checked in the number of regions with different level of reddening
that the operational procedure defined in equations 
(\ref{eqn:clumpcut1})-(\ref{eqn:clumpcut5}) properly picks up clump
giants in a wide range of color-magnitude diagrams.

Clump region defined by conditions
(\ref{eqn:clumpcut1})-(\ref{eqn:clumpcut5})
is plotted in Figure~\ref{fig:cmd} and surrounded by bold green lines.
In addition to the boundary of clump region, we also indicate
cut (\ref{eqn:clumpcut1}) by blue lines, cuts
(\ref{eqn:clumpcut2}) and (\ref{eqn:clumpcut3}) by red lines, and cut (\ref{eqn:clumpcut4})
by dotted magenta line, with arrows indicating
its non-universal, CMD-specific character. The clump events are marked
as filled triangles. The events inside bold green lines marked as open
triangles belong to fields in 300 series, and are excluded by cut
(\ref{eqn:clumpcut5}). Alcock et al.\ (1997a) clump region is presented
for comparison.

\begin{figure}[t]
\epsfig{figure=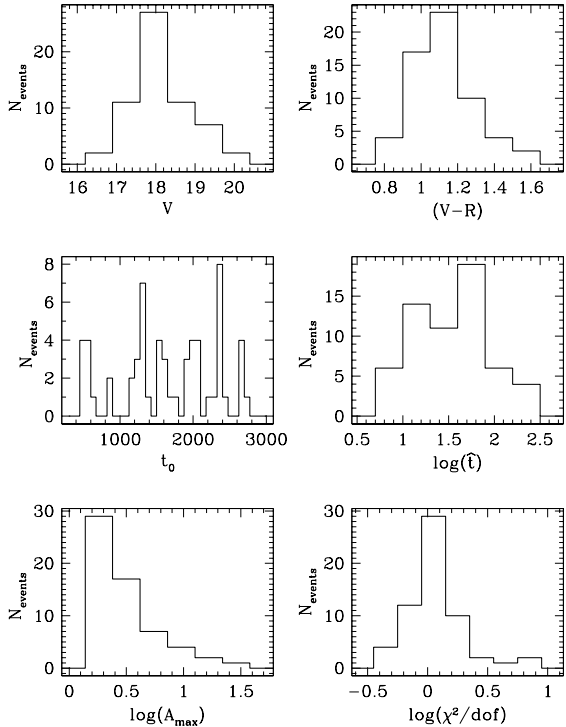, width = 0.48\textwidth}
\caption{Basic properties of the clump sample.\label{fig:panel1}}
\end{figure}

The properties of the 62 lensed clump giant stars as well as the microlensing 
fit parameters are given in Tables~\ref{tab:clumpevents} and
\ref{tab:taubyevent}. In these and the following tables $V$ and
$(V-R)$ will refer to baseline magnitudes -- we drop subscript
``base'' to avoid clutter.
We note that two events are marked with letters (`a' or `b') 
indicating they are duplicates of 
another event (duplicate also marked `a' or `b').
In general, this is caused either by the same 
physical event being detected in an overlapping field, or by a nearby star 
receiving flux from the actual event due to incorrect flux sharing in the 
photometric PSF fitting code. Both duplicate events listed in Tables 
\ref{tab:clumpevents} and \ref{tab:taubyevent} result from field overlaps.   
For the optical depth calculation we will count both 
of the duplicate events since field overlaps increase 
both the total number of stars monitored and the number of events detected
proportionally, and since the field overlap and number of duplicates
is fairly small.
Finally we note that 5 of the clump giant microlensing events are 
potential binary events, marked with a $^\ddagger$ symbol.

In Figure~\ref{fig:panel1} we graphically summarize the basic
properties of clump events. Most of the panels are self-explanatory.
Let us only remark about the distribution of the number of events
versus the time of maximum amplification $t_0$. The peaks in the
histograms show 7 observing seasons analyzed here. The number of
detected events varies mostly due to Poisson noise with the noticeable
exceptions of the second and last seasons. The numbers of events are
particularly low there due to an observing strategy that targeted
very few fields in search for short-duration events.

\begin{figure}[t]
\epsfig{figure=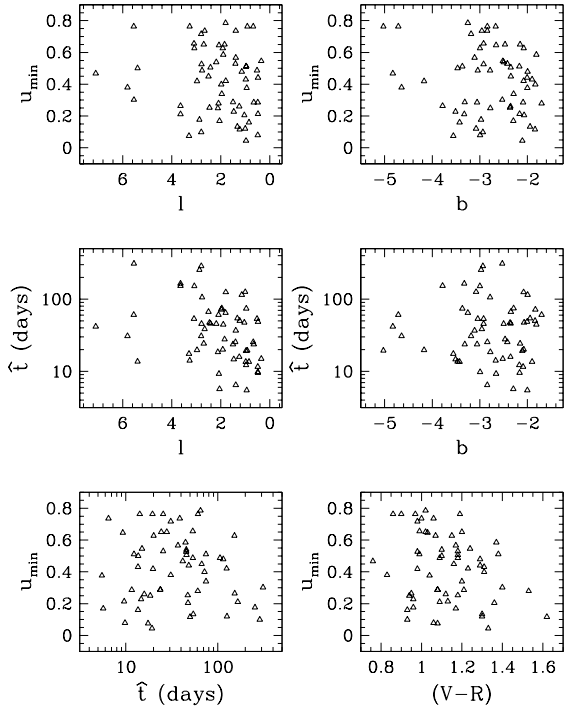, width = 0.48\textwidth}
\caption{Microlensing properties of events as a function of position,
duration, or color.\label{fig:panel2}}
\end{figure}

In Figure~\ref{fig:panel2} we test for possible systematic effect
in microlensing parameters. In particular, in the upper four panels
we plot impact parameters and Einstein diameter crossing times
as a function of Galactic coordinates.
One sees no dependence of parameters upon position.
The left lower panel suggests
that there is no correlation between the impact parameter and
duration, and the right lower panel shows that the impact parameter is
not related to the event color.

Finally, we note that many of the events selected here were alerted
on by us earlier in the experiment. In addition, the MACHO fields overlap 
with fields from other microlensing experiments. We cross-referenced 
the clump sample with our own alert events and with events from EROS,
MOA, and OGLE experiments. We found no counterparts in the MOA set (as
expected from non-overlapping observing seasons). Alternative
designations from the MACHO Alert system and counterparts from EROS
and OGLE are listed in Table~\ref{tab:crossReferences}.

\section{Is Blending Negligible for the Clump Sample?}
\label{sec:blending}

We make an analytic estimate of the effect of blending
on our clump analysis using the scaled luminosity function 
for Baade's Window from Holtzman et al.\ (1998).
We find that even under very
conservative assumptions, the number of highly blended clump events
should not exceed 25\%. Moreover, this level of contamination affects
the optical depth at the level not exceeding 10\%, and for some 
photometric code behaviors the contamination may be completely 
negligible. The details are discussed in Appendix~A.
To check whether the sample of selected clump giants meets our expectations
of being weakly affected by blending we applied several tests.

\begin{figure}[t]
\subfigure{\epsfig{figure=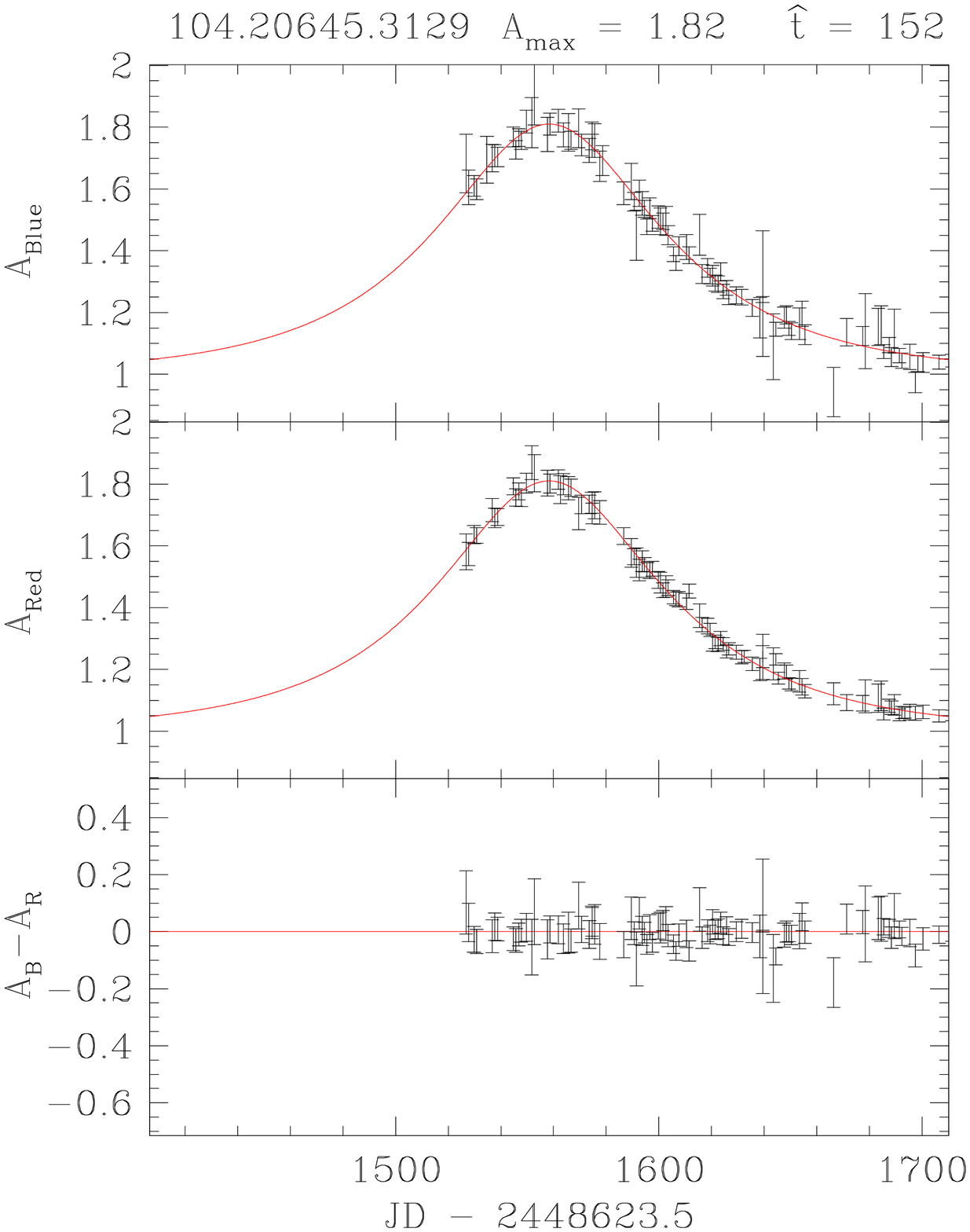, width = 0.40\textwidth }}
\hfill
\subfigure{\epsfig{figure=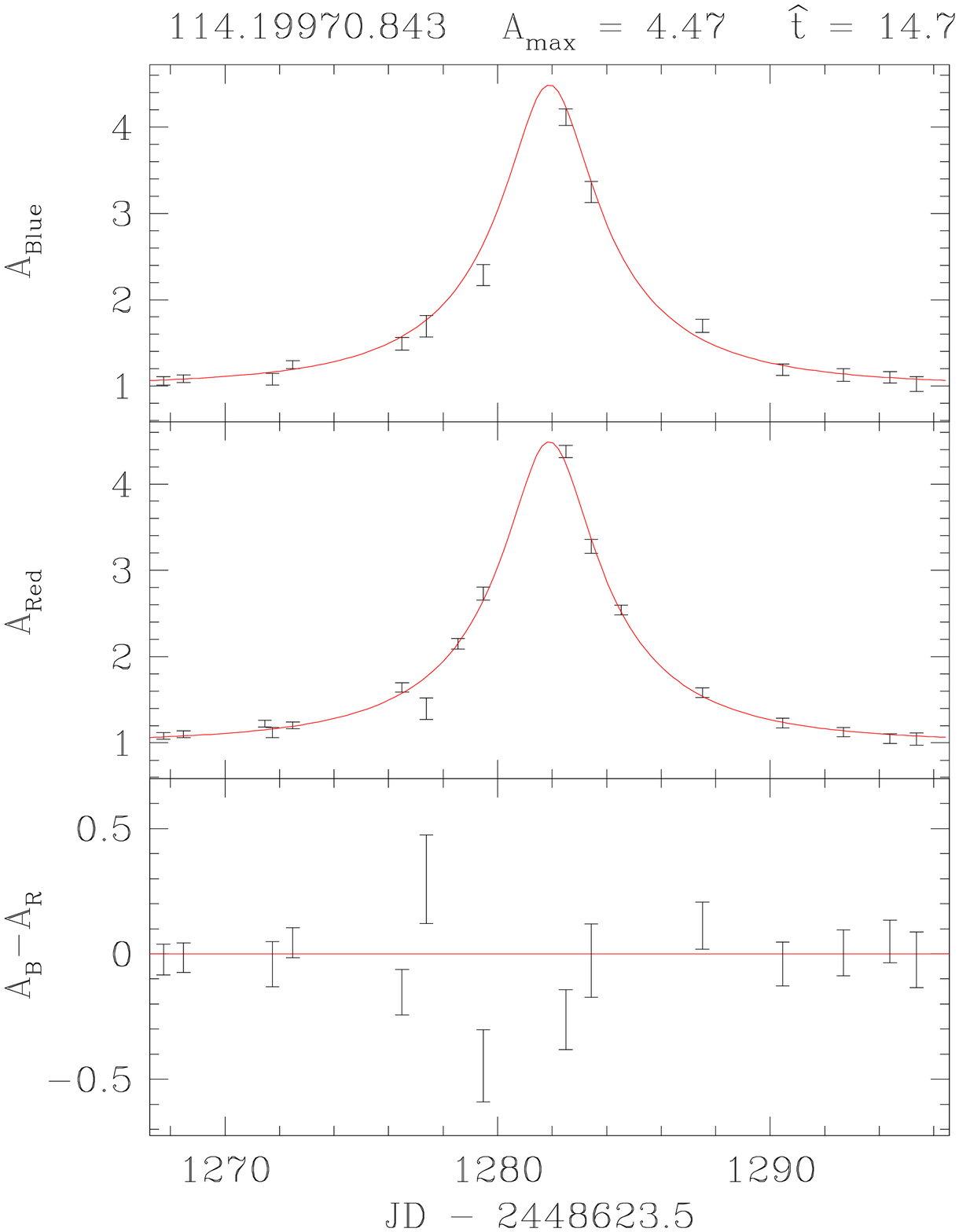,  width=0.40\textwidth }}
\caption{The upper panel shows the event with no color signal, and the
lower panel the event with a weak signal. On each panel,
the upper box shows the amplification in the blue filter
(the baseline flux has $A = 1$), the middle box the amplification in
the red filter, and the lower box the difference between the two
which is an indication of color. Shown are only the peak regions
within $\hat{t}$ of the time with maximum amplification $t_0$.
\label{fig:colorlc}}
\end{figure}

First, we reviewed the color light curves of all events with
particular emphasis on the peak region. For clump giants, blends 
with bluer stars are more likely than the blends with stars 
of identical color and so this test should be sensitive to the
majority of potential contaminants
The light curves of all clump
events showing blue and red magnitudes and the difference between them
for simultaneous observations are shown in Figure~\ref{fig:colorlc}. Here
we display two example events, whereas the entire sample is
available in the electronic version and will be available on the World Wide 
Web upon the acceptance of this paper\footnote{See {\tt http://wwwmacho.mcmaster.ca}}.
The red lines superposed on the data are the best fits obtained under the
assumptions of no blending, and in most cases they properly represent
observational points.
We examine the light curves for any deviations from the unblended
fits. We note that events 109.20640.360, 113.19192.365, 
176.18826.909, 401.47994.1182, 401.48052.861, \linebreak 
401.48167.1934,
402.47856.561, and 403.47491.770 either lack or have extremely sparse
color coverage in the peak region. Therefore, any blending-related
information that can be extracted from their light curves is very
limited. 
The events for which there is at least a slight possibility of
achromatic signal
are listed in Table~\ref{tab:lcdeviations}. Several events have
deviation from fitted light
curves that are due to their exotic character: either binarity or
parallax effect. Only 11 out of 62 events are not immediately
explained by known effects. In addition, 3 of them have either asymmetric or
caustic type signal which is more likely an indication of their binary 
character than blending. Therefore, only 8 events (13\%) can be
suspected of significant blending based on the visual inspection of 
their light curves.

\begin{figure}[t]
\epsfig{figure=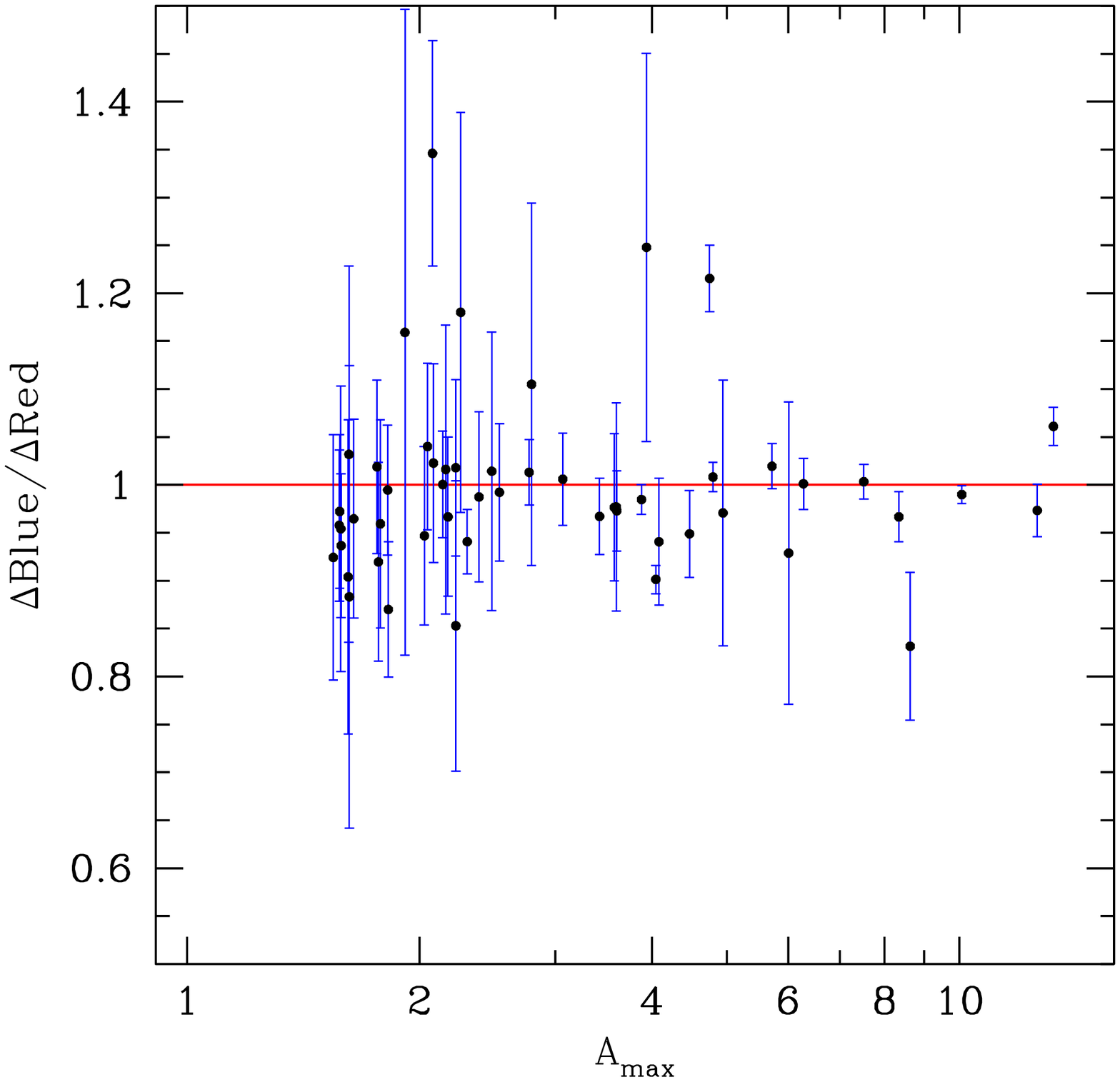, width = 0.48\textwidth}
\caption{Ratios of $\Delta V/\Delta R$ for the 53 events with observations in the peak of the lightcurve. \label{fig:colorshift}}
\end{figure}

Next we examined the color close to the center of the peak ($t - t_0 <
\that/\sqrt{2} $) compared to the baseline of the lightcurve in a more
quantitative fashion.  We start with a region of width $\that/8$
centered on $t_0$ and expand it until we have at least 2 points with simultaneous observations in each filter.  Using these points in each lightcurve we calculate the difference in flux between the peak and the baseline in each filter ($\Delta V$ or $\Delta R$).
In Figure \ref{fig:colorshift} we plot the ratio $\Delta V/\Delta R $ versus $\amax$.  
For blending-free microlensing events $\Delta V/\Delta R $ should be unity since gravitational lensing is wavelength independent.
Of the 53 events that had simultaneous observations in both filters in the peak, 20.7\% were more than $ 1\sigma $ away from unity.  
This is consistent within errors with no blending in our sample.  There are two events (108.18951.593 and 403.47610.576) 
which differ by more than 6 sigma from unity, which might be
indicative of blending. One (108.18951.593) is likely blended with
a faint star of a very different color, and
examination of the lightcurve of the second one shows that it has many
points far off the microlensing fit (403.47610.576), probably
indicating intrinsic variability or problems with the photometry of
this star. If we ignore these two events we have 17.6\% more than 
$1 \sigma$ from unity and the distribution is roughly Gaussian.

\begin{figure}[t]
\center
\epsfig{figure=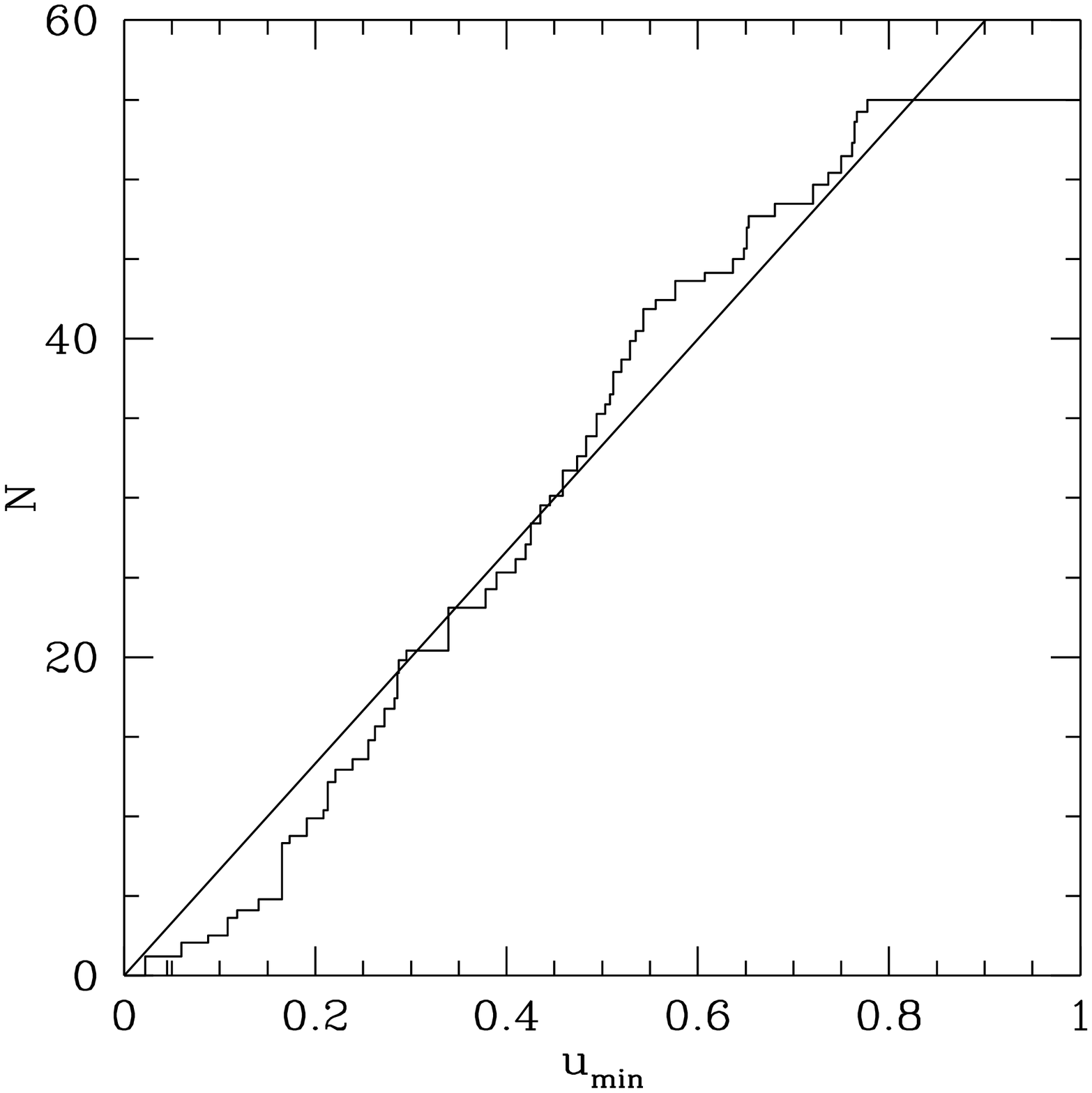, width=0.48\textwidth} 
\caption{Cumulative distribution of impact parameters weighted by
inverse efficiencies (renormalized to 55, the total number of unique
non-binary clump giant events).  
Our cuts limit the impact parameter to 0.826. \label{fig:ucum}} 
\end{figure}

As our third test, in Figure~\ref{fig:ucum}, we plot the cumulative distribution of the impact parameters $u_{\rm min}$ and a uniform distribution (straight line) of $\umin$ between 0 and 0.826.  
This distribution is what is predicted for microlensing events with magnifications between infinity and 1.5 as imposed by our selection criteria.  
We find good agreement, with the
Kolmogorov-Smirnov $D=0.107$ for 55 events 
(60 unique events minus 5 binaries).  
Under the hypothesis that all events are microlensing, this gives a 
probability of 53\% of obtaining a value of $D$ this large or larger.  
For this analysis we made a correction for the efficiency of each event,
but even without making this correction we get excellent agreement 
($D=0.103$ for a probability of 58\%). This suggests that $u_{\rm min}$
values from standard microlensing fits are not seriously biased by
blending.

\begin{figure}[t]
\center
\subfigure{\epsfig{figure=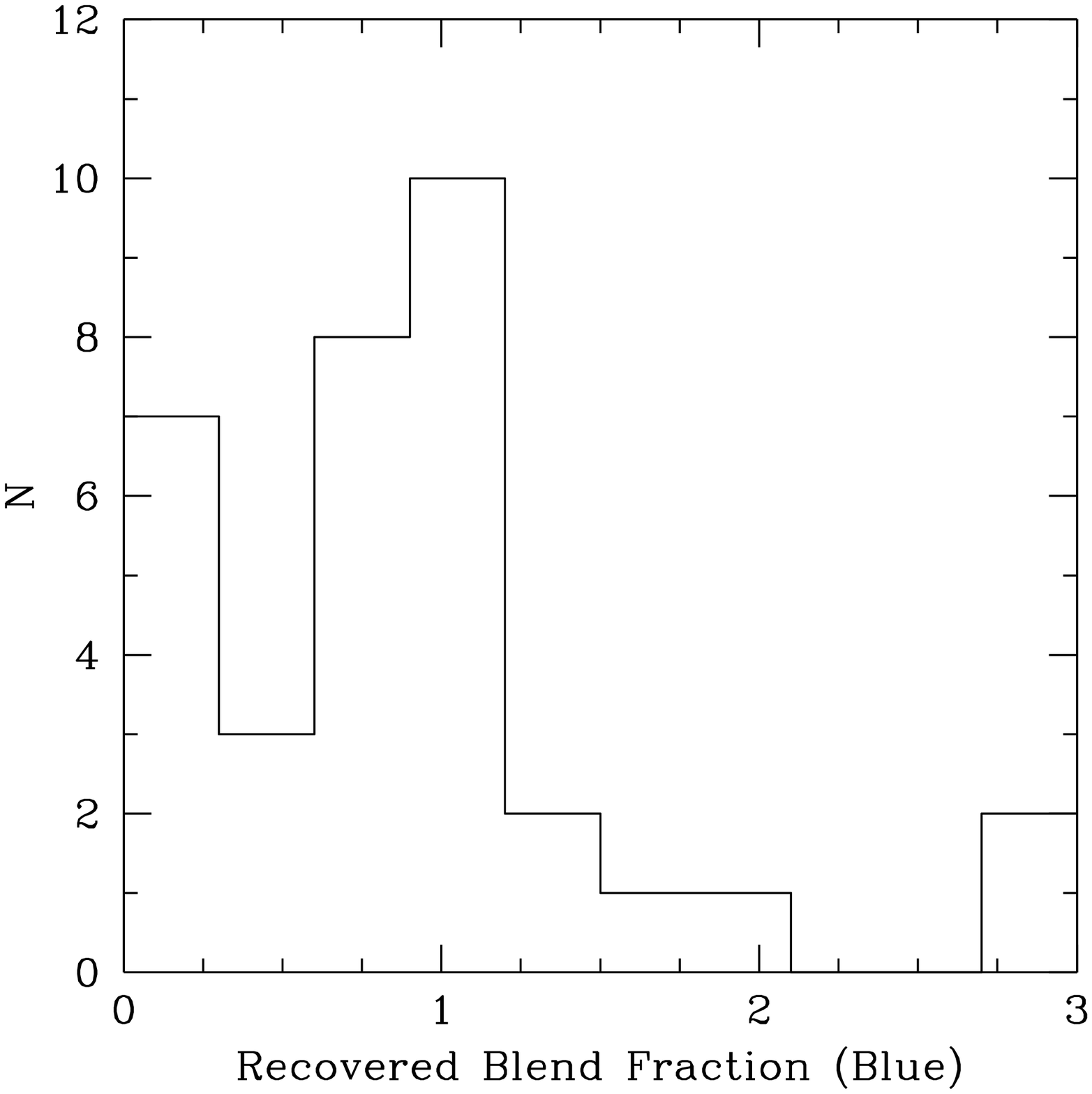, width = 0.45\textwidth }}
\hfill
\subfigure{\epsfig{figure=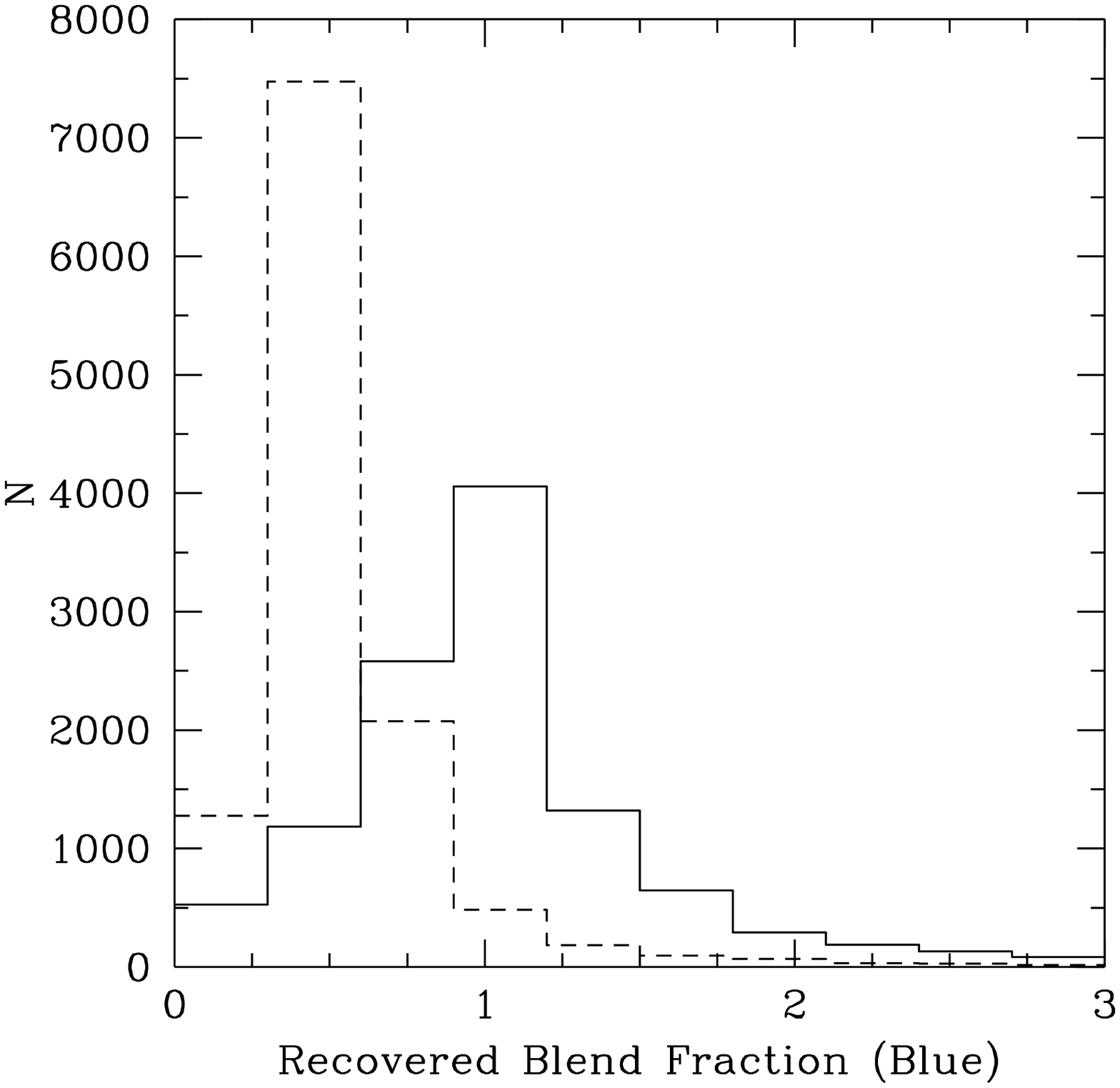,  width=0.45\textwidth }}
\caption{Upper panel presents blue blend fraction distribution for the 
clump sample from CGR (as defined in \S 5). The lower panel shows the recovered
blend fractions for the sample of simulated unblended events (solid line)
and events with input blend fraction of 0.5 (dashed line).
\label{fig:blendfrac}} 
\end{figure}

Finally, we perform so-called blend fits.
That is we fit every flux curve with a formula
\be
F(t) = F_{\rm base} \left[ (1-f) + f A(t) \right],
\label{eqn:blendform}
\ee
where $A(t)$ is the amplification of standard point lens microlensing, 
$F_{\rm base}$ is the baseline
flux of an object unaffected by the microlensing magnification, and $f$ is
the blending fraction.
For a completely unblended source one expects $f = 1$.
We fit all curves using MINUIT 
(CERN Lib. 2003), and report the results together with
the parabolic errors in Table~\ref{tab:blend_params}.
In our fitting setup, the blending fractions for an event can take 
different values in blue and red filters.
The upper panel of Figure~\ref{fig:blendfrac} shows the distribution 
of blend fractions
for the events from the CGR (a group of 9 fields used for the
determination of the average optical depth, which is defined in \S 5).
As expected, many events have blend fractions close to 1.0, i.e.
consistent with no blending. 
At the same time, however, the results look quite alarming. The
distribution of blend fractions is very broad and there is a number
of events with small blend fractions that indicate heavy blending.
To understand the meaning of this result we perform Monte
Carlo simulations of the
expected distribution of recovered blend fractions for {\em completely
unblended} clump events from a sample identical to the one presented
in the upper panel of Figure~\ref{fig:blendfrac}. We create
light curves of unblended events with durations and $u_{\min}$ values
identical to the ones from the real sample. We produce a few hundred
light curve realizations for each event and compute for them blend fit
parameters. Thus for each real event we obtain an expected
distribution of recovered blend fractions if the
original event was unblended. We sum those distributions together to
obtain the expected blend fraction distribution for the real 
sample. We plot the resultant distribution in the lower panel of
Figure~\ref{fig:blendfrac} (solid line). Therefore, even if there
is no blending in our clump sample, the blend fit procedure will
predict many blended events. This problem has been studied before and
is a result of sampling and data quality (Wo\'{z}niak and Paczy\'{n}ski 1997).
The shapes of the distributions from Figure~\ref{fig:blendfrac}
are similar, which suggests that our null
hypothesis that the clump sample is unblended may be correct.
For comparison, we also plot the recovered blend fractions for the
sample with input blend fractions of 0.5 (dashed line). It is clear 
that the peak follows the true blend fraction, which again suggests 
that our sample is mostly unblended. In addition, the distribution 
resulting from the input blend fractions of 0.5 is narrower 
than the true one and lacks extended wing toward large $f$ values.
However, we also note that the real sample seems to have some excess 
of events with very small blend fractions, inconsistent with neither
of the two distributions shown in the lower panel.

\begin{figure}[t]
\center
\subfigure{\epsfig{figure=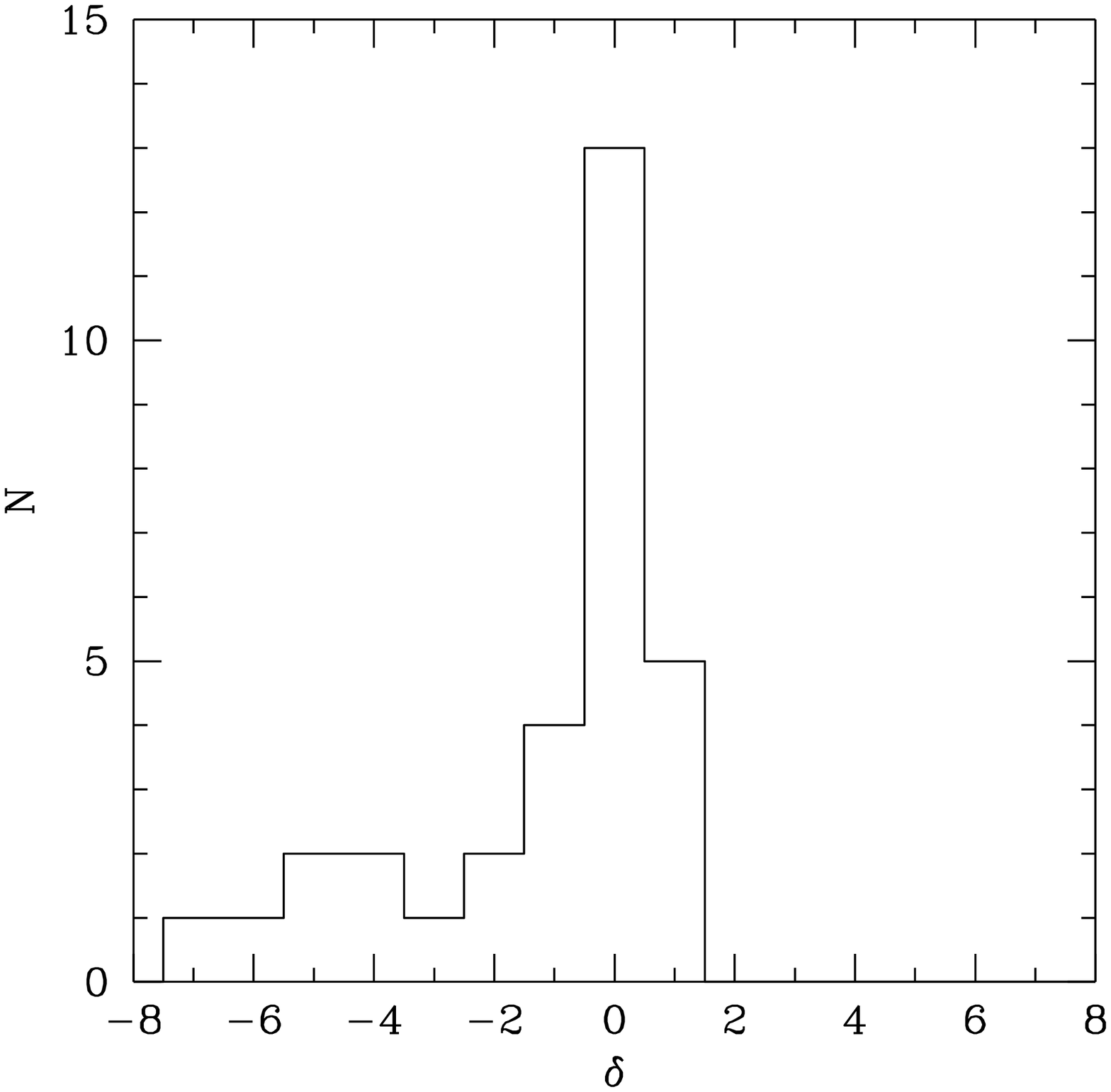, width = 0.45\textwidth }}
\hfill
\subfigure{\epsfig{figure=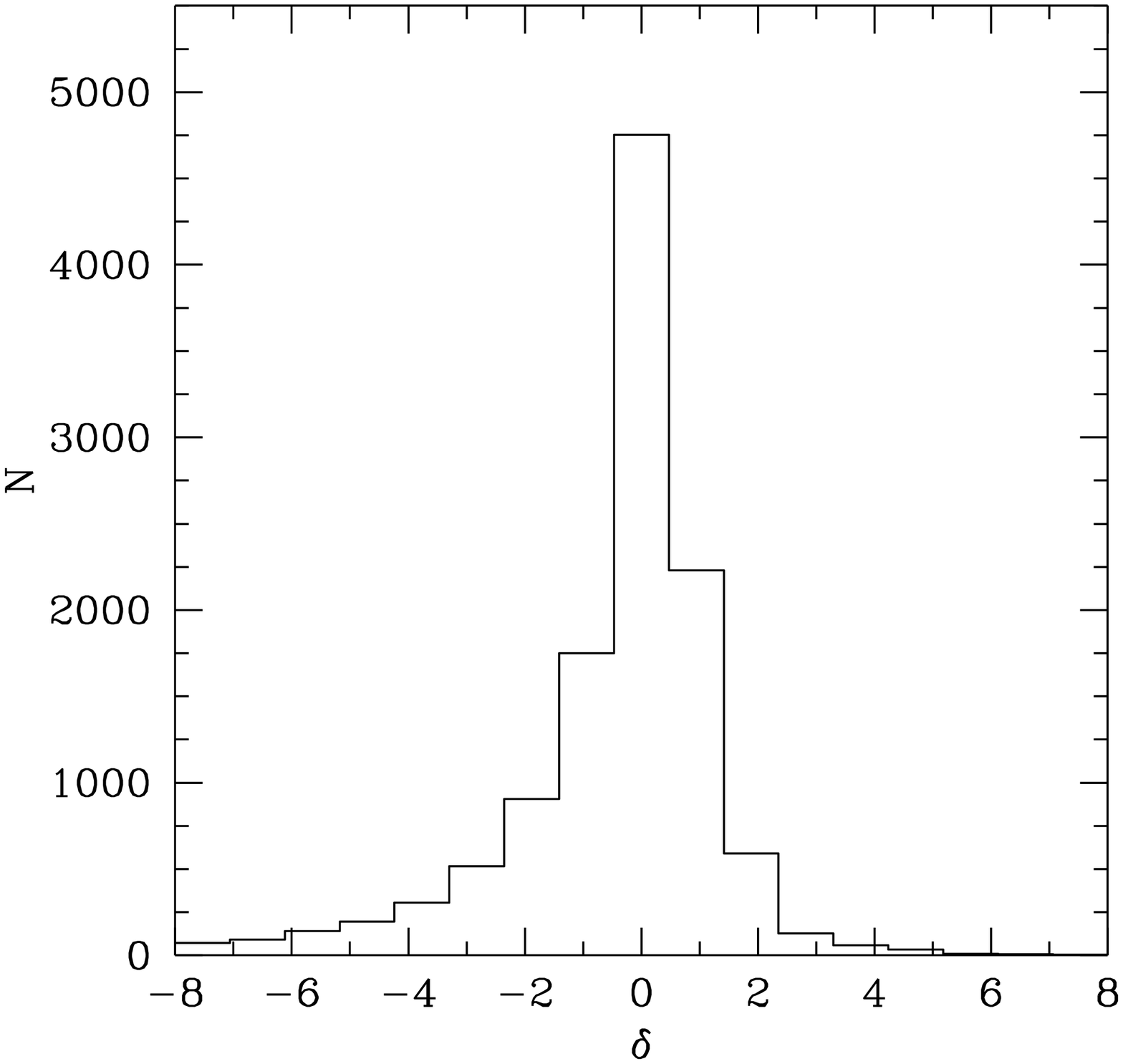,  width=0.45\textwidth }}
\caption{
Upper panel presents the distribution of $\delta$ in blue filter for
the clump sample from CGR (selected in the same way as for 
Figure~\ref{fig:blendfrac}). The lower panel shows the corresponding
histogram for the simulated unblended events.\label{fig:deltahist}} 
\end{figure}

To further test possible contamination of the clump sample by highly
blended events, we notice that the blending fractions themselves do
not tell the entire story. There is no reason to reject a
non-blending hypothesis as long as the blend fractions are consistent
with f = 1. Therefore, we compute the deviation $\delta$ of a blend fraction
from f = 1 in units of the error in the blend fraction:
\be
\delta = \frac{f-1.0}{\sigma_f}, \label{eqn:delta}
\ee
where $\delta$ assumes negative values for $f<1.0$ and positive values
for $f>1.0$.\footnote{The blend fractions with $f>1.0$ seem unphysical, but
even in the case of negligible observational errors they can result
from local fluctuations in the sky level. Such
fluctuations will typically produce $f \lsim 1.2$. However, in the
presence of noise $f>1.2$ solutions
should not be rejected as long as they are consistent with physically
motivated $f$ values. All solutions in our sample meet this basic
condition.} 
It is obvious from the definition that $\delta$ will not significantly
differ from 0 for events consistent with no blending. The events with
large deviations can be classified as suspicious.
We plot the distributions of $\delta_{\rm blue}$ for the real and simulated
samples in Figure~\ref{fig:deltahist}.
Again, the distributions are not very different from
each other, but we cannot exclude the possibility of significant
blending for some events in our original sample of 60 unique clump
events.
We conclude that, given the quality of our data,
the blend fit parameters we find for events with a single lens
cannot be used to make a definitive determination of 
the amount of blending in our sample. 
Nonetheless, since our ultimate goal is to obtain a reliable estimate
of the optical depth, we use blend fit results to select a subset of 
our sample, which we design to be fully consistent with no blending. This
extremely conservative subset consists only of the events with 
$|\delta_{\rm blue}|<2.0$. 
It is encouraging that this subsample contains very few 
events from Table~\ref{tab:lcdeviations}.
The analysis of this verification sample conducted 
in the next section will
provide an essential check on the systematic error in our primary 
optical depth determination based on the entire clump sample assumed
to be blending-free.

Finally, we notice that blending fits to caustic crossing events with 
binary lenses are
substantially better constrained than the fits to events with single
lenses. Our clump sample contains 4 binaries that have been previously
analyzed by Alcock et al.\ (2000c): 108.18951.593 (97-BLG-28),
118.18141.731 (94-BLG-4), 401.48408.649 (98-BLG-14), and
402.47864.1576 (97-BLG-41).
The first three have fits in Alcock et al.\ (2000c, Table 3),
and 97-BLG-41 has two possible solutions: binary with a planet 
suggested by Bennett et al.\ (1999) and rotating binary discussed
by Albrow et al.\ (2000). 
Out of 16 bulge binaries that have blend fits in Alcock et al.\ (2000c)
only 6 are fully consistent with no blending (94-BLG-4,
96-BLG-4, 97-BLG-1, 97-BLG-28, 98-BLG-14 ("dashed solution"), and
108-E). On the other hand, the four {\em clump} binaries for
which we have fits have blend fractions very close to 1 and are
consistent with
no blending (see their blend fractions in Table~\ref{tab:binaryblendfr}).
The probability that the clump binaries share the same blending
characteristics as the entire sample of binaries is
$P(4) \equiv {4 \choose 4} (6/16)^4 (10/16)^0 = 0.020$.
Therefore, we have a $2.3 \, \sigma$ indication that our clump sample
may be affected by blending to negligible extent.

\section{The Optical Depth}
\label{sec:optical_depth}
We propose to use an estimator of the optical depth which is different
from, but closely related to, the one we used previously.
The estimator used in our previous papers is:
\be
\label{eqn:tauthat}
\tau = \frac{\pi}{4E} \sum_{i=1}^{N_{\rm events}} \frac{\that_i}
{\epsilon(\that_i)},
\ee
where $\that$ is time to cross the Einstein ring diameter,
$E$ is the total exposure in star-days (equal to the number of stars times time span of observations), and
$\epsilon(\that)$ is an efficiency for detecting an event with a given $\that$.

A formula which gives the same answer in the limit of large numbers of
events, but which we find more computationally efficient is:
\be
\label{eqn:taute}
\tau = \frac{1}{E} \sum_{i=1}^{N_{\rm events}} 
\frac{t_i}{\epsilon(t_{i},A_{{\rm max}, i})},
\ee
where $t_i$ is the measured time for which the magnification is above
$A=1.34$, that is the time the line-of-sight is within 
the Einstein ring radius.
Appendix B gives the derivation of both of these formulae and makes the
case for the use of equation~(\ref{eqn:taute}).  Note that due to our
cut $A_{\rm max}\geq 1.5$,
we must multiply the value derived from equation~(\ref{eqn:taute}) by 
a correction factor of 1.09. This is explained in Appendix B.

There are five likely binary events in our sample. Not only is the
parameter determination more difficult in their case, but our
detection efficiency pipeline described below is designed for the 
events with a single lens. Therefore, it is not immediately clear
how to include binaries in the analysis. They cannot be simply
omitted because this would lead to an underestimate of the optical
depth.
For simplicity we used for binaries $t_i$
as found from single lens fit.  We found this quite accurately measured
the time above $A=1.34$, and agrees quite well with what is found from
the binary lens fits (Alcock et al.\ 2000c)\footnote{We notice that the
values of $\that$ found
from the single lens fits are not very useful for these events.}. 
Even if make a combined systematic error of 50\% in the estimate of the time 
above $A=1.34$ and detection efficiency for binary
events, this will result in only about 4\% error in the optical depth
because binary events constitute only 1/12 of the entire sample.  
Another problem could occur if our cuts selected against binaries
due to non-standard shapes of their light curves. Taking the
fraction of binaries for all events (selected with much more relaxed
cut) from the companion paper, we find that the expectation
number of binaries in the clump sample is six. We see five.
If we count one missing binary as a source of bias in the optical
depth, then we predict the optical depth underestimate to be at the
level of $(1/60) \times 100$\% $= 1.5$\%. Both effects are small
compared to the statistical error in the optical depth.

\begin{figure*}[t]
\subfigure[]{\epsfig{figure=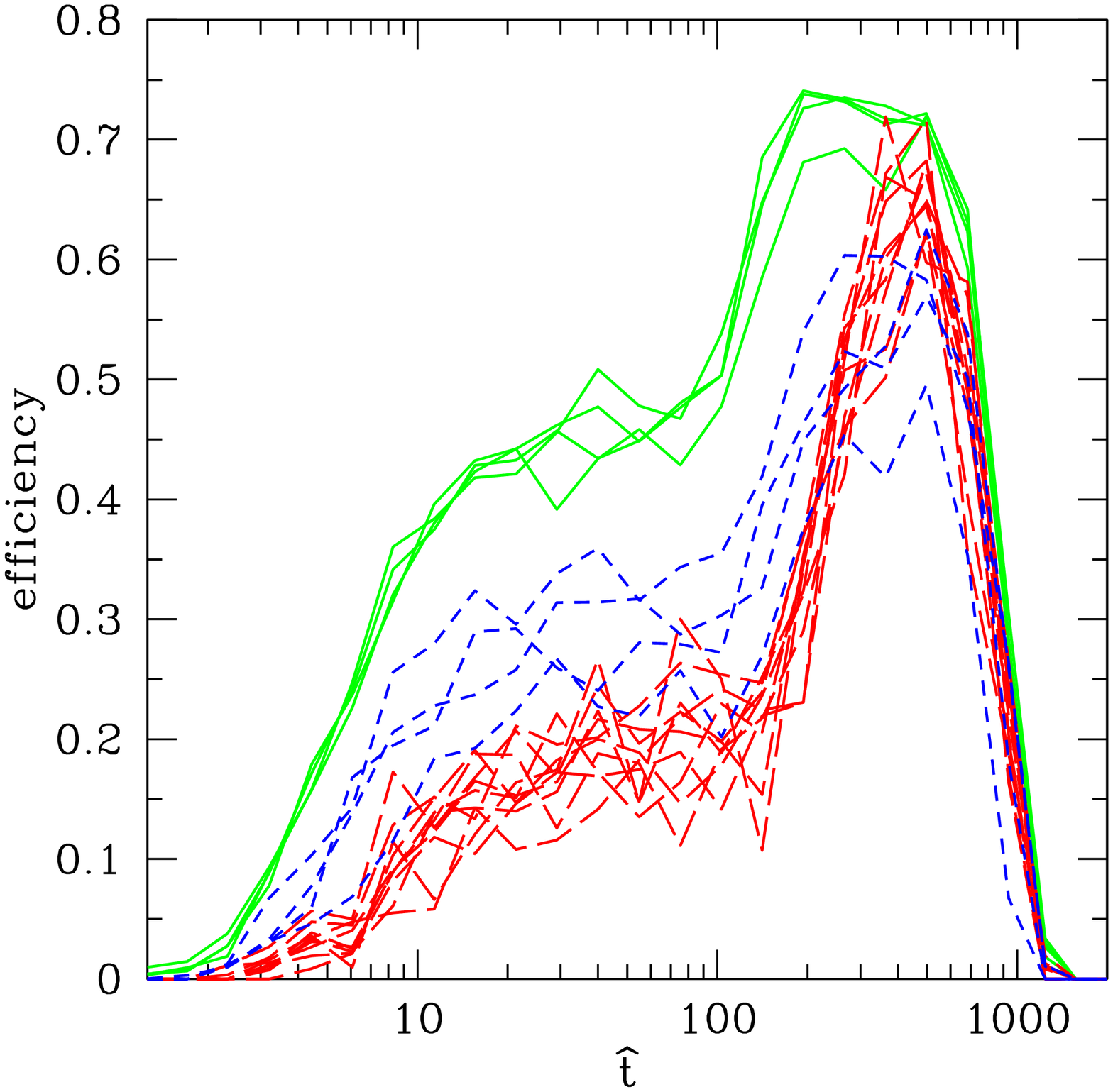, width=0.47\textwidth}}
\hfill
\subfigure[]{\epsfig{figure=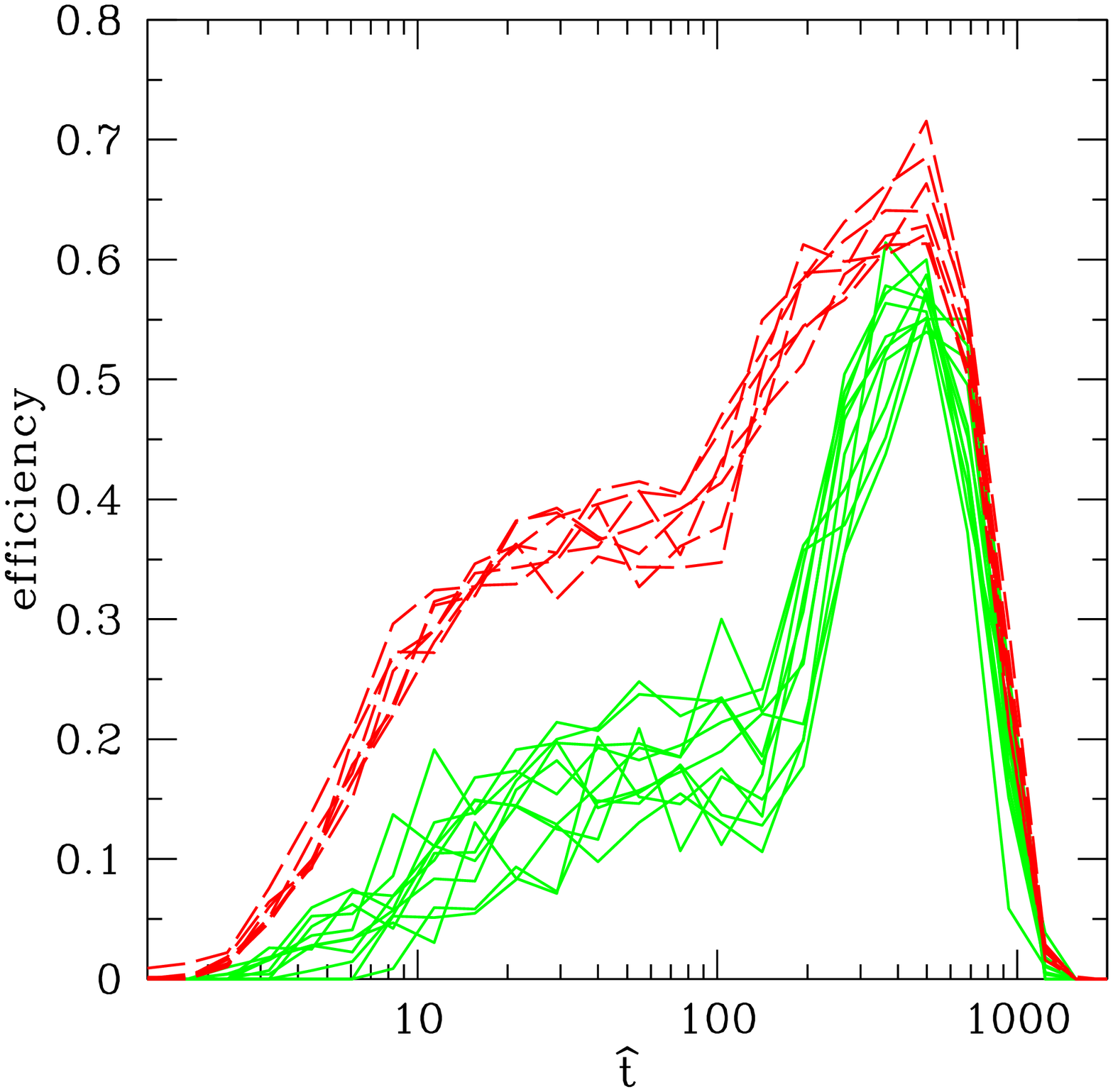, width=0.47\textwidth}}
\hfill
\subfigure[]{\epsfig{figure=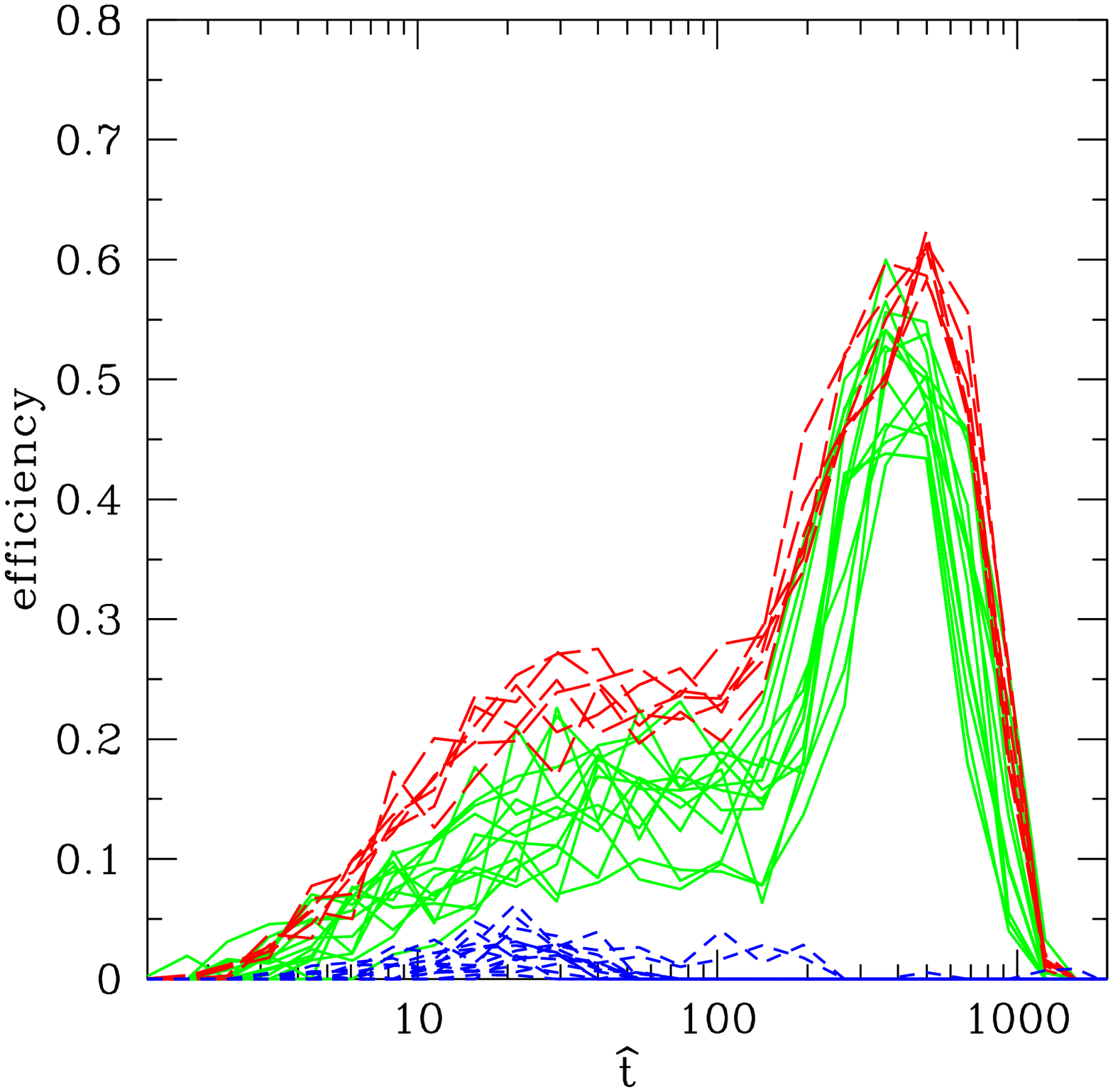, width=0.47\textwidth}}
\hfill
\subfigure[]{\epsfig{figure=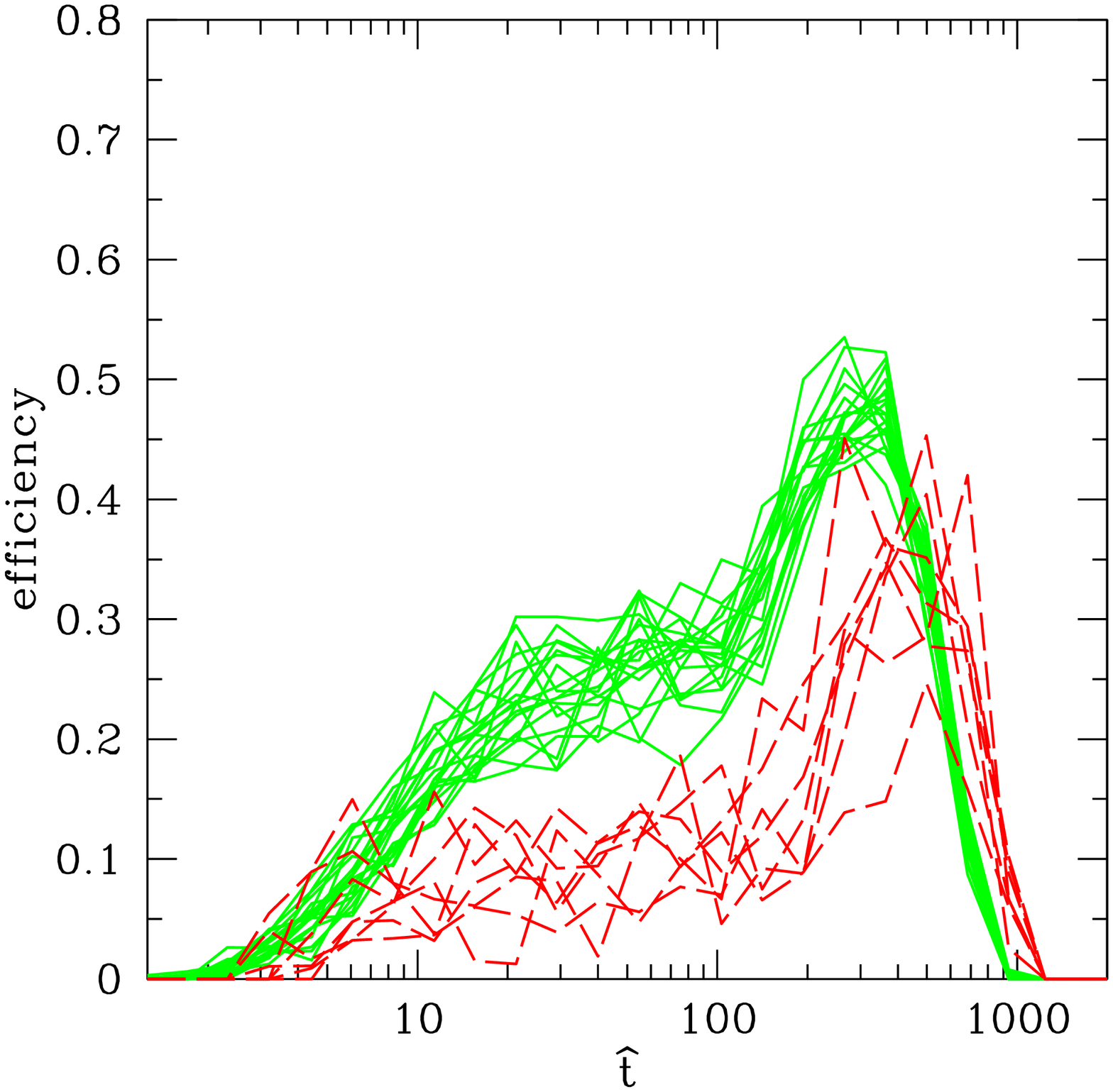, width=0.47\textwidth}}
\caption{Clump detection efficiencies versus $\that$ for all fields. {\bf (a)} solid green - 108, 113, 118, 119; red long dashed -  116, 124, 125, 136, 142, 143, 146, 148, 149, 155; blue short dashed - 110, 159, 161, 162  
{\bf (b)} solid green - 111, 122, 131, 132, 137, 150, 152, 158, 163, 171, 174; red dashed - 101, 104, 105, 109, 114, 120, 128  
{\bf (c)} solid green -  127, 133, 138, 144, 147, 151, 153, 154, 156,  157, 160, 166, 168; red long dashed - 102, 103, 115, 121, 167; blue short dashed - 106, 107, 112, 117, 123, 126, 129, 130, 135, 141, 164, 165, 169, 170  
{\bf (d)} solid green - 176-180, 301-311, 401-403; red dashed - 134, 139, 140, 145, 172, 173, 175.  Note that fields 176-180 and fields 401-403 efficiencies are low because they were not observed for the whole 7 year period.  As explained earlier, fields 301-311 are not included in the analysis but are included here for completeness. 
\label{fig:eff}}
\end{figure*}

The sampling efficiencies were obtained with the pipeline that has been
previously applied to the LMC data (for a description see 
Alcock et al.\ 2000b). 
In brief, a random subset of 1\% of all lightcurves was selected and
artificial microlensing light curves with different parameters were 
added to these lightcurves.  Then the same analysis used
to select real events was applied to this set. 
Efficiencies were computed as the number of events recovered divided by
the number of events input as a function of both the input duration
and the input maximum magnification of the input event.
Since the number of clump giants in each 1\% sample field is not large,
several passes were made through each 1\% data base, until enough statistics
were gathered.  We averaged over $\amax$ to report efficiencies for each field 
as a function the event duration ($\that$), and we also did special runs 
using the measured $t_i$ and $A_{{\rm max}, i}$ for 
each real clump giant microlensing event.
This last analysis gave us $\epsilon(t_{i},A_{{\rm max}, i})$ for use with $t_i$ in 
equation~(\ref{eqn:taute}) above.
Figure~\ref{fig:eff} shows the results of our efficiency calculations
for the fields with events, and Table~\ref{tab:taubyevent} shows 
the efficiency calculated for each event.  Relative sampling efficiencies
for all 94 bulge fields are given in Table~\ref{tab:fielddat}.  The results in fields in the 300 series are not used in the optical depth analysis and they may be affected to higher extent by systematic effects.

Using the above method and the clump events from Table~\ref{tab:taubyevent},
we present in Table~\ref{tab:taubyfield} optical depths for each field 
that contains clump giant events. The center position
of each field, the number of clump giant source stars (in the text
designated shortly by $N_f$), and the
number of recovered microlensing events are also given.  The number of days,
$T_{\rm obs}=2530$, used to find each field's exposure $E=N_f T_{\rm obs}$ was
set by the efficiency calculation\footnote{The number of days
$T_{\rm obs}$ is equal to the sum of the allowed $t_0$ span ($2850d-419d =
2431d$) and two $\sim
50$-day buffer regions added at the beginning and the end of the
observing interval.}.  Since
the number of events in each field is so small the error in the
optical depth is dominated by Poisson statistics. We compute these
errors using the formula given by Han \& Gould (1995).

\begin{figure*}[th]
\center
\subfigure[]{\epsfig{figure=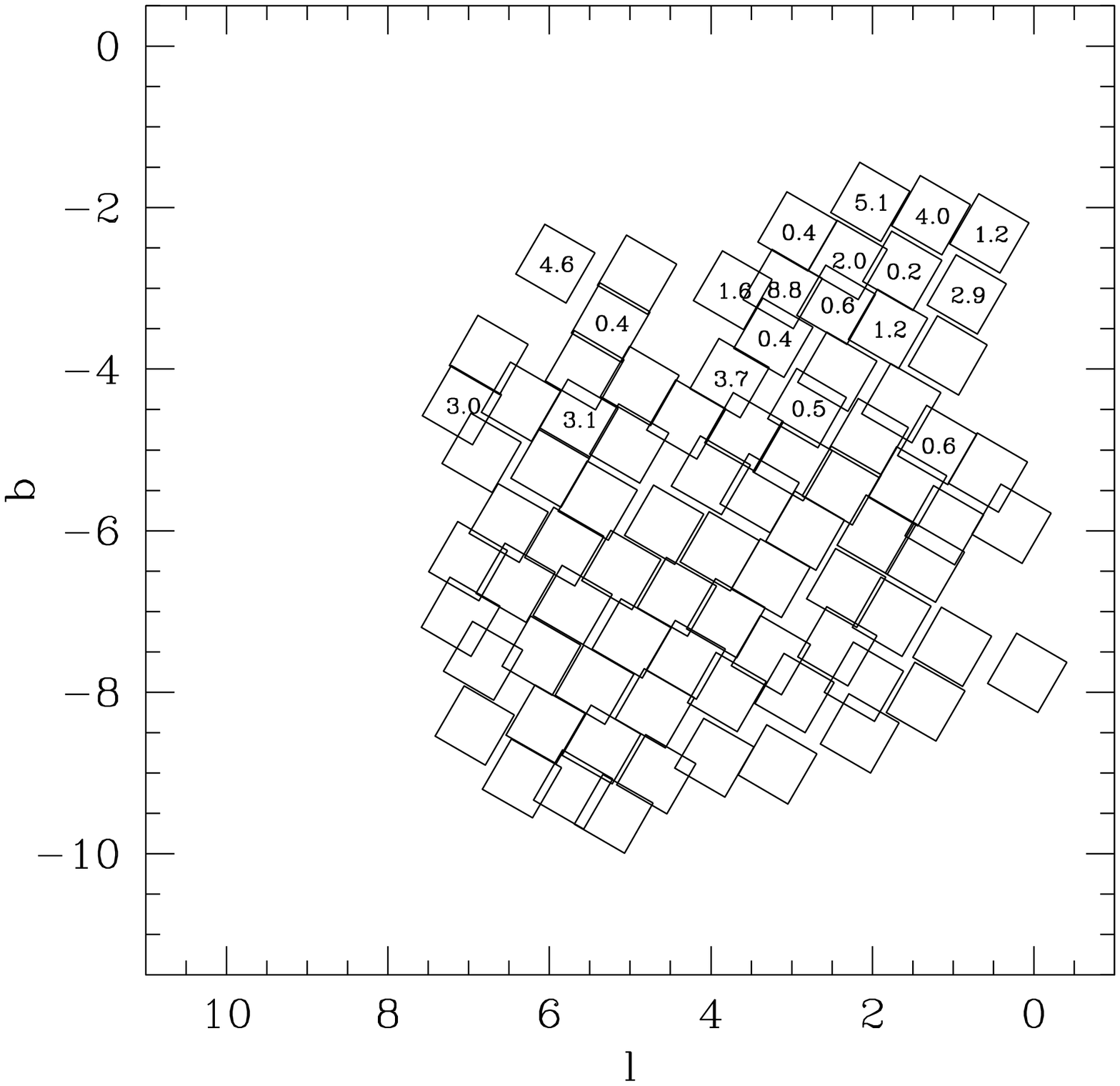, width=0.48\textwidth}} 
\subfigure[]{\epsfig{figure=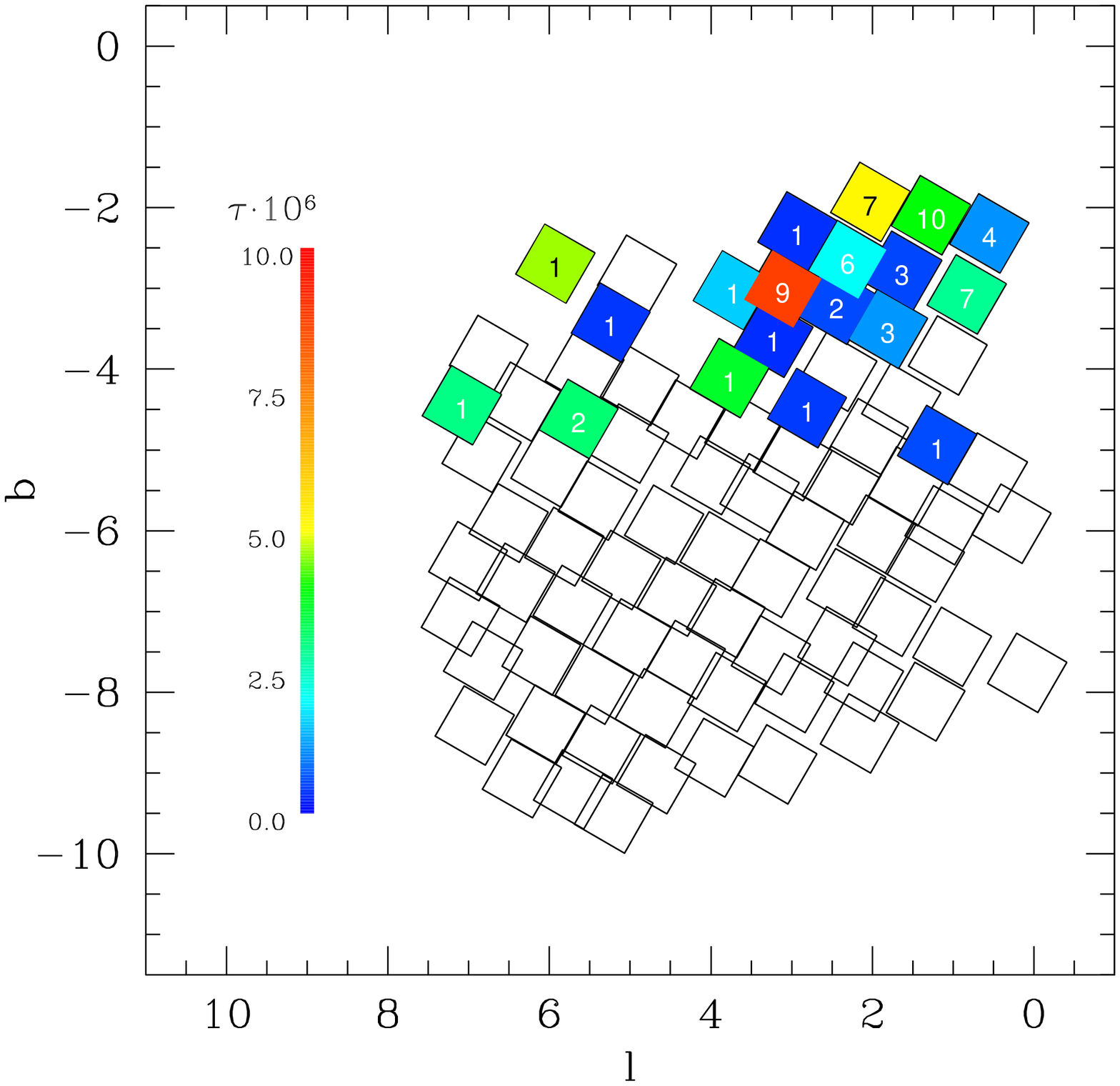, width=0.48\textwidth}} 
\caption{(a) Microlensing optical depth ($\times 10^6$) toward each
field that contains a clump event. (b) Optical depth in each field
with a detected clump event and the number of events in that
field. Only analyzed fields are shown in both panels.}
\label{fig:fieldtau}
\end{figure*}

\begin{figure}[t]
\center
\epsfig{figure=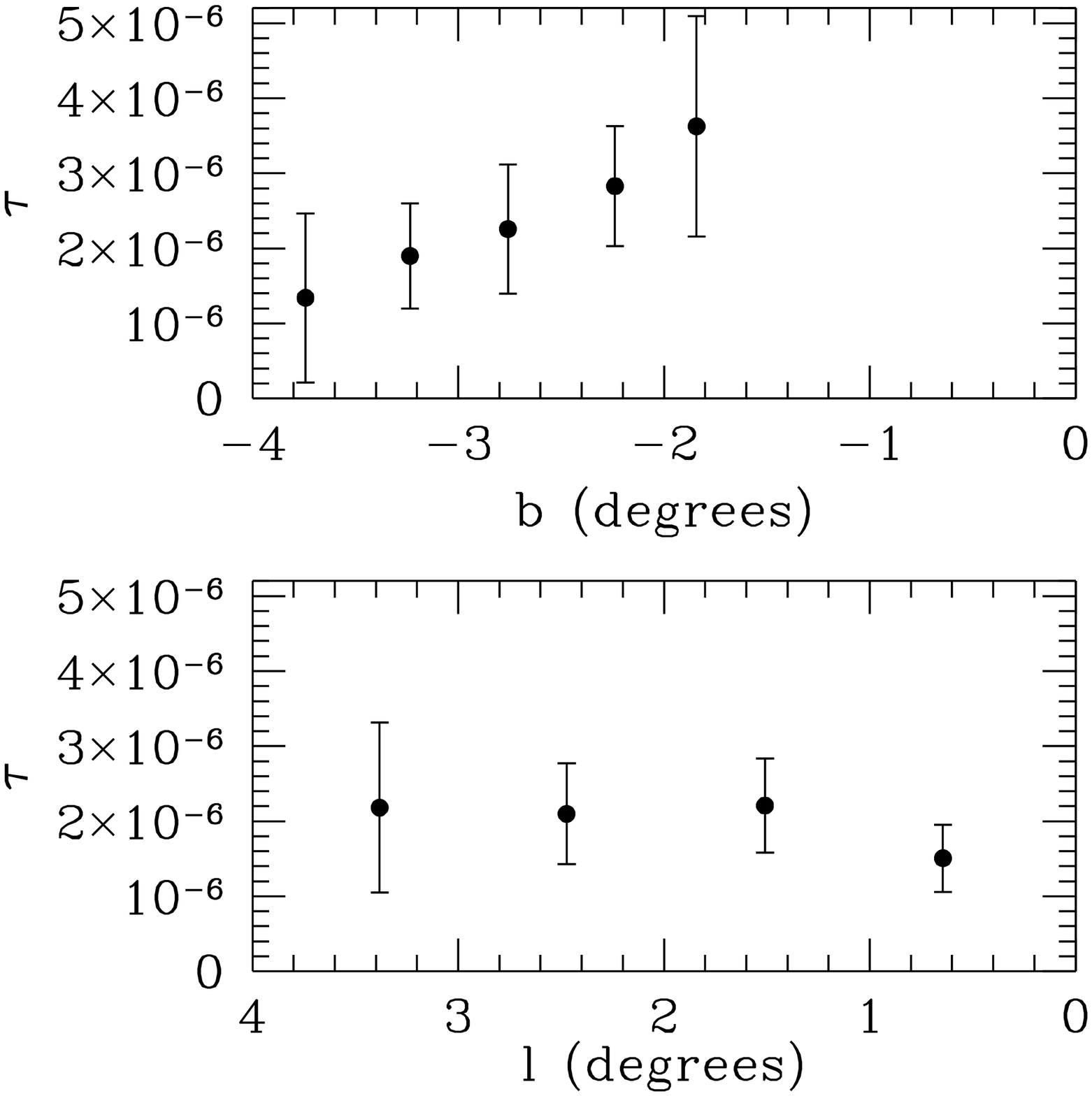, width=0.48\textwidth} 
\caption{Average optical depth in latitude and longitude
strips for events within $5.5^{\circ}$ from the Galactic center. 
Gradient in $b$ is much steeper than the one in $l$.\label{fig:taugrad}} 
\end{figure}

We show in Figure~\ref{fig:fieldtau} the spatial distribution  
of optical depths on the sky.
In the left panel we give the optical depth value at the center of
each field with clump events, and in the right panel we present the
color representation of the same results and indicate the number of 
detected events.
We note the anomalous character of field 104, with 
an optical depth of $(8.8 \pm 3.7) \times 10^{-6}$,
almost two sigma above the average.
The results for individual fields are very uncertain but taken as an
ensemble they suggest that the 
optical depth has a substantial gradient in Galactic latitude. 
This property is more apparent
in Figure~\ref{fig:taugrad}, where we binned the data on a 0.5 deg
scale in the $b$-direction and 1.0 deg scale in the $l$-direction.
We then fit straight lines to these binned data.
Based on 57 clump giant events selected from observations of
the 1260000 clump giant source stars within 5.5 degrees of the 
Galactic center we find the optical depth gradients of 
$(1.06 \pm 0.71) \times 10^{-6} {\rm deg}^{-1}$ in the 
$b$-direction and $(0.29 \pm 0.43) \times 10^{-6} {\rm deg}^{-1}$ in the
$l$-direction. It is clear from the errors that the gradient in
the $l$-direction is consistent with 0 and with different small
values close to 0 (both positive and negative).  

Many fields do not contain any clump events. In those cases we derive
an upper limit on the optical depth from
\be
\tau_f^{\rm lim, CL} = \nu^{\rm CL} \left( \frac{\pi}{4 N_f T_{\rm obs}}
\frac{\sum_{i=1}^{N_{\rm events}}
{\hat{t}_i}/{\epsilon_i(\hat{t}_i)}}{\sum_{i=1}^{N_{\rm events}}
{\epsilon_f(\hat{t}_i)}/{\epsilon_{i}(\hat{t}_i)}} \right),                    \label{eqn:upperlim3}
\ee
where $\epsilon_f$ is the efficiency in field $f$, $\epsilon_{i}$
is the efficiency for $i$-th event in its field of origin,
and $\nu^{\rm CL}$ is the multiplication factor dependent
on the confidence level.
Equation~(\ref{eqn:upperlim3}) is derived in Appendix C.
We use $\nu^{1 \sigma} = 1.8379$
when we compute the upper limits on the optical depth given in 
Table~\ref{tab:upperlim}.
Those upper limits were computed without any direct reference to
the Galactic model. However, they rely on the assumption that the 
intrinsic duration
distribution in empty-fields is identical to the one in fields with
events, and this approximation can break down for some Galactic models.
In addition, let us note that the limits from Table~\ref{tab:upperlim}
are rather conservative in the sense that the true values may be
lower. This is caused by the fact that we used durations of all
events when we applied equation~(\ref{eqn:upperlim3}). We show in the
next section that field 104 tends to have longer events. If events in
field 104 were excluded from the summation in
equation~(\ref{eqn:upperlim3}), then the limits would go down a little.

Since the number of events in individual fields is typically very
small, microlensing studies customarily give average optical depth
based on many fields to beat the Poisson noise. The main problem
with this approach is the fact that the interpretation of the
average optical depth and position is difficult due to 
non-linear variations of $\tau$ over the area of the survey.
Decreasing the size of the region for averaging results in
reappearance of the Poisson noise issue. Here we deal with these
competing effects by selecting a very compact region that contains
2/3 of all clump events. We call this region Central Galactic Region
(CGR) to stress its close proximity to the Galactic center (but we
would like to clarify that there are no MACHO fields that include
the Galactic center itself). CGR covers about 4.5 sq.\ degrees and
is composed of 9 fields: 108, 109, 113, 114, 118, 119, 401, 402,
403. Using 42 events in CGR, we find an average optical depth of 
$\tau_{\rm CGR} = 2.17^{+0.47}_{-0.38} \times 10^{-6}$ 
at $(l,b) = (1 \hbox{$.\!\!^\circ$} 50, -2 \hbox{$.\!\!^\circ$}68)$
\footnote{The unweighted average position for 9 CGR fields is
$(l,b) = (1 \hbox{$.\!\!^\circ$} 55, -2 \hbox{$.\!\!^\circ$} 82)$}.  
For this set of fields, an exposure is $E= 1.869 \ten{9}$ star-days 
(2530 days times 738850 clump giant stars).  
In this case the error was found by Monte Carlo simulations as described by
Alcock et al.\ (1997b), and is somewhat larger than the simple Poisson error 
(which would be $\pm 0.34 \ten{-6}$), but consistent with the 
formula of Han and Gould (1995), which gives $\pm 0.42 \ten{-6}$.  
We note that the result found using equation~(\ref{eqn:tauthat}) and an 
interpolated efficiency as a function of $\that$ is nearly identical
($2.16^{+0.46}_{-0.38} \times 10^{-6}$). 
Field 104, with the largest
optical depth, falls just outside of the CGR. To test how
sensitive the results are to the detailed shape of the region selected
for the determination of the average optical depth,
we create a rectangular region ``CGR+3'', which is composed of CGR
plus the three adjacent fields at the same galactic latitudes 
(fields: 176, 104, 105). The optical depth in such an extended region 
is $\tau_{\rm CGR+3} = 2.37^{+0.47}_{-0.39} \times 10^{-6}$ 
at $(l,b) = (1 \hbox{$.\!\!^\circ$} 84, -2 \hbox{$.\!\!^\circ$}73)$, 
entirely consistent with our principal result.
Also, we note for completeness
that when we use all 2.44 million clump giants and all 62 events the optical 
depth over all 83 analyzed fields is: 
$\tau_{\rm all} = (1.21 \pm 0.21) \ten{-6}$ at $(l,b) = (3
\hbox{$.\!\!^\circ$} 18, -4 \hbox{$.\!\!^\circ$} 30)$, but this number 
is hard to interpret given non-linear (and often rapid) variations 
of the optical depth with Galactic position
mentioned earlier.

Finally, we return to the question of blending and its effect on
the optical depth. To test the sensitivity of the optical depth
to possible systematic errors, we analyze the verification sample
of events. To avoid possible systematic errors we first exclude
all binaries. Then we keep only the events that have
$|\delta_{\rm blue}| < 2.0$, which makes them very consistent with 
no blending. We list the events that pass this additional cut in 
Table~{\ref{tab:verificationevents}}. We list events
that belong to CGR on the left and the events in the
other fields on the right. 

The analysis on this sample proceeds in a fashion that is similar
to the main analysis but has a noticeable difference. Our verification
sample is constructed by utilizing the cut that uses blend fit 
results. Applying the same procedure to our massive efficiency
simulations is beyond the scope of this paper. Therefore, we
use the previously determined efficiencies and the following
estimator of the optical depth:
\be
\tau^{\rm ver} = \frac{N^{\rm orig}_{\rm events}}{N^{\rm ver}_{\rm
events}} \left( \frac{1}{E} \sum_{i=1}^{N^{\rm ver}_{\rm events}} 
\frac{t_i}{\epsilon(t_{i},A_{{\rm max}, i})}\right), 
\label{eqn:tauver}
\ee
where $N^{\rm orig}_{\rm events}$ and $N^{\rm ver}_{\rm
events}$ are the numbers of events in a given field or region
counted in the original and verification sample, respectively.
In designing the optical depth estimator given in equation
~(\ref{eqn:tauver})
we assumed that blending (or other systematic effect we would like to
test) does not change the number of detected
events, but it may substantially affect the event parameters and
may render the sampling efficiencies inappropriate. However, both
the parameters and efficiencies can still be used for the verification
sample. The factor ${N^{\rm orig}_{\rm events}}/{N^{\rm ver}_{\rm
events}}$ is introduced to compensate for the fact that we
remove some events without replacing their optical
depth contributions with anything else.

The optical depth values in individual fields derived using
equation~({\ref{eqn:tauver}}) for our verification sample are given
in Table~{\ref{tab:vertau}}. The first four columns give the field
number, ${N^{\rm ver}_{\rm events}}$, the optical depth derived
using equation~({\ref{eqn:tauver}}), and its error. For comparison, 
columns 5 and 6 give ${N^{\rm orig}_{\rm events}}$ and the optical
depth based on equation~({\ref{eqn:taute}}) derived earlier.
The results from the verification sample agree very well with our previous
estimates. As the last step we limit the verification sample to the
CGR\footnote{It may seem surprising that only 59\% or 22 events out of
37 non-binary events in the CGR pass $|\delta_{\rm blue}| < 2.0$
cut. We note however that the
corresponding ratio for our simulated {\em unblended} sample from the lower
panel of Figure~\ref{fig:deltahist} is 80\%.  This difference
is at least partly explained
by the fact that the Monte Carlo simulations do not include exotic
events (e.g., parallax events) which are present in the real sample and
can cause the blend fits to falsely indicate blending.}.
The average optical depth based on this verification sample of
22 events is $\tau^{\rm ver}_{\rm CGR} = 2.42^{+0.75}_{-0.58} \times 10^{-6}$, 
in excellent agreement with our main result.

In summary, we see no evidence that our optical depth results are
strongly affected by blending. There are two likely explanations
of why this is the case. First, it is possible that the clump sample
is to a large extent unaffected by blending. Second, substantial blending
is a result of a large number of unresolved sources. In such
situation, the bias in the recovered event duration is countered
by the bias in the inferred number of sources.
At the moment, we do not have enough information to distinguish
between these two possibilities, but our optical depth determination
benefits from their fortunate properties.

\section{Clustering of Events and Concentration of Long Events in Field 104}
\label{sec:clustering}
The optical depth for field 104 is larger than any other field
and we would like to investigate whether this is a statistical
fluctuation or an indication of something unusual about this region
of the Galaxy.
The large optical depth is related to the fact that
9 of the 62 clump giant events are in field 104,
and because there is also a high concentration of
long-duration events in this field.
Four out of the 10
events longer than $\that = 100$ days are in field 104, including the
2 of the 3 longest.  

We investigate how statistically significant this concentration is.
We do not account for the change in the detection efficiency of
events with different durations in different fields, so 
we place only a lower limit on significance of the difference between
duration distributions in field 104 and the others. The
efficiency for detecting long events is similar in most fields
with events observed for 7 years, because this does not depend strongly 
on the sampling pattern. The detection of short events 
will be lower in a sparsely sampled fields. Therefore,
the number of short events in some of the fields used for comparison may be
relatively too small with respect to a frequently-sampled field 104, but this 
is only going to lower the significance of the $\hat{t}$ distribution 
difference computed here.
In conclusion, the analysis of event durations {\em uncorrected} for 
efficiencies should provide a lower limit on the difference between field 
104 and all the remaining clump giant fields.

For the significance test, we use the Wilcoxon's number-of-element-inversions 
statistic. 
First, we divide events into two populations: events in field 104 and all
the remaining ones.
We test two types of possible separations into those two populations:
A. we select all events in field 104 to the first sample (9 events) and all
the remaining events except duplicates from field 109 to the second
sample (51 events),
B. we select all events in fields 104 and not in any other field to
the first sample (7 events) and all the events in the remaining fields
the second sample (53 events).
In both cases the entire sample of 60 unique clump giants can be
recovered as the sum of populations in 104 and not in 104.

Case A and B are then each tested separately according to the
following procedure.
We order the events in the combined sample of two populations from the 
shortest to the longest. Then we count
how many times one would have to exchange the events from field 104 with
the others to have all the 104 field events at the beginning of the list. 
If $N_1$ and $N_2$ designate numbers
of elements in the first and second sample, respectively, then for $
N_1 \geq 4$, $N_2 
\geq 4$, and $(N_1+N_2) \geq 20$, the Wilcoxon's statistic is approximately
Gaussian distributed with an average of $N_1 N_2/2$ and a dispersion $\sigma$
of $\sqrt{N_1 \, N_2 \, (N_1+N_2+1) / 12}$.

In case A, the Wilcoxon's statistic is equal to 324,
whereas the expected number is 230 with an error of about 48. Therefore
the events in 104 differ (are longer) by $1.96 \sigma$ from the other fields.
In case B, the Wilcoxon's statistic is equal to 282,
whereas the expected number is 186 with an error of about 43. Therefore
the events in 104 differ (are longer) by $2.22 \sigma$ from the other fields.
Both divisions of the clump sample into field 104 and the remaining
fields lead to the conclusion that events in fields 104 are
inconsistent at the 2 $\sigma$ level with being drawn from the same 
parent population as events in the remaining fields. 
We obtain similar discrepancy between the time scales of events 
in field 104 and the remaining fields when we use all 318 unique 
events from selection criteria c listed in Table 3 of the companion
paper.
This result is somewhat less significant than the one derived by 
Popowski et al.\ (2001a). We note, however, that similar analysis
of an almost independent Alcock et al.\ (2000a) DIA sample also
suggests anomalous character of duration distribution in field
104 (Popowski 2002).
We conclude that the unusual character of field 104 should be
investigated with additional observations but 
we cannot completely exclude the possibility that this effect is due to a 
statistical fluctuation.

In addition, there seem to be clusters of events (of all durations) on the sky
in fields 104, 108, and 402.  If microlensing events are clustered on the sky above random chance 
it has important consequences.  It could indicate clustering of lenses,
perhaps in some bound Galactic substructure or very young star forming region.  The long duration of
the events could indicate that the substructure contained heavier
objects such as black holes.  Alternatively, it could indicate
a concentration of orbits through the Milky Way and a place where the
orbital speeds were slower than normal.

In order to test the significance of the apparent clustering of microlensing
events we devised a Monte Carlo test in which we simulated 10000 microlensing
experiments each detecting 60 unique clump giant events.
In each experiment we found densest cluster of 3 events and the densest
cluster of 4 events and recorded the diameter of the cluster.
We then compared the diameters of the densest 3-cluster and densest 4-cluster
in the actual data to the density distributions formed by our 
Monte Carlo experiments.  This gives us the probability of finding by
chance a 3-cluster (or 4-cluster) as small as we found in the actual data. 
In both the data and in the simulations we remove any star that is found two
or more times (e.g., in an overlap region), since these give a false measure 
of dense clustering.

In performing the simulations we selected 60 clump giant stars at random from
the 1 percent data base so that the spatial 
distribution of lensing sources matched
that in the data.  We also weighted each event by the 50-day efficiency
of that field (Table~\ref{tab:fielddat}), 
so that fields with low sampling efficiencies were
properly under-represented.  Finally we considered various gradients in
optical depth across our fields. As one limit we use a uniform optical 
depth across our fields
and as the other we use a steep linear gradient which results in
the optical depth changing from a maximum at
the Galactic center to 0 at $|b|=4^\circ$.  This later case, is roughly 
consistent
with, but somewhat steeper than, the gradient shown in Figure~{\ref{fig:taugrad}}.
The actual gradient should be between these two extremes.
The second case increases the chance of randomly finding a cluster
with respect to the first case.

With no optical depth gradient we find a probability of 1.9\% of finding
a 3-cluster as dense as in the data, and 4.6\% of finding a 4-cluster this
dense.  For the steep optical depth gradient, we find probabilities of
8.7\% and 27\% respectively.  Unfortunately these results are not
very conclusive.  The clustering of microlensing events we find in our
data is marginally significant, but certainly not compelling.
We note for completeness that the densest clump 3-cluster in real
sample had a diameter of 0.0453 degrees and the densest 4-cluster 
had a diameter of 0.11 degrees.

We also performed the clustering analysis on the entire sample of 318 unique
microlensing events selected with criteria ``c".
Here the densest 3- and 4-clusters in real data had the diameters
of 0.0272 and 0.0497 degrees, respectively. Note that for random
surface distribution of points the size of the
smallest cluster scales with the number of events $N$ as $N^{-1/2}$.
We conclude that the above diameters are very consistent with the clump only
result since they produce rescaled diameters of 0.0626 and 0.114 for 3- and
4-clusters respectively when multiplied by $\sqrt{318/60} = 2.302$.
For the entire sample of 318  unique event and for uniform optical
depth, the probability of finding a 3-cluster as dense
as in the data is 7.3\% and the probability of finding a 4-cluster this
dense is 3.6\%.  For the steep optical depth gradient the probabilities
are 36\%  and 32\% respectively.  
So for the entire sample of microlensing events
we also find no strong evidence of clustering on the sky.

\section{Conclusions and Discussion}
\label{sec:discussion}

Using 7 years of MACHO survey data, we presented a new determination of
the optical depth to microlensing towards the Galactic bulge.  We selected the
sample of 62 microlensing events (60 unique) on clump giant sources
(out of more than 500 total) and performed a detailed efficiency
analysis. Using a
subsample of 42 clump events detected during monitoring of about
739000 clump giant sources concentrated in just 4.5 deg$^2$, we found
$\tau =  2.17^{+0.47}_{-0.38} \ten{-6}$ at $(l,b) = 
(1 \hbox{$.\!\!^\circ$} 50, -2 \hbox{$.\!\!^\circ$} 68)$,
consistent with theoretical expectations.

We can make a comparison with previous MACHO collaboration measurements
by using our current clump sample in the fields analyzed previously.
For example in the 24 fields used by Alcock et al.\ (1997a), 
we find 37 clump events
giving $\tau = 1.36^{+0.35}_{-0.28} \ten{-6}$,
which is two sigma below the earlier result of 
$\tau = 3.9^{+1.8}_{-1.2} \ten{-6}$.
Over the 8 fields in
our difference imaging study (Alcock et al.\ 2000a), 
our current, almost independent sample 
gives $\tau = 2.20^{+0.67}_{-0.52} \ten{-6}$ at
$(l,b) = (2 \fdg 68,-3 \fdg 35)$,
which is in very good agreement with that earlier result of
$\tau = 2.43 ^{+0.38}_{-0.38} \ten{-6}$ (the agreement would be 
somewhat worse if we introduced the fudge factor discussed in
the Introduction). 

Our new value of
$\tau = 2.17^{+0.47}_{-0.38} \ten{-6}$ 
is also very consistent with the recent
EROS collaboration (Afonso et al.\ 2003) measurement. They found 
$\tau = (0.94 \pm 0.29) \ten{-6}$ 
at $(l,b) = (2\hbox{$.\!\!^\circ$}5, -4\hbox{$.\!\!^\circ$}0)$
from 16 clump giant events and an efficiency calculation similar to
ours.
The values are different but they are reported at different position.
The difference becomes entirely insignificant if we take
into account either the numerical value of the gradient 
reported in \S 5 or read off the approximate optical depth value at 
$b=-4.0$ from Figure~{\ref{fig:taugrad}}.
In addition, our new optical depth is in very good agreement with 
recent theoretical prediction of the Galactic bulge optical depth 
($\tau = 1.63 \ten{-6}$) by Han \& Gould (2003).

\begin{figure*}[t]
\subfigure{\epsfig{figure=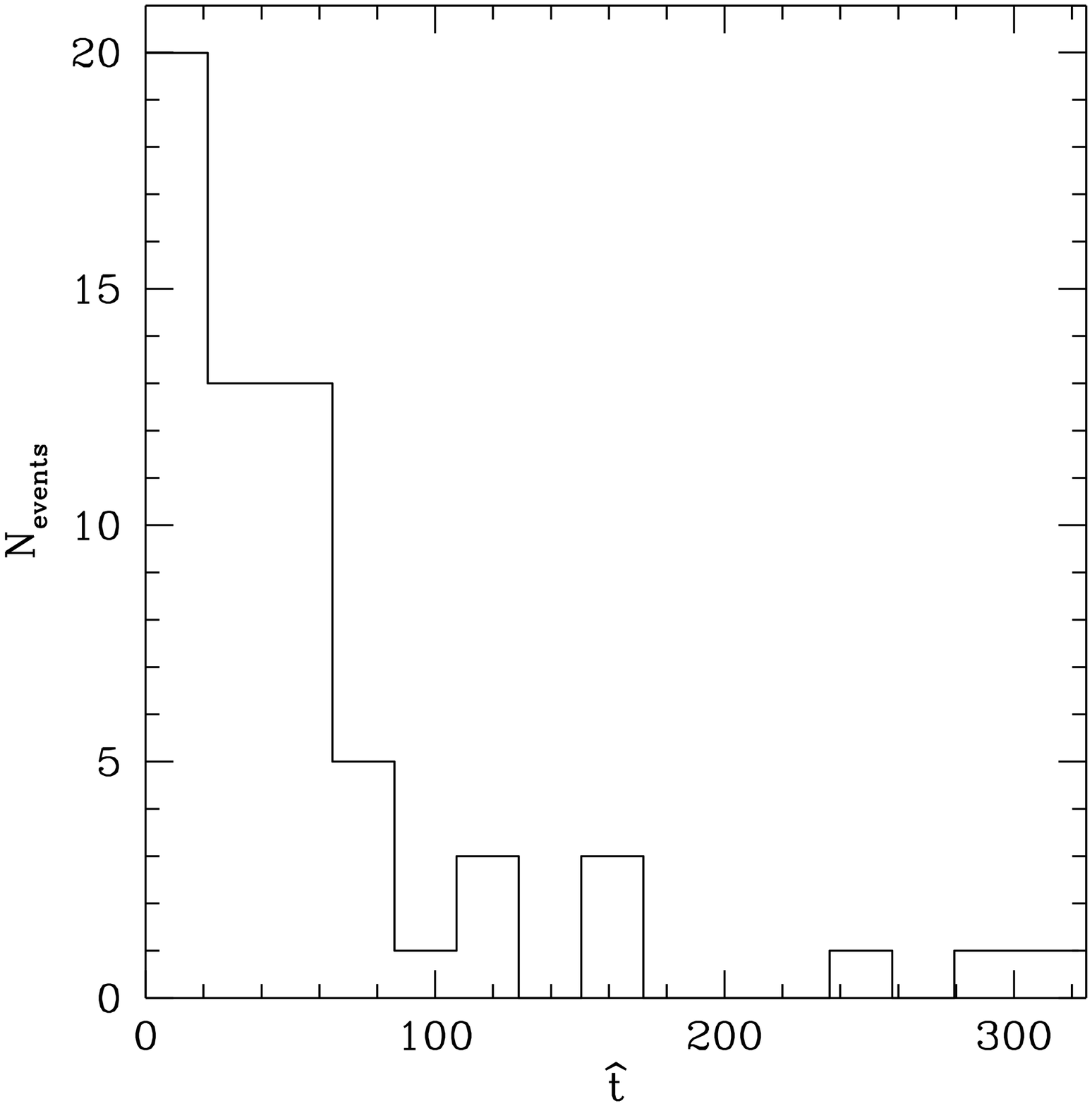, width = 0.48\textwidth }}
\subfigure{\epsfig{figure=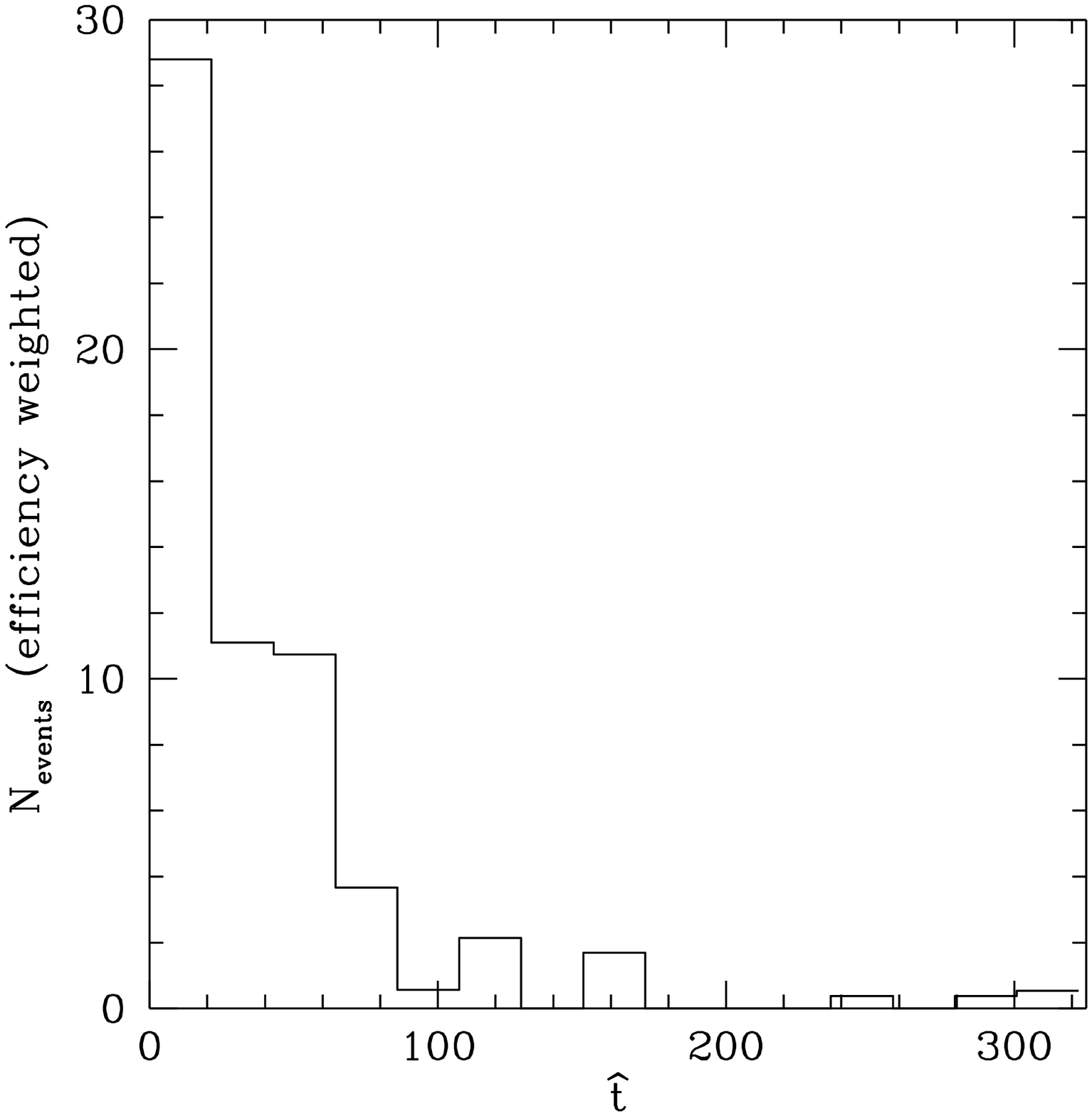,  width=0.48\textwidth }}
\caption{Left panel shows the histogram of the distribution of Einstein 
crossing times. The right panel shows the same distribution with each event weighted by its inverse efficiency and renormalized. \label{fig:thathist}}
\end{figure*}

We note that about 41\% of the optical depth is in the events longer than 
100 days (10 out of 62 events), and about 63\% of the optical 
depth is in events longer than 50 days (21 out of 62 events). 
This may be at odds with some models of 
the Galactic structure and kinematics (e.g., Han \& Gould 1996),
but consistent with others (Evans \& Belokurov 2002) which include streaming
motion along the bar.   
The distributions of event durations are shown in Figure~\ref{fig:thathist}
where we plot histograms for uncorrected and efficiency-corrected cases. 
We note that the mean Einstein diameter crossing time of
all our clump giant events is $\VEV{\that} = 56 \pm 64$ days 
and for our fields close to the Galactic center
$\VEV{\that}_{CGR} = 39 \pm 31$ days.  
To facilitate proper comparison with theoretical models we have to
weigh each event by its inverse efficiency (Table~\ref{tab:taubyevent}).
Then the averages become:
$\VEV{\that}({\rm eff}) = 40 \pm 50$ days for all fields and 
$\VEV{\that}_{CGR}({\rm eff}) = 30 \pm 29$ days for fields in the 
CGR\footnote{Because the distribution is non-Gaussian we also give 
the median and quartiles, which are (23.7, 12.0, 47.8) for all
clump events and (18.7, 9.5, 46.9) for clump events in the CGR.}.
Evans \& Belokurov (2002) find average Einstein crossing times
range from around 100 days in models that include bar streaming motion
to 30 days in models without bar streaming.
Our values are too uncertain to distinguish between these kinematic
possibilities and are consistent with both types of models.

{\bf Note added.} After this paper was completed, Sumi et al.\ (2005) 
presented 
determination of the microlensing optical depth toward the Galactic bulge
based on OGLE-II data. Using a sample of 32 events from the extended red
clump giant region of the color-magnitude diagram, they obtained
$\tau = 2.37^{+0.53}_{-0.43} \times 10^{-6}$ at 
$(l,b)=(1 \hbox{$.\!\!^\circ$} 16, -2 \hbox{$.\!\!^\circ$} 75)$.
Their optical depth is fully consistent with our value
of $\tau =  2.17^{+0.47}_{-0.38} \ten{-6}$ at $(l,b) = 
(1 \hbox{$.\!\!^\circ$} 50, -2 \hbox{$.\!\!^\circ$} 68)$
which is based on 42 clump giant events.

\acknowledgments
We thank the referee, Przemys{\l}aw Wo\'{z}niak, for many valuable
comments and suggestions.
KG and CT were supported in part by the DoE under grant
DEFG0390ER40546.
DM is supported by FONDAP Center for Astrophysics 15010003.
This work was performed under the auspices of the U.S. Department of Energy,
National Nuclear Security Administration by the University of California,
Lawrence Livermore National Laboratory under contract No. W-7405-Eng-48.

\appendix

\section{Theoretical Blending Estimates}
\label{sec-blend-th}

\subsection{Fraction of Faint Star Blend Events}
\label{sec-blend-frac}

A simple theoretical estimate of the effect of blending on our clump 
giant analysis can be made with the help of the HST luminosity 
function for Baade's Window from Holtzman et al.\ (1998). We multiply
this luminosity function by a factor of 1.5 to account for the fact that
our Central Galactic Region (CGR; defined in \S 5) is at a lower 
Galactic latitude than Baade's Window. Faint stars within a radius 
of approximately full width at half maximum of the point spread function
of an image will be close enough to the 
clump giant star to be blended with it. Both the baseline level and
the microlensing signal for
each event are dominated by images with seeing better than the
median, so we make the conservative choice of the median seeing
in the MACHO blue band images, namely $2.1"$, to use for our blending 
estimates. The Holtzman et al.\ (1998) luminosity function multiplied
by 1.5 implies a total of $\sim 16$ stars with an unreddened absolute
$V$ magnitude $M_V > 1.3$ within our seeing disk, and combined
brightness of these
stars is $M_V = 3.13$, or $2.13$ magnitudes fainter than a typical
clump giant. Due to turnoff feature, $M_V = 3.13$ is too bright 
for a single bulge main
sequence star, but three $M_V = 4.32$ stars could combine to give
the same total brightness. 
The blending effect of these stars will be approximately equivalent
to the blending effect from the luminosity function under consideration.

We consider an object that consists of a single clump giant with blend fraction
$f_{\rm cg} = 0.877$ blended with three main sequence stars, each with
blend fraction, $f_{\rm ms} = 0.041$.
In order to pass our detection threshold of $A^{\rm obs}_{\rm max}
\geq 1.5$ for fits without blending (apparent {\em obs}erved)
which corresponds to $u^{\rm obs}_{\rm cut} = 0.827$,
the blended clump giant 
must reach an actual brightness of $A_{\rm max,cg} \geq 1.57$,
corresponding $u_{\rm cut,cg} = 0.771$. For each of the
three faint, main sequence stars, the requirement of 
$A^{\rm obs}_{\rm max} \geq 1.5$ implies a peak magnification of
$A_{\rm max,ms} = 13.2$ or $u_{\rm cut,ms} = 0.076$. The blending
will also cause a reduction in the fit $\that$ values for a fit
without blending compared to the actual values. While the 
threshold magnification is $A^{\rm obs}_{\rm max} \geq 1.5$ for
detection, the typical event has $A^{\rm obs}_{\rm max} = 2.5$, so
we will use this magnification for our estimates of the blending effect
on $\that$. For the clump giant,
the unblended fit will yield a $\that^{\rm obs}_{\rm cg}$ value about 
a factor of 1.06 smaller
than actual value, $\that_{\rm cg}$ and for the faint main sequence stars, 
the true $\that_{\rm ms}$
value will be reduced by a factor of 8.2 according to a calculation
similar to that shown in Fig.~2 of Wo\'{z}niak \& Paczy\'{n}ski
(1997). This results in
a negligible change in the detection efficiency for the clump giant
by a factor of $\xi_{\rm cg} = 1.00$,
but for a main sequence star event with a true $\that_{\rm ms} \sim
50\,$days, the detection efficiency is reduced by about a factor of 
$\xi_{\rm ms} =1.5$. Using the
fact that microlensing events are uniformly distributed in $\umin$,
we can find the ratio of the numbers of faint blended events to clump giant events,
\begin{equation}
\label{eqn:blendrat}
\frac{N_{\rm ms}}{N_{\rm cg}} = \frac{3\,  
u_{\rm cut,ms}/\xi_{\rm ms}}{u_{\rm cut,cg}/\xi_{\rm cg}} = 0.197 \ .
\end{equation}
So, this estimate indicates that $\sim 83.5$\% of apparent clump 
giant events should have actual clump giant sources, while 
$\sim 16.5\,$\% of the apparent clump giant events will be
due to much fainter main sequence stars. This estimate has
not included a contribution from lensing of faint blended stars
that are part of an apparent clump giant that is actually a blend of
stars that are all fainter than a clump giant. Such events are 
rare because of the steep drop in the luminosity function just 
below the clump and the fact that most of the stars that are
slightly fainter than the clump are much bluer foreground disk
main sequence stars that are much less likely to be lensed.
So, it is unlikely that the fraction of blended events could
reach 25\%.

\subsection{Blending Effect on Microlensing Optical Depth}
\label{sec-blend-tau}

The general formula for the microlensing optical depth is
\begin{equation}
\label{eqn:taugen}
\tau = \frac{\pi}{4E} \sum_{i=1}^{N_{\rm events}} \frac{\that_i}
{\epsilon(\that_i)},
\end{equation}
where $E=N T_{\rm obs}$ is the total exposure which depends on
the number of source stars, $N$, and the time span of the
observations, $T_{\rm obs}$.
Let us assume that the number of clump giants in the bulge fields have
been estimated properly and so $N$ does not have to be analyzed
in what follows, and $T_{\rm obs}$ is fixed by the experimental setup.
Now, we seek to calculate the correct optical depth 
for non-contaminated clump giant sample that would have been observed
if there had been no blending.
\begin{equation}
\label{eqn:tautrue}
\tau^{\rm true}_{\rm cg} = \frac{\pi}{4E} \sum_{i=1}^{N^{\rm would-be}_{\rm cg}}
\frac{\that_{{\rm cg},i}}{\epsilon(\that_{{\rm cg},i})},
\end{equation}
where $N^{\rm would-be}_{\rm cg}$ is the number of clump events that
would have been observed if there had been no blending.
Note that:
\begin{equation}
\label{eqn:numwouldbe}
\frac{N^{\rm would-be}_{\rm cg}}{N_{\rm cg}} = \frac{u^{\rm obs}_{\rm
cut}}{u_{\rm cut,cg}/\xi_{\rm cg}},
\end{equation}
where typically we do not expect $N_{\rm cg} = N_{\rm events}$ (see
equations \ref{eqn:blendrat} and \ref{eqn:numsum}).

In reality, we effectively evaluate the following estimator in the
clump region of the color-magnitude diagram:
\begin{equation}
\label{eqn:tauobs}
\tau^{\rm obs} = \tau^{\rm obs}_{\rm cg} + \tau^{\rm obs}_{\rm ms} =
\frac{\pi}{4E} \sum_{i=1}^{N_{\rm cg}}
\frac{\that^{\rm obs}_{{\rm cg},i}}{\epsilon(\that^{\rm obs}_{{\rm
cg},i})} + \frac{\pi}{4E} \sum_{j=1}^{N_{\rm ms}}
\frac{\that^{\rm obs}_{{\rm ms},j}}{\epsilon(\that^{\rm obs}_{{\rm
ms},j})},
\end{equation}
where $\tau^{\rm obs}_{\rm cg}$ and $\tau^{\rm obs}_{\rm ms}$ are the
contributions to the observed optical depth coming from true
lightly-blended clump
giants and from main sequence blends, respectively. Therefore:
\begin{equation}
\label{eqn:numsum}
N_{\rm cg} + N_{\rm ms} = N_{\rm events},
\end{equation}
and we know from equation (\ref{eqn:blendrat}) that:
\begin{equation}
\label{eqn:numms}
N_{\rm ms} = \frac{3\,  u_{\rm cut,ms}/\xi_{\rm ms}}{u_{\rm cut,cg}/\xi_{\rm cg}} N_{\rm cg}.
\end{equation}

Let us introduce the following notation: $\eta_{\rm cg} = \that^{\rm obs}_{\rm
cg}/\that_{\rm cg}$ and $\eta_{\rm ms} = \that^{\rm obs}_{\rm
ms}/\that_{\rm ms}$.
Then we have:
\begin{equation}
\label{eqn:tautrue2}
\tau^{\rm true}_{\rm cg} = \frac{\pi}{4E} \frac{u^{\rm obs}_{\rm
cut}}{u_{\rm cut,cg}/\xi_{\rm cg}} N_{\rm cg} \left< \frac{\that_{\rm
cg}}{\epsilon(\that_{\rm cg})} \right>,
\end{equation}
\begin{equation}
\label{eqn:tauobscg}
\tau^{\rm obs}_{\rm cg} = \frac{\pi}{4E} \frac{\xi_{\rm cg}}{\eta_{\rm
cg}} N_{\rm cg} \left< \frac{\that_{\rm
cg}}{\epsilon(\that_{\rm cg})} \right>,
\end{equation}
\begin{equation}
\label{eqn:tauobsms}
\tau^{\rm obs}_{\rm ms} = \frac{\pi}{4E} \frac{\xi_{\rm ms}}{\eta_{\rm
ms}} \frac{3\,  u_{\rm cut,ms}/\xi_{\rm ms}}{u_{\rm cut,cg}/\xi_{\rm cg}} 
N_{\rm cg} \left< \frac{\that_{\rm ms}}{\epsilon(\that_{\rm ms})} \right>,
\end{equation}
where notation $\left< . \right>$ indicates an average over events.
To make further progress we will assume that:
\begin{equation}
\label{eqn:timescaleequity}
\left< \frac{\that_{\rm ms}}{\epsilon(\that_{\rm ms})} \right> =
\left< \frac{\that_{\rm cg}}{\epsilon(\that_{\rm cg})} \right>,
\end{equation}
but the conclusions will be almost identical even if equation
(\ref{eqn:timescaleequity}) does not hold precisely.
Combining equations
(\ref{eqn:tautrue2}), (\ref{eqn:tauobscg}), (\ref{eqn:tauobsms}),
(\ref{eqn:tauobs}), and (\ref{eqn:timescaleequity}) we find:
\begin{equation}
\label{eqn:tauratio}
\frac{\tau^{\rm obs}}{\tau^{\rm true}_{\rm cg}} =
\frac{\displaystyle{\frac{1}{\eta_{\rm cg}}} u_{\rm cut,cg} + \displaystyle{\frac{3}{\eta_{\rm ms}}}
u_{\rm cut,ms}}{u^{\rm obs}_{\rm cut}}.
\end{equation}

We can apply equation~(\ref{eqn:tauratio}) to the situation
from \S~\ref{sec-blend-frac}, where
we considered the effect of blending
on typical events with $A^{\rm obs}_{\rm max} \approx 2.5$.
For the estimate of the fraction of blended events, we implicitly
assumed that the blended clump giant and main sequence stars
all stand out above the background, so that the photometry code
identifies a blended star that has the combined brightness of
the clump giant and the main sequence stars. However, the
unresolved main sequence stars also contribute to the background
brightness that is attributed to the sky background by the 
photometry code. In areas of the bulge fields where giant stars 
are absent, our photometry code finds
``objects" that are generally brighter than the top of the bulge
main sequence, and blends of multiple main sequence stars.
Thus, the faint main sequence stars contribute to the ``sky
background" if they occur in regions of lower than average
star density, and they can contribute to stellar ``objects" identified
by our photometry code if they are in high density regions. When
a clump giant is present, it will be blended with both the faint stars
that make up part of the background and with the stars that would
have combined to make up part of a stellar ``object" if the clump
giant happens to be in a location with a higher than average
main sequence star density. So, to bracket the real situation,
we will consider two cases in which the clump giant is blended
with the same 3 main sequence stars discussed in 
\S~\ref{sec-blend-frac}:
\begin{enumerate}
\item The faint main sequence stars and the clump giant will
         combine to form a single stellar ``object" with their
         combined brightness.
\item The faint main sequence stars will only contribute to the
          background ``sky" brightness, and the stellar ``object"
          seen by the photometry code will have the same 
          brightness as the clump giant.
\end{enumerate}

In case 1, $\that$ reduction factors for the fits assuming no blending
are $\eta_{\rm cg} = 1.06$ and $\eta_{\rm ms} = 8.2$, whereas the $\umin$
thresholds are $u_{\rm cut,cg} = 0.771$, $u_{\rm cut,ms} = 0.076$,
and $u^{\rm obs}_{\rm cut} = 0.827$. Therefore, using
equation~(\ref{eqn:tauratio}) we obtain:
\begin{equation}
\label{eqn:tauratiocase1}
\frac{\tau^{\rm obs}}{\tau^{\rm true}_{\rm cg}} = 0.913.
\end{equation}
In case 2, $\that$ reduction factors for the fits assuming no blending
are $\eta_{\rm cg} = 1.0$ and $\eta_{\rm ms} = 7.4$, whereas the $\umin$
thresholds are $u_{\rm cut,cg} = u^{\rm obs}_{\rm cut} = 0.827$ and
 $u_{\rm cut,ms} = 0.086$.
Therefore, equation~(\ref{eqn:tauratio}) yields:
\begin{equation}
\label{eqn:tauratiocase2}
\frac{\tau^{\rm obs}}{\tau^{\rm true}_{\rm cg}} = 1.042.
\end{equation}
Thus, our two estimates of $\tau^{\rm obs} $ bracket the true value.

There are some additional blending effects that we did not include
in this estimate. There will be some ``objects" that pass our clump
giant cuts only because of blending. Typically, these will be dominated
by light from a single star that is slightly below the clump giant cut,
and the inclusion of these false clump giants will be partly compensated
by the loss of true clump giants that appear brighter than our clump
region due to blending. Overall, we expect this effect to cause a slight
increase in $\tau^{\rm obs}/\tau^{\rm true}_{\rm cg}$, but the overall 
systematic error in the optical depth due to blending is probably 
less than 10\%. 

The selection of the clump giant sample minimizes the effects of blending
on the optical depth measurement because clump giants are primarily blended
with stars that are much fainter. This means that blending has only
a small effect on the apparent $\that^{\rm obs}_{\rm cg}$ values for
the true clump giant
events that dominate the sample, while the faint star blends have 
such a large duration shift
that they cause little bias of the microlensing optical depth.

\section{Optical depth estimation}

The optical depth is usually defined as the probability of a microlensing 
event happening at a given time on any given star.  
An ``event" is usually defined as occurring when a lens comes within
one Einstein radius of the line-of-sight to the source.  This is easily
determined observationally since the formula for magnification is
\be 
A(u) = \frac{u^2+2}{u (u^2+4)^{1/2}},
\label{eqn:aofu}
\ee
where $u= l/R_E$ is the distance of the lens from the line-of-sight
in units of the Einstein radius $R_E$, and 
\be
R_E = 2 \left( \frac{GMD_l(D_s-D_l)}{c^2 D_s} \right)^{1/2},
\label{eq:RE}
\ee
where $M$ is the mass of the lens, $D_s$ is the observer-source
distance, and $D_l$ is the observer-lens distance.
Thus $u\leq 1$ or $A\geq 1.34$ means the line-of-sight is within one 
Einstein radius.

To estimate the probability of lensing occurring we must use the observed
microlensing events.  
First consider the case of a 100\% efficient microlensing experiment.
When there are many stars observed,
one would like
to find the probability of lensing per star at a given time 
by dividing the number of stars being lensed at a given time
by the total number of observed 
stars.  Observationally, however, one cannot tell whether or not a 
given star is being lensed;  one only
measures stars brightening and then fading as the line-of-sight enters
and leaves the Einstein radius.
In this case one can estimate the lensing probability by the fraction of 
the total available observing time that a given star spends lensed with 
$A \geq 1.34$.  With many stars, the optical depth is thus estimated
as $\tau = \sum_i t_i/E$, where $t_i$ is the time inside the Einstein
radius (i.e. $A>1.34$), and
the exposure $E=N_*T_{\rm obs}$ is the number of
observed stars times the duration of the observations.
The quantity $t_i$ is easily found from the two microlensing fit parameters:
$\that_i$, the Einstein diameter crossing time, and $A_{{\rm max}, i}$, the
maximum magnification:
\be
t_i = \that_i \sqrt{1-u_{{\rm min}, i}^2},
\label{eqn:ti}
\ee
where $u_{\rm min, i}$ is found from $A_{{\rm max}, i}$ by 
inverting equation~(\ref{eqn:aofu}) above.

For an imperfect observational program only a fraction $\epsilon$ of
microlensing events will be detected.  Calculating this efficiency
as a function of event duration and magnification allows a correction to
be made for each detected event giving
the formula we use:
\be
\tau = \frac{1}{E} \sum_{i=1}^{N_{\rm events}} \frac{t_i}{\epsilon(t_i,
A_{{\rm max}, i})}.
\label{eqn:appendixtau}
\ee

If one wished to use the fit
$\that_i$ and avoid using $t_i$ one could
replace $t_i$ in the formula above with an estimate of the typical
value of $\that$.  This could be done by noting that every impact parameter, 
$u$, between 0 and 1 is equally likely (given perfect efficiency)
and so averaging over all impact parameters
for a given Einstein ring gives $\VEV{t_i} = \frac{\pi}{4} \VEV{ \that _i}$.
Since for a large number of events the sum over $t_i$ is effectively
an average over $t_i$, one could replace $t_i$ with $\frac{\pi}{4} \that_i$
inside the sum.
The resulting formula is equation~(\ref{eqn:tauthat}), and has been used 
in the past.

One difference between the two formulas is that the new formula allows
weighting each event with an efficiency calculated as a function
of both $\hat{t}_{i}$ and $u_{{\rm min}, i}$. When using the 
sum over $\hat{t}_i$ one usually weights by the efficiency only as a
function of $\that$.

\begin{figure}[t]
\center
\epsfig{figure=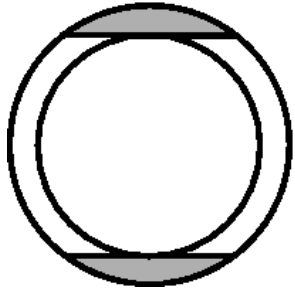, width=1.5in}
\caption{Area of Einstein rings used for calculating optical depths, shaded regions are missed by our cuts.  \label{fig:exp_109}}
\end{figure}

There is one subtlety when calculating efficiencies for each event, rather
than as a function of $\that$.  Since we used a cut $\amax \geq 1.5$
and optical depth is defined as the probability of finding events 
with $\amax\geq 1.34$, the above formula needs to be multiplied by a 
correction factor of 1.09 to account
for events with $ 1.34 \leq \amax < 1.5 $.  These low amplification events are 
not detected so can not be included in the efficiency calculation which 
only inputs simulated events for the measured values of $t$ and $\amax$.  
When efficiencies are calculated as a function of $\that$, events with such 
values are input, but not detected, resulting in a lower efficiency 
(and therefore higher optical depth) for that value of $\that$.  
One can invert equation (\ref{eqn:aofu}) and find that $ u(1.5) = 0.83 $, and 
$ u(1.34) = 1 $.  With the assumption that the lens is moving at 
uniform velocity through the line of sight, the correction factor becomes 
simply the ratio of area within the Einstein ring to the area in 
the ring with an impact parameter less than 0.83.  
In Figure~\ref{fig:exp_109} one can see the Einstein ring and the ring at 
$ u = 0.83 $ ( $ \amax = 1.5 $).
The areas missed with our cuts are the small sections on 
the top and bottom of the circle.

\section{Upper limits on optical depth}

Optical depth in a given field can be estimated as a number of
events in a given field multiplied by the contribution of a typical
event.
Therefore, one can define upper limits on the optical depth in
fields with no events as:
\be
\tau_f^{\rm lim, CL} = \nu^{\rm CL} \tau_f^{\rm 1 \, event},
\label{eqn:upperlim1a}
\ee
where
$\tau_f^{\rm 1 \, event}$ would be a typical contribution of one event to
the optical depth in field $f$ and $\nu^{\rm CL}$
is a multiplication factor dependent on the confidence level (CL).
Equation~(\ref{eqn:upperlim1a}) can be expressed as:
\be
\tau_f^{\rm lim, CL} = \nu^{\rm CL} \frac{\pi}{4 N_f T_{\rm obs}} \left<
\frac{\hat{t}}{\epsilon(\hat{t})} \right>_f, \label{eqn:upperlim1b}
\ee
where the angle brackets indicate an average in field $f$.
Let us first note that for each field, we have efficiencies
$\epsilon(\hat{t})$, which come from our simulations. However, since
we have no events in the fields of interest we cannot estimate the
right side of equation~(\ref{eqn:upperlim1b}) based exclusively on the data
in field $f$. Therefore, we are forced to assume that duration
distribution in the entire area of our bulge survey is universal
and proceed as follows. Based on the observed duration distribution in
fields with events and efficiencies in those fields we can recover the
intrinsic duration distribution of events. This distribution can
then be mapped using the efficiencies in field $f$ to recover the 
hypothetical duration distribution that would be observed in field
$f$. Each $\hat{t}_i$ from our clump sample contributes to the
hypothetical distribution in field $f$ with a weight
$w_{f,i} \equiv {\epsilon_f(\hat{t}_i)}/{\epsilon_{i}(\hat{t}_i)}$,
where $\epsilon_f$ is the efficiency in field $f$, and $\epsilon_{i}$
is the efficiency in the event's field of origin.
We now may express:
\begin{eqnarray}
\left<
\frac{\hat{t}}{\epsilon(\hat{t})} \right>_f & = & \frac{\sum_{i=1}^{N_{\rm
events}} w_{f,i}
\left ({\hat{t}_i}/{\epsilon_f(\hat{t}_i)} \right)
}{\sum_{i=1}^{N_{\rm events}}
w_{f,i}} \nonumber \\
& = & \frac{\sum_{i=1}^{N_{\rm
events}}
{\hat{t}_i}/{\epsilon_i(\hat{t}_i)}}{\sum_{i=1}^{N_{\rm events}}
{\epsilon_f(\hat{t}_i)}/{\epsilon_{i}(\hat{t}_i)}} \label{eqn:upperlim2}
\end{eqnarray}
Substitution of equation~(\ref{eqn:upperlim2}) to
equation~(\ref{eqn:upperlim1b}) leads to:
\be
\tau_f^{\rm lim, CL} = \nu^{\rm CL} \frac{\pi}{4 N_f T_{\rm obs}}
\frac{\sum_{i=1}^{N_{\rm events}}
{\hat{t}_i}/{\epsilon_i(\hat{t}_i)}}{\sum_{i=1}^{N_{\rm events}}
{\epsilon_f(\hat{t}_i)}/{\epsilon_{i}(\hat{t}_i)}},                             \label{eqn:upperlim3app}
\ee
which is the formula that we use in \S 5.
Let us explain the role of the multiplication
factor $\nu^{\rm CL}$.
The number of observed events is Poisson
distributed. To constrain optical depth we simply ask what is the
upper limit on the number
of expected events given 0 detected events for the
Poisson distribution.
We can obtain the necessary number 
of expected events $\nu^{\rm CL}$ that rules out $\tau_f^{\rm lim,
CL}$ or larger at CL from
the following equation:
\be
\exp({-\nu^{\rm CL}}) = \frac{1.0 - {\rm CL}}{2} \label{eqn:nucl}
\ee
For $1 \sigma$ (68.17\%) CL, one obtains $\nu^{1 \sigma} = 1.8379$.

\clearpage

\begin{deluxetable}{ll}
\tablewidth{0in}
\tablecomments{Designations $V$ and $(V-R)$ indicate baseline
quantities, i.e. the ones in the limit of no
microlensing-induced amplification.}
\tablecaption{Selection Criteria\label{tab:cuts}}
\tablehead{ \colhead{Selection} & \colhead{Description} }
\startdata
\multicolumn{2}{l}{{\bf microlensing fit parameter cuts:}} \\
$ N_{(V-R)}>0 $ & require color information \\
$ \chi^2_{out} < 3.0, \; \chi^2_{out} > 0.0 $ & require high quality baselines \\
$N_{amp} \geq 8 $ & require 8 points in the amplified region \\
$N_{rising} \geq 1, \; N_{falling}\geq 1$ & require at least one point in the rising and falling part of the peak \\
$A_{max} > 1.5 $ & magnification threshold \\
($A_{max}-1) > 2.0(\sigma_R + \sigma_B)$ & signal to noise cut on amplification \\
$ \!\!\! \begin{array}{l}
(\sigma_R+\sigma_B) < (0.05\,\delta\chisq)/(N_{amp}\,\chisq_{in}\,\chisq_{out})  \\
(N_{amp}\,\chisq_{out}\,f_{chrom})/(\mathit{befaft}\,\delta\chisq)<0.0003  \\
(N_{amp}\,\chisq_{in})/(\mathit{befaft}\,\delta\chisq)<0.00004 
\end{array} $
&
$ \left\} \raisebox{1.8ex}[4.6ex]{} \right.$ remove spurious photometric signals caused by nearby saturated stars \\
$\delta\chisq/fc2 > 320.0 $ & good overall fit to microlensing lc shape \\
$\delta\chisq/\chisq_{peak} \geq 400.0 $ & same quality fit in peak as whole lc\\ 
$0.5(rcrda+bcrda) \leq 143.0 $ & source star not too crowded \\
$\xi^{auto}_B/\xi^{auto}_R <2.0 $ & remove long period variables \\
$t_0 > 419.0,\, t_0 < 2850.0 $ & constrain the peak to period of observations \\
$\hat{t} < 1700 $ & limit event duration to $\sim$half the span of observations \\
\hline
\multicolumn{2}{l}{{\bf clump giant cuts:}}\\
$V \geq 15,\, V\leq 20.5 $ &  select bright stars with reliabale photometry \\
$V \geq 4.2(V-R)+12.4 $ & define bright boundary of extinction strip\\
$V \leq 4.2(V-R)+14.2 $ & define faint boundry of extinction strip  \\
$(V-R) \geq(V-R)_{boundary} $ & avoid main sequence contamination \\
exclude fields $301-311$ & avoid disk contamination \\
\enddata
\end{deluxetable}

\begin{deluxetable}{lccccccccccc}
\tabletypesize{\small}
\rotate
\tablecaption{Event parameters}
\tablecomments{Designations:\\
$^\ddagger$: binary event,\\
$^a$ and $^b$: each superscript marks two members of a pair that
represents a single microlensing event in two overlapping fields.
\label{tab:clumpevents}}
\tablewidth{0pt}
\tablehead{\colhead{field.tile.seq} & \colhead{RA (J2000)} &
\colhead{DEC (J2000)} & \colhead{$l$} & \colhead{$b$} & \colhead{$V$}
& \colhead{$(V-R)$} & \colhead{$(V-R)_{\rm boundary}$} &  \colhead{$
t_0 $} & \colhead{$ \hat{t} $} &\colhead{ $ A_{\rm max} $} &\colhead{ $\frac{\chi^2}{n_{dof}}$} }

\startdata
$ 101.21689.315 $ 	& 18 06 58.32 & $-27$ 27 45.7	 & 3.639 &  $-3.327$  & 17.50 & 1.09 &	0.81\phm{81} &	593.5 & 165 & 4.80 &1.19 \\
$ 102.22466.140 $  	& 18 08 47.04 & $-27$ 40 47.3	 & 3.643 &  $-3.783$  & 16.81 & 0.95 &	0.81\phm{81} &	1275.6 & 153 & 3.88 &0.95 \\
$ 104.19992.858 $ 	& 18 02 53.76 & $-27$ 57 50.0	 & 2.761 &  $-2.784$  & 18.59 & 1.09 &	0.85\phm{81} &	1169.7 & 106 & 2.23 &2.15 \\
$ 104.20119.6312^{a} $ 	& 18 03 20.64 & $-28$ 08 14.6	 & 2.658 &  $-2.955$  & 17.99 & 1.06 &	0.78\phm{81} &	1975.3 & 38.7 & 1.62 &1.62 \\
$ 104.20251.50 $ 	& 18 03 34.08 & $-28$ 00 19.1	 & 2.797 &  $-2.933$  & 17.33 & 0.93 & 	0.82\phm{81} &	493.9 & 287 & 10.1 &1.75 \\
$ 104.20251.1117 $ 	& 18 03 29.04 & $-28$ 00 31.0	 & 2.785 &  $-2.919$  & 18.22 & 0.98 &	0.82\phm{81} &	463.2 & 45.5 & 2.09 &0.93 \\
$ 104.20259.572 $ 	& 18 03 33.36 & $-27$ 27 47.2	 & 3.269 &  $-2.665$  & 18.16 & 1.19 &	0.91\phm{81} &	538.5 & 14.2 & 1.58 &1.56 \\
$ 104.20382.803 $ 	& 18 03 53.28 & $-27$ 57 35.6	 & 2.872 &  $-2.973$  & 17.79 & 0.96 &	0.84\phm{81} &	1765.8 & 254 & 5.72 &4.40 \\
$ 104.20515.498 $ 	& 18 04 09.60 & $-27$ 44 35.2	 & 3.090 &  $-2.919$  & 17.63 & 0.99 &	0.81\phm{81} &	2061.7 & 53.6 & 1.76 &1.42 \\
$ 104.20640.8423^{b} $ 	& 18 04 33.60 & $-28$ 07 31.8	 & 2.800 &  $-3.183$  & 17.15 & 0.98 &	0.79\phm{81} &	2374.9 & 30.8 & 1.65 &2.23 \\
$ 104.20645.3129 $ 	& 18 04 26.16 & $-27$ 47 35.2	 & 3.077 &  $-2.997$  & 17.68 & 1.07 &	0.83\phm{81} &	1558.3 & 152 & 1.82 &0.77 \\
$ 105.21813.2516 $ 	& 18 07 06.96 & $-27$ 52 34.3	 & 3.292 &  $-3.555$  & 17.41 & 1.08 &	0.83\phm{81} &	1928.5 & 17.5 & 13.2 &1.14 \\
$ 108.18947.3618 $ 	& 18 00 29.04 & $-28$ 19 20.3	 & 2.186 &  $-2.498$  & 17.72 & 1.10 &	0.89\phm{81} &	1287.9 & 45.6 & 2.05 &0.94 \\
$ 108.18951.593^{\ddagger} $ 	& 18 00 33.84 & $-28$ 01 10.6	 & 2.458 &  $-2.363$  & 18.07 & 1.18 &	0.95\phm{81} &	1991.6 & 47.3 & 4.05 &7.02 \\
$ 108.18951.1221 $ 	& 18 00 25.92 & $-28$ 02 35.2	 & 2.423 &  $-2.350$  & 18.94 & 1.18 &	0.95\phm{81} &	583.6 & 45.8 & 2.16 &0.92 \\
$ 108.18952.941 $ 	& 18 00 36.00 & $-27$ 58 30.0	 & 2.501 &  $-2.348$  & 18.90 & 1.16 &	1.02\phm{81} &	1324.9 & 66.9 & 2.39 &1.11 \\
$ 108.19074.550 $ 	& 18 00 52.32 & $-28$ 29 52.1	 & 2.076 &  $-2.659$  & 17.54 & 1.03 &	0.88\phm{81} &	2050.6 & 9.30 & 1.78 &0.97 \\
$ 108.19334.1583 $ 	& 18 01 26.40 & $-28$ 31 14.2	 & 2.117 &  $-2.779$  & 17.67 & 0.94 &	0.86\phm{81} &	1231.4 & 18.5 & 4.09 &0.38 \\
$ 109.20119.1051^{a} $ 	& 18 03 20.64 & $-28$ 08 14.6	 & 2.658 &  $-2.955$  & 18.09 & 0.97 &	0.78\phm{81} &	1977.6 & 28.7 & 1.62 &1.01 \\
$ 109.20640.360^{b} $ 	& 18 04 33.60 & $-28$ 07 32.5	 & 2.799 &  $-3.183$  & 17.33 & 1.05 & 	0.79\phm{81} &	2375.1 & 28.7 & 1.67 &0.72 \\
$ 110.22455.842 $ 	& 18 08 51.36 & $-28$ 27 11.2	 & 2.971 &  $-4.169$  & 18.28 & 0.98 &	0.76\phm{81} &	1900.1 & 19.8 & 2.54 &0.84 \\
$ 113.18552.581 $ 	& 17 59 36.00 & $-28$ 36 24.1	 & 1.842 &  $-2.470$  & 17.60 & 1.02 &	0.86\phm{81} &	603.7 & 27.9 & 1.77 &1.11 \\
$ 113.18804.1061 $ 	& 18 00 03.36 & $-29$ 11 04.2	 & 1.390 &  $-2.843$  & 18.28 & 1.00 &	0.84\phm{81} &	1166.6 & 6.50 & 1.62 &0.96 \\
$ 113.19192.365 $ 	& 18 01 06.96 & $-29$ 18 53.6	 & 1.391 &  $-3.109$  & 17.80 & 1.17 &	0.96\phm{81} &	2092.6 & 36.9 & 1.97 &0.52 \\
$ 114.19712.813 $ 	& 18 02 11.52 & $-29$ 19 20.6	 & 1.500 &  $-3.317$  & 18.12 & 1.08 &	0.87\phm{81} &	1643.4 & 23.7 & 3.58 & 0.62 \\
$ 114.19846.777 $ 	& 18 02 36.72 & $-29$ 01 41.9	 & 1.801 &  $-3.252$  & 17.98 & 1.02 &	0.80\phm{81} &	545.4 & 65.0 & 1.55 &1.86 \\
$ 114.19970.843 $ 	& 18 02 54.48 & $-29$ 26 29.4	 & 1.472 &  $-3.511$  & 17.99 & 0.96 &	0.85\phm{81} &	1281.9 & 14.7 & 4.47 &0.65 \\
$ 118.18014.320 $ 	& 17 58 25.20 & $-29$ 47 59.6	 & 0.678 &  $-2.840$  & 17.01 & 0.97 &	0.89\phm{81} &	877.6 & 25.5 & 1.58 &1.05 \\
$ 118.18141.731^{\ddagger} $ 	& 17 58 36.72 & $-30$ 02 19.3	 & 0.491 &  $-2.995$  & 17.92 & 1.06 &	0.94\phm{81} &	884.1 & 9.80 & 12.6 &0.65 \\
$ 118.18271.738 $ 	& 17 58 52.56 & $-30$ 02 08.2	 & 0.522 &  $-3.043$  & 18.25 & 1.18 &	0.96\phm{81} &	456.4 & 53.7 & 2.23 &0.69 \\
$ 118.18402.495 $ 	& 17 59 13.92 & $-29$ 55 53.4	 & 0.651 &  $-3.058$  & 17.63 & 1.08 &	0.93\phm{81} &	454.1 & 23.8 & 3.60 &0.96 \\
$ 118.18797.1397 $ 	& 18 00 06.96 & $-29$ 38 06.0	 & 1.004 &  $-3.078$  & 19.34 & 1.30 &	0.97\phm{81} &	2013.1 & 126 & 8.35 &1.07 \\
$ 118.19182.891 $ 	& 18 01 09.84 & $-29$ 56 19.0	 & 0.852 &  $-3.425$  & 17.97 & 0.93 &	0.84\phm{81} &	2367.0 & 13.6 & 6.28 &1.20 \\
$ 118.19184.939 $ 	& 18 01 10.32 & $-29$ 48 55.4	 & 0.960 &  $-3.366$  & 18.07 & 0.99 &	0.83\phm{81} &	2692.8 & 74.4 & 2.14 &1.13 \\
$ 121.22032.133 $ 	& 18 07 46.80 & $-30$ 39 41.4	 & 0.915 &  $-5.025$  & 16.37 & 0.86 &	0.75\phm{81} &	1287.3 & 19.5 & 1.58 &2.51 \\
$ 158.27444.129 $ 	& 18 20 10.32 & $-25$ 08 25.8	 & 7.097 &  $-4.827$  & 16.97 & 0.76 &	0.63\phm{81} &	2376.4 & 41.7 & 2.31 &1.80 \\
$ 162.25865.442 $ 	& 18 16 30.72 & $-26$ 26 58.2	 & 5.549 &  $-4.712$  & 17.52 & 0.90 &	0.73\phm{81} &	1225.3 & 60.9 & 1.58 &1.32 \\
$ 162.25868.405 $ 	& 18 16 46.08 & $-26$ 11 43.1	 & 5.801 &  $-4.643$  & 17.33 & 0.83 &	0.77\phm{81} &	1222.8 & 30.8 & 2.77 &1.39 \\
$ 176.18826.909 $ 	& 18 00 15.36 & $-27$ 42 47.9	 & 2.691 &  $-2.153$  & 18.77 & 1.20 &	1.01\phm{81} &	2364.7 & 24.3 & 1.77 &1.17 \\
$ 178.23531.931 $ 	& 18 11 17.52 & $-25$ 59 51.4	 & 5.391 &  $-3.467$  & 18.71 & 1.10 &	0.92\phm{81} &	1590.3 & 13.7 & 2.18 &1.08 \\
$ 180.22240.202 $ 	& 18 07 57.84 & $-25$ 24 48.2	 & 5.542 &  $-2.528$  & 18.92 & 1.40 &	0.98\phm{81} &	1286.9 & 311 & 3.42 &1.15 \\
$ 401.47991.1840 $ 	& 17 57 07.44 & $-28$ 13 44.0	 & 1.899 &  $-1.810$  & 19.46 & 1.37 &	1.12\phm{81} &	1953.0 & 44.7 & 1.92 &0.61 \\
$ 401.47994.1182 $ 	& 17 57 07.92 & $-28$ 00 27.7	 & 2.091 &  $-1.701$  & 19.65 & 1.53 &	1.23\phm{81} &	2362.7 & 60.6 & 3.68 &1.05 \\
$ 401.48052.861 $ 	& 17 57 23.76 & $-28$ 10 24.6	 & 1.977 &  $-1.834$  & 18.75 & 1.31 &	1.13\phm{81} &	2291.5 & 72.0 & 2.65 &0.69 \\
$ 401.48167.1934 $ 	& 17 57 57.36 & $-28$ 27 09.7	 & 1.796 &  $-2.080$  & 18.42 & 1.23 &	1.09\phm{81} &	2179.0 & 125 & 2.53 &0.89 \\
$ 401.48229.760^{\ddagger} $ 	& 17 58 13.44 & $-28$ 20 32.3	 & 1.921 &  $-2.076$  & 18.07 & 1.15 &	1.00\phm{81} &	2694.5 & 20.1 & 1.82 &1.18 \\
$ 401.48408.649^{\ddagger} $ 	& 17 59 08.88 & $-28$ 24 54.7	 & 1.959 &  $-2.289$  & 17.70 & 1.20 &	0.92\phm{81} &	2325.4 & 74.6 & 3.07 &1.30 \\
$ 401.48469.789 $ 	& 17 59 22.32 & $-28$ 20 21.8	 & 2.050 &  $-2.294$  & 18.23 & 1.17 &	0.92\phm{81} &	1554.8 & 5.70 & 5.96 &0.42 \\
$ 402.47678.1666 $ 	& 17 55 22.56 & $-29$ 04 12.7	 & 0.978 &  $-1.901$  & 19.19 & 1.31 &	1.07\phm{81} &	2393.1 & 13.6 & 2.48 &0.88 \\
$ 402.47737.1590 $ 	& 17 55 41.52 & $-29$ 09 00.7	 & 0.944 &  $-2.001$  & 18.49 & 1.04 &	0.99\phm{81} &	1552.3 & 5.50 & 2.79 &1.03 \\
$ 402.47742.3318 $ 	& 17 55 42.96 & $-28$ 49 17.4	 & 1.230 &  $-1.840$  & 20.19 & 1.62 &	1.10\phm{81} &	1377.1 & 50.1 & 8.64 &1.21 \\
$ 402.47796.1893 $ 	& 17 56 10.80 & $-29$ 10 44.4	 & 0.972 &  $-2.107$  & 19.52 & 1.33 &	1.03\phm{81} &	1678.7 & 19.3 & 22.5 &1.47 \\
$ 402.47798.1259 $ 	& 17 56 11.04 & $-29$ 06 14.0	 & 1.038 &  $-2.070$  & 19.72 & 1.36 &	1.10\phm{81} &	1583.7 & 47.6 & 4.94 &1.38 \\
$ 402.47799.1736 $ 	& 17 56 07.20 & $-28$ 58 32.5	 & 1.142 &  $-1.994$  & 19.59 & 1.29 &	1.16\phm{81} &	2410.0 & 115 & 2.26 &0.47 \\
$ 402.47856.561 $ 	& 17 56 24.96 & $-29$ 12 56.5	 & 0.966 &  $-2.170$  & 18.81 & 1.38 &	1.15\phm{81} &	1566.6 & 12.3 & 2.14 & 1.63 \\
$ 402.47862.1576^{\ddagger} $ 	& 17 56 20.64 & $-28$ 47 42.0	 & 1.323 &  $-1.945$  & 19.47 & 1.30 &	1.15\phm{81} &	2028.5 & 54.3 & 7.52 &6.13 \\
$ 402.48158.1296 $ 	& 17 57 47.52 & $-29$ 03 54.4	 & 1.247 &  $-2.346$  & 18.69 & 1.12 &	1.01\phm{81} &	2325.7 & 15.9 & 3.94 & 0.98 \\
$ 402.48280.502 $ 	& 17 58 24.96 & $-28$ 57 46.4	 & 1.404 &  $-2.422$  & 18.17 & 1.18 &	0.90\phm{81} &	1342.5 & 25.5 & 2.08 &1.14 \\
$ 403.47491.770 $ 	& 17 54 38.64 & $-29$ 33 13.0	 & 0.480 &  $-2.007$  & 17.94 & 1.29 &	1.04\phm{81} &	2633.9 & 48.5 & 2.43 &0.99 \\
$ 403.47550.807 $ 	& 17 55 00.00 & $-29$ 35 03.8	 & 0.492 &  $-2.089$  & 17.95 & 1.21 &	1.04\phm{81} &	2774.3 & 11.6 & 3.60 &0.87 \\
$ 403.47610.576 $ 	& 17 55 17.04 & $-29$ 37 40.8	 & 0.486 &  $-2.164$  & 17.54 & 1.13 &	0.98\phm{81} &	2657.9 & 9.60 & 4.75 &2.56 \\
$ 403.47845.495 $ 	& 17 56 22.80 & $-29$ 55 16.3	 & 0.351 &  $-2.517$  & 18.20 & 1.24 &	1.07\phm{81} &	1945.8 & 15.0 & 2.03 &0.88 \\
\enddata
\end{deluxetable}

\begin{deluxetable}{lccccccc}

\tableheadfrac{0.05}
\tablewidth{0pt}
\tablecaption{Optical Depth Contribution by Event\label{tab:taubyevent}}
\tablehead{ \colhead{field.tile.seq} & \colhead{$ t_i $} & \colhead{$
\hat{t} $} & \colhead{$ A_{{\rm max}, i} $} & \colhead{$\epsilon(t_i,A_{{\rm
max}, i}) $} & \colhead{$\epsilon(\hat{t}) $} & \colhead{$ \tau(t_i) $} &\colhead{ $ \tau(\hat{t}) $} }

\startdata
$ 101.21689.315 $ & $ 161 $ & $ 165 $ & $ 4.80 $ & $ 0.68 $ & $ 0.55 $ & $ 1.62 $ & $ 1.47 $ \\
$ 102.22466.140 $ & $ 147 $ & $ 153 $ & $ 3.88 $ & $ 0.39 $ & $ 0.32 $ & $ 3.67 $ & $ 3.34 $ \\
$ 104.19992.858 $ & $ 93.4 $ & $ 107 $ & $ 2.23 $ & $ 0.55 $ & $ 0.43 $ & $ 1.12 $ & $ 1.19 $ \\
$ 104.20119.6312 $ & $ 25.7 $ & $ 38.3 $ & $ 1.62 $ & $ 0.48 $ & $ 0.38 $ & $ 0.36 $ & $ 0.47 $ \\
$ 104.20251.50 $ & $ 285 $ & $ 287 $ & $ 10.1 $ & $ 0.77 $ & $ 0.63 $ & $ 2.44 $ & $ 2.18 $ \\
$ 104.20251.1117 $ & $ 38.7 $ & $ 45.6 $ & $ 2.09 $ & $ 0.46 $ & $ 0.39 $ & $ 0.56 $ & $ 0.56 $ \\
$ 104.20259.572 $ & $ 9.22 $ & $ 14.2 $ & $ 1.58 $ & $ 0.41 $ & $ 0.33 $ & $ 0.15 $ & $ 0.20 $ \\
$ 104.20382.803 $ & $ 250 $ & $ 254 $ & $ 5.73 $ & $ 0.78 $ & $ 0.62 $ & $ 2.11 $ & $ 1.95 $ \\
$ 104.20515.498 $ & $ 40.4 $ & $ 53.5 $ & $ 1.76 $ & $ 0.48 $ & $ 0.39 $ & $ 0.56 $ & $ 0.65 $ \\
$ 104.20640.8423 $ & $ 21.4 $ & $ 30.8 $ & $ 1.65 $ & $ 0.47 $ & $ 0.4 $ & $ 0.30 $ & $ 0.37 $ \\
$ 104.20645.3129 $ & $ 118 $ & $ 152 $ & $ 1.82 $ & $ 0.66 $ & $ 0.52 $ & $ 1.19 $ & $ 1.38 $ \\
$ 105.21813.2516 $ & $ 17.5 $ & $ 17.5 $ & $ 13.2 $ & $ 0.40 $ & $ 0.31 $ & $ 0.35 $ & $ 0.32 $ \\
$ 108.18947.3618 $ & $ 38.4 $ & $ 45.6 $ & $ 2.05 $ & $ 0.56 $ & $ 0.48 $ & $ 0.38 $ & $ 0.38 $ \\
$ 108.18951.593 $ & $ 45.8 $ & $ 47.3 $ & $ 4.05 $ & $ 0.57 $ & $ 0.47 $ & $ 0.44 $ & $ 0.39 $ \\
$ 108.18951.1221 $ & $ 39.5 $ & $ 45.8 $ & $ 2.16 $ & $ 0.57 $ & $ 0.48 $ & $ 0.38 $ & $ 0.38 $ \\
$ 108.18952.941 $ & $ 59.7 $ & $ 66.9 $ & $ 2.39 $ & $ 0.58 $ & $ 0.46 $ & $ 0.56 $ & $ 0.57 $ \\
$ 108.19074.550 $ & $ 7.07 $ & $ 9.26 $ & $ 1.78 $ & $ 0.41 $ & $ 0.34 $ & $ 0.09 $ & $ 0.11 $ \\
$ 108.19334.1583 $ & $ 17.9 $ & $ 18.5 $ & $ 4.09 $ & $ 0.53 $ & $ 0.42 $ & $ 0.18 $ & $ 0.17 $ \\
$ 109.20119.1051 $ & $ 19.4 $ & $ 28.7 $ & $ 1.62 $ & $ 0.44 $ & $ 0.4 $ & $ 0.29 $ & $ 0.34 $ \\
$ 109.20640.360 $ & $ 20.4 $ & $ 28.7 $ & $ 1.67 $ & $ 0.45 $ & $ 0.4 $ & $ 0.29 $ & $ 0.34 $ \\
$ 110.22455.842 $ & $ 18.0 $ & $ 19.8 $ & $ 2.54 $ & $ 0.41 $ & $ 0.33 $ & $ 0.46 $ & $ 0.46 $ \\
$ 113.18552.581 $ & $ 21.2 $ & $ 27.9 $ & $ 1.77 $ & $ 0.53 $ & $ 0.46 $ & $ 0.21 $ & $ 0.23 $ \\
$ 113.18804.1061 $ & $ 4.44 $ & $ 6.53 $ & $ 1.62 $ & $ 0.29 $ & $ 0.27 $ & $ 0.08 $ & $ 0.09 $ \\
$ 113.19192.365 $ & $ 30.4 $ & $ 36.9 $ & $ 1.97 $ & $ 0.59 $ & $ 0.49 $ & $ 0.27 $ & $ 0.28 $ \\
$ 114.19712.813 $ & $ 22.7 $ & $ 23.7 $ & $ 3.57 $ & $ 0.44 $ & $ 0.36 $ & $ 0.36 $ & $ 0.33 $ \\
$ 114.19846.777 $ & $ 40.0 $ & $ 64.9 $ & $ 1.55 $ & $ 0.46 $ & $ 0.39 $ & $ 0.61 $ & $ 0.84 $ \\
$ 114.19970.843 $ & $ 14.3 $ & $ 14.7 $ & $ 4.47 $ & $ 0.45 $ & $ 0.32 $ & $ 0.22 $ & $ 0.23 $ \\
$ 118.18014.320 $ & $ 16.3 $ & $ 25.5 $ & $ 1.58 $ & $ 0.54 $ & $ 0.44 $ & $ 0.18 $ & $ 0.24 $ \\
$ 118.18141.731 $ & $ 9.74 $ & $ 9.77 $ & $ 12.6 $ & $ 0.45 $ & $ 0.34 $ & $ 0.12 $ & $ 0.12 $ \\
$ 118.18271.738 $ & $ 46.9 $ & $ 53.7 $ & $ 2.23 $ & $ 0.56 $ & $ 0.46 $ & $ 0.49 $ & $ 0.49 $ \\
$ 118.18402.495 $ & $ 22.8 $ & $ 23.8 $ & $ 3.60 $ & $ 0.54 $ & $ 0.43 $ & $ 0.24 $ & $ 0.23 $ \\
$ 118.18797.1397 $ & $ 125 $ & $ 126 $ & $ 8.35 $ & $ 0.73 $ & $ 0.58 $ & $ 1.00 $ & $ 0.91 $ \\
$ 118.19182.891 $ & $ 13.4 $ & $ 13.6 $ & $ 6.28 $ & $ 0.49 $ & $ 0.39 $ & $ 0.16 $ & $ 0.15 $ \\
$ 118.19184.939 $ & $ 64.0 $ & $ 74.4 $ & $ 2.14 $ & $ 0.56 $ & $ 0.48 $ & $ 0.66 $ & $ 0.65 $ \\
$ 121.22032.133 $ & $ 12.5 $ & $ 19.4 $ & $ 1.58 $ & $ 0.27 $ & $ 0.21 $ & $ 0.59 $ & $ 0.85 $ \\
$ 158.27444.129 $ & $ 36.8 $ & $ 41.7 $ & $ 2.31 $ & $ 0.22 $ & $ 0.2 $ & $ 2.96 $ & $ 2.59 $ \\
$ 162.25865.442 $ & $ 39.5 $ & $ 60.9 $ & $ 1.58 $ & $ 0.33 $ & $ 0.24 $ & $ 1.73 $ & $ 2.63 $ \\
$ 162.25868.405 $ & $ 28.5 $ & $ 30.8 $ & $ 2.77 $ & $ 0.30 $ & $ 0.24 $ & $ 1.38 $ & $ 1.33 $ \\
$ 176.18826.909 $ & $ 18.4 $ & $ 24.3 $ & $ 1.77 $ & $ 0.26 $ & $ 0.24 $ & $ 0.37 $ & $ 0.39 $ \\
$ 178.23531.931 $ & $ 11.9 $ & $ 13.7 $ & $ 2.18 $ & $ 0.25 $ & $ 0.19 $ & $ 0.43 $ & $ 0.46 $ \\
$ 180.22240.202 $ & $ 297 $ & $ 312 $ & $ 3.42 $ & $ 0.58 $ & $ 0.44 $ & $ 4.62 $ & $ 4.62 $ \\
$ 401.47991.1840 $ & $ 36.2 $ & $ 44.7 $ & $ 1.92 $ & $ 0.29 $ & $ 0.24 $ & $ 0.55 $ & $ 0.60 $ \\
$ 401.47994.1182 $ & $ 58.2 $ & $ 60.6 $ & $ 3.68 $ & $ 0.31 $ & $ 0.25 $ & $ 0.83 $ & $ 0.78 $ \\
$ 401.48052.861 $ & $ 66.0 $ & $ 72.0 $ & $ 2.65 $ & $ 0.34 $ & $ 0.25 $ & $ 0.88 $ & $ 0.93 $ \\
$ 401.48167.1934 $ & $ 113 $ & $ 125 $ & $ 2.54 $ & $ 0.38 $ & $ 0.33 $ & $ 1.34 $ & $ 1.23 $ \\
$ 401.48229.760 $ & $ 15.7 $ & $ 20.1 $ & $ 1.82 $ & $ 0.24 $ & $ 0.20 $ & $ 0.29 $ & $ 0.33 $ \\
$ 401.48408.649 $ & $ 70.1 $ & $ 74.6 $ & $ 3.07 $ & $ 0.32 $ & $ 0.25 $ & $ 0.98 $ & $ 0.96 $ \\
$ 401.48469.789 $ & $ 5.59 $ & $ 5.67 $ & $ 6.01 $ & $ 0.098 $ & $ 0.059 $ & $ 0.25 $ & $ 0.31 $ \\
$ 402.47678.1666 $ & $ 12.3 $ & $ 13.7 $ & $ 2.48 $ & $ 0.26 $ & $ 0.20 $ & $ 0.17 $ & $ 0.18 $ \\
$ 402.47737.1590 $ & $ 5.06 $ & $ 5.46 $ & $ 2.8 $ & $ 0.13 $ & $ 0.086 $ & $ 0.15 $ & $ 0.17 $ \\
$ 402.47742.3318 $ & $ 49.8 $ & $ 50.1 $ & $ 8.64 $ & $ 0.32 $ & $ 0.26 $ & $ 0.59 $ & $ 0.52 $ \\
$ 402.47796.1893 $ & $ 19.3 $ & $ 19.3 $ & $ 22.7 $ & $ 0.29 $ & $ 0.24 $ & $ 0.25 $ & $ 0.22 $ \\
$ 402.47798.1259 $ & $ 46.6 $ & $ 47.6 $ & $ 4.95 $ & $ 0.31 $ & $ 0.26 $ & $ 0.56 $ & $ 0.49 $ \\
$ 402.47799.1736 $ & $ 101 $ & $ 115 $ & $ 2.26 $ & $ 0.38 $ & $ 0.32 $ & $ 0.99 $ & $ 0.98 $ \\
$ 402.47856.561 $ & $ 10.5 $ & $ 12.3 $ & $ 2.14 $ & $ 0.24 $ & $ 0.20 $ & $ 0.16 $ & $ 0.17 $ \\
$ 402.47862.1576 $ & $ 53.8 $ & $ 54.3 $ & $ 7.52 $ & $ 0.34 $ & $ 0.26 $ & $ 0.6 $ & $ 0.55 $ \\
$ 402.48158.1296 $ & $ 15.4 $ & $ 15.9 $ & $ 3.94 $ & $ 0.29 $ & $ 0.21 $ & $ 0.20 $ & $ 0.21 $ \\
$ 402.48280.502 $ & $ 21.7 $ & $ 25.5 $ & $ 2.08 $ & $ 0.3 $ & $ 0.24 $ & $ 0.27 $ & $ 0.28 $ \\
$ 403.47491.770 $ & $ 43.5 $ & $ 48.5 $ & $ 2.43 $ & $ 0.31 $ & $ 0.26 $ & $ 0.57 $ & $ 0.53 $ \\
$ 403.47550.807 $ & $ 11.1 $ & $ 11.6 $ & $ 3.60 $ & $ 0.21 $ & $ 0.16 $ & $ 0.20 $ & $ 0.20 $ \\
$ 403.47610.576 $ & $ 9.37 $ & $ 9.59 $ & $ 4.76 $ & $ 0.2 $ & $ 0.14 $ & $ 0.19 $ & $ 0.19 $ \\
$ 403.47845.495 $ & $ 12.6 $ & $ 15.0 $ & $ 2.03 $ & $ 0.25 $ & $ 0.20 $ & $ 0.20 $ & $ 0.22 $ \\
\enddata
\end{deluxetable}

\begin{deluxetable}{llllll}
\tabletypesize{\small}
\tablecaption{Alternative designations of our events\label{tab:crossReferences}}
\tablewidth{0pt}

\tablehead{\colhead{field.tile.seq} & \colhead{RA (J2000)} &
\colhead{DEC (J2000)} & \colhead{MACHO Alert} & \colhead{EROS} &
\colhead{OGLE} }

\startdata
$ 101.21689.315 $ & 18 06 58.32 & $ -27 $ 27 45.7 & 101-B &  &  \\
$ 102.22466.140 $ & 18 08 47.04 & $ -27 $ 40 47.3 & 95-BLG-13 &  &  \\ 
$ 104.19992.858 $ & 18 02 53.76 & $ -27 $ 57 50.0 &  &  &  \\ 
$ 104.20119.6312 $ & 18 03 20.64 & $ -28 $ 08 14.6 & 97-BLG-34 &  &  \\ 
$ 104.20251.50 $ & 18 03 34.08 & $ -28 $ 00 19.1 & 104-C &  &  \\ 
$ 104.20251.1117 $ & 18 03 29.04 & $ -28 $ 00 31.0 & 104-D &  &  \\ 
$ 104.20259.572 $ & 18 03 33.36 & $ -27 $ 27 47.2 & 104-A &  &  \\ 
$ 104.20382.803 $ & 18 03 53.28 & $ -27 $ 57 35.6 & 96-BLG-12 & \#16, BLG-12 &  \\ 
$ 104.20515.498 $ & 18 04 09.60 & $ -27 $ 44 35.2 & 97-BLG-58 & \#11, BLG-11 & BUL\_SC35-144974 \\ 
$ 104.20640.8423 $ & 18 04 33.60 & $ -28 $ 07 31.8 &  & \#7,  BLG-13
& 1998-BUL-23 \\ 
$ 104.20645.3129 $ & 18 04 26.16 & $ -27 $ 47 35.2 & 96-BLG-1 &  &  \\ 
$ 105.21813.2516 $ & 18 07 06.96 & $ -27 $ 52 34.3 & 97-BLG-10 &  &  \\ 
$ 108.18947.3618 $ & 18 00 29.04 & $ -28 $ 19 20.3 &  &  &  \\ 
$ 108.18951.593 $ & 18 00 33.84 & $ -28 $ 01 10.6 & 97-BLG-28 &  &  \\ 
$ 108.18951.1221 $ & 18 00 25.92 & $ -28 $ 02 35.2 & 108-A &  &  \\ 
$ 108.18952.941 $ & 18 00 36.00 & $ -27 $ 58 30.0 & 95-BLG-32 &  &  \\ 
$ 108.19074.550 $ & 18 00 52.32 & $ -28 $ 29 52.1 & 97-BLG-59 &  &  \\ 
$ 108.19334.1583 $ & 18 01 26.40 & $ -28 $ 31 14.2 & 95-BLG-14 &  &  \\ 
$ 109.20119.1051 $ & 18 03 20.64 & $ -28 $ 08 14.6 &  &  &  \\ 
$ 109.20640.360 $ & 18 04 33.60 & $ -28 $ 07 32.5 &  & \#7, BLG-13  & 1998-BUL-23 \\ 
$ 110.22455.842 $ & 18 08 51.36 & $ -28 $ 27 11.2 & 97-BLG-5 & \#4, BLG-28 &  \\ 
$ 113.18552.581 $ & 17 59 36.00 & $ -28 $ 36 24.1 & 113-A &  &  \\ 
$ 113.18804.1061 $ & 18 00 03.36 & $ -29 $ 11 04.2 & 95-BLG-4 &  &  \\ 
$ 113.19192.365 $ & 18 01 06.96 & $ -29 $ 18 53.6 &  &  &  \\ 
$ 114.19712.813 $ & 18 02 11.52 & $ -29 $ 19 20.6 & 96-BLG-19 &  &  \\
$ 114.19846.777 $ & 18 02 36.72 & $ -29 $ 01 41.9 & 114-A &  &  \\ 
$ 114.19970.843 $ & 18 02 54.48 & $ -29 $ 26 29.4 & 95-BLG-24 &  &  \\ 
$ 118.18014.320 $ & 17 58 25.20 & $ -29 $ 47 59.6 & 94-BLG-3 &  &  \\ 
$ 118.18141.731 $ & 17 58 36.72 & $ -30 $ 02 19.3 & 94-BLG-4 &  &  \\ 
$ 118.18271.738 $ & 17 58 52.56 & $ -30 $ 02 08.2 & 118-D &  &  \\ 
$ 118.18402.495 $ & 17 59 13.92 & $ -29 $ 55 53.4 & 118-C &  &  \\ 
$ 118.18797.1397 $ & 18 00 06.96 & $ -29 $ 38 06.0 & 97-BLG-26 & \#14, BLG-4 &  \\ 
$ 118.19182.891 $ & 18 01 09.84 & $ -29 $ 56 19.0 & 98-BLG-33 &  & 1998-BUL-22 \\ 
$ 118.19184.939 $ & 18 01 10.32 & $ -29 $ 48 55.4 & 99-BLG-12 & \#12, BLG-5 & 1999-BUL-07 \\ 
$ 121.22032.133 $ & 18 07 46.80 & $ -30 $ 39 41.4 &  &  &  \\ 
$ 158.27444.129 $ & 18 20 10.32 & $ -25 $ 08 25.8 &  &  &  \\ 
$ 162.25865.442 $ & 18 16 30.72 & $ -26 $ 26 58.2 &  &  &  \\ 
$ 162.25868.405 $ & 18 16 46.08 & $ -26 $ 11 43.1 & 95-BLG-8 &  &  \\ 
$ 176.18826.909 $ & 18 00 15.36 & $ -27 $ 42 47.9 &  &  &  \\ 
$ 178.23531.931 $ & 18 11 17.52 & $ -25 $ 59 51.4 & 178-A &  &  \\ 
$ 180.22240.202 $ & 18 07 57.84 & $ -25 $ 24 48.2 &  &  &  \\ 
$ 401.47991.1840 $ & 17 57 07.44 & $ -28 $ 13 44.0 & 97-BLG-19 &  &  \\ 
$ 401.47994.1182 $ & 17 57 07.92 & $ -28 $ 00 27.7 & 98-BLG-26 &  &  \\ 
$ 401.48052.861 $ & 17 57 23.76 & $ -28 $ 10 24.6 & 98-BLG-10 &  &  \\ 
$ 401.48167.1934 $ & 17 57 57.36 & $ -28 $ 27 09.7 &  &  &  \\ 
$ 401.48229.760 $ & 17 58 13.44 & $ -28 $ 20 32.3 & 99-BLG-25 &  &  \\ 
$ 401.48408.649 $ & 17 59 08.88 & $ -28 $ 24 54.7 & 98-BLG-14 &  & BUL\_SC20-395103  \\ 
$ 401.48469.789 $ & 17 59 22.32 & $ -28 $ 20 21.8 &  &  &  \\ 
$ 402.47678.1666 $ & 17 55 22.56 & $ -29 $ 04 12.7 & 98-BLG-39 &  &  \\
$ 402.47737.1590 $ & 17 55 41.52 & $ -29 $ 09 00.7 &  &  &  \\ 
$ 402.47742.3318 $ & 17 55 42.96 & $ -28 $ 49 17.4 &  &  &  \\ 
$ 402.47796.1893 $ & 17 56 10.80 & $ -29 $ 10 44.4 & 402-B &  &  \\ 
$ 402.47798.1259 $ & 17 56 11.04 & $ -29 $ 06 14.0 &  &  &  \\ 
$ 402.47799.1736 $ & 17 56 07.20 & $ -28 $ 58 32.5 & 98-BLG-37 &  &  \\ 
$ 402.47856.561 $ & 17 56 24.96 & $ -29 $ 12 56.5 &  &  &  \\ 
$ 402.47862.1576 $ & 17 56 20.64 & $ -28 $ 47 42.0 & 97-BLG-41 &  &  \\ 
$ 402.48158.1296 $ & 17 57 47.52 & $ -29 $ 03 54.4 & 98-BLG-19 &  & 1998-BUL-17 \\ 
$ 402.48280.502 $ & 17 58 24.96 & $ -28 $ 57 46.4 &  &  &  \\ 
$ 403.47491.770 $ & 17 54 38.64 & $ -29 $ 33 13.0 & 99-BLG-7 &  & 1999-BUL-02 \\ 
$ 403.47550.807 $ & 17 55 00.00 & $ -29 $ 35 03.8 & 99-BLG-54 &  & 1999-BUL-41 \\ 
$ 403.47610.576 $ & 17 55 17.04 & $ -29 $ 37 40.8 &  &  &  \\ 
$ 403.47845.495 $ & 17 56 22.80 & $ -29 $ 55 16.3 & 97-BLG-17 &  &  \\ 
\enddata

\end{deluxetable}

\begin{deluxetable}{ll}

\tableheadfrac{0.05}
\tablecaption{Events with light curve deviations\label{tab:lcdeviations}}
\tablewidth{0pt}

\tablehead{ \colhead{field.tile.seq} & \colhead{Comments}}
\startdata
101.21689.315 & possible parallax \\
104.20119.6312/109.20119.1051 & parallax or xallarap \\ 
104.20251.50 & parallax \\
104.20382.803 & parallax \\
104.20640.8423 & --- \\
105.21813.2516 & possible red blend \\
108.18951.593 & binary \\
114.19970.843 & weak signal \\
118.18141.731 & binary \\
118.18797.1397 & asymmetry in blue peak \\
158.27444.129 & --- \\
180.22240.202 & post-peak blue points preferentially low \\
401.48229.760 & binary \\
401.48408.649 & binary \\
402.47742.3318 & --- \\
402.47799.1736 & coherent steep variation in blue \\
402.47862.1576 & binary \\
402.48280.502 & possible blue blend \\
403.47610.576 & possible red blend \\
\enddata
\end{deluxetable}

\begin{deluxetable}{lcccc}
\tablecaption{Event parameters from blend fits \label{tab:blend_params}}
\tablewidth{0pt}
\tablecomments{Known binaries and two strong parallax
events (104.20251.50 and 104.20382.803) are not included.}
\tablehead{\colhead{field.tile.seq} & \colhead{$\hat{t}$} &
\colhead{$u_{\rm min}$} & \colhead{$f_{r_M}$} & \colhead{$f_{b_M}$}}

\startdata
$ 101.21689.315 $ 	& $ 196.6 \pm 4.1$ 	& $ 0.1609 \pm 0.0051 $ & $ 0.726 \pm 0.026 $ 	& $ 0.736 \pm 0.026 $ \\
$ 102.22466.140 $ 	& $ 148.0 \pm 6.9$ 	& $ 0.276 \pm 0.019 $ 	& $ 1.059 \pm 0.089 $ 	& $ 1.049 \pm 0.088 $ \\
$ 104.19992.858 $ 	& $ 103 \pm 19$ 	& $ 0.52 \pm 0.17 $ 	& $ 1.09 \pm 0.56 $ 	& $ 1.11 \pm 0.57 $ \\
$ 104.20119.6312 $ 	& $ 46 \pm 11$  	& $ 0.56 \pm 0.22 $ 	& $ 0.65 \pm 0.40 $ 	& $ 0.58 \pm 0.37 $ \\
$ 104.20251.1117 $ 	& $ 68.7 \pm 8.0$ 	& $ 0.274 \pm 0.048 $ 	& $ 0.413 \pm 0.089 $ 	& $ 0.413 \pm 0.089 $ \\
$ 104.20259.572 $ 	& $ 14.2 \pm 1.5$ 	& $ 0.77 \pm 0.11 $ 	& $ 1.01 \pm 0.28 $ 	& $ 1.010 \pm 0.27 $ \\
$ 104.20515.498 $ 	& $ 43 \pm 10$  	& $ 0.92 \pm 0.33 $ 	& $ 1.8 \pm 1.3 $ 	& $ 1.9 \pm 1.4 $ \\
$ 104.20640.8423 $ 	& $ 78 \pm 14$  	& $ 0.178 \pm 0.045 $ 	& $ 0.150 \pm 0.043 $ 	& $ 0.134 \pm 0.038 $ \\
$ 104.20645.3129 $ 	& $ 133 \pm 13$ 	& $ 0.78 \pm 0.13 $ 	& $ 1.47 \pm 0.44 $ 	& $ 1.48 \pm 0.44 $ \\
$ 105.21813.2516 $ 	& $ 17.84 \pm 0.72$ 	& $ 0.0741 \pm 0.0049 $ 	& $ 0.952 \pm 0.062 $ 	& $ 1.01 \pm 0.066 $ \\
$ 108.18947.3618 $ 	& $ 46.2 \pm 4.2$ 	& $ 0.529 \pm 0.080 $ 	& $ 0.95 \pm 0.22 $ 	& $ 1.01 \pm 0.23 $ \\
$ 108.18951.1221 $ 	& $ 31 \pm 22$  	& $ 0.92 \pm 0.98 $ 	& $ 2.7 \pm 5.8 $ 	& $ 2.9 \pm 6.2 $ \\
$ 108.18952.941 $ 	& $ 63.3 \pm 4.3$ 	& $ 0.491 \pm 0.053 $ 	& $ 1.14 \pm 0.18 $ 	& $ 1.14 \pm 0.18 $ \\
$ 108.19074.550 $ 	& $ 12.7 \pm 2.0$ 	& $ 0.36 \pm 0.10 $ 	& $ 0.44 \pm 0.16 $ 	& $ 0.42 \pm 0.15 $ \\
$ 108.19334.1583 $ 	& $ 16.7 \pm 2.5$ 	& $ 0.294 \pm 0.067 $ 	& $ 1.22 \pm 0.34 $ 	& $ 1.19 \pm 0.33 $ \\
$ 109.20119.1051 $ 	& $ 3.92 \pm 0.30$ 	& $ 8.75 \pm 0.64 $ 	& $ 1930 \pm 550 $ 	& $ 1840 \pm 520 $ \\
$ 109.20640.360 $ 	& $ 25.7 \pm 9.6$ 	& $ 0.84 \pm 0.49 $ 	& $ -- $ 	& $ 1.4 \pm 1.5 $ \\
$ 110.22455.842 $ 	& $ 16.5 \pm 3.4$ 	& $ 0.55 \pm 0.17 $ 	& $ 1.49 \pm 0.71 $ 	& $ 1.50 \pm 0.72 $ \\
$ 113.18552.581 $ 	& $ 614800 \pm 102700$ 	& $ (1.23 \pm 0.20)\times 10^{-5} $  &   $(1.03 \pm 0.17)\times 10^{-5} $ 	& $ (9.00 \pm 1.50)\times 10^{-6} $ \\
$ 113.18804.1061 $ 	& $ 19 \pm 14$  	& $ 0.17 \pm 0.16 $ 	& $ 0.13 \pm 0.14 $ 	& $ 0.14 \pm 0.15 $ \\
$ 113.19192.365 $ 	& $ 30 \pm 18$  	& $ 0.86 \pm 0.84 $ 	& $ 1.5 \pm 2.9 $ 	& $ 2.0 \pm 3.7 $ \\
$ 114.19712.813 $ 	& $ 24.0 \pm 2.0$ 	& $ 0.283 \pm 0.036 $ 	& $ 0.98 \pm 0.16 $ 	& $ 0.97 \pm 0.16 $ \\
$ 114.19846.777 $ 	& $ 122 \pm 14$ 	& $ 0.284 \pm 0.052 $ 	& $ 0.226 \pm 0.051 $ 	& $ 0.211 \pm 0.048 $ \\
$ 114.19970.843 $ 	& $ 16.8 \pm 1.5$ 	& $ 0.181 \pm 0.025 $ 	& $ 0.79 \pm 0.12 $ 	& $ 0.73 \pm 0.11 $ \\
$ 118.18014.320 $ 	& $ 15.7 \pm 8.6$ 	& $ 1.6 \pm 1.3 $ 	& $ 5 \pm 11 $ 	& $ 5 \pm 11 $ \\
$ 118.18271.738 $ 	& $ 93 \pm 17$  	& $ 0.213 \pm 0.057 $ 	& $ 0.35 \pm 0.11 $ 	& $ 0.33 \pm 0.10 $ \\
$ 118.18402.495 $ 	& $ 26 \pm 1.8$ 	& $ 0.246 \pm 0.027 $ 	& $ 0.83 \pm 0.11 $ 	& $ 0.85 \pm 0.11 $ \\
$ 118.18797.1397 $ 	& $ 121.3 \pm 2.5$ 	& $ 0.1274 \pm 0.0039 $ & $ 1.069 \pm 0.034 $ 	& $ 1.043 \pm 0.034 $ \\
$ 118.19182.891 $ 	& $ 12.48 \pm 0.91$ 	& $ 0.195 \pm 0.030 $ 	& $ 1.21 \pm 0.19 $ 	& $ 1.22 \pm 0.20 $ \\
$ 118.19184.939 $ 	& $ 88.5 \pm 5.4$ 	& $ 0.388 \pm 0.037 $ 	& $ 0.673 \pm 0.087 $ 	& $ 0.675 \pm 0.088 $ \\
$ 121.22032.133 $ 	& $ 15.0 \pm 9.3$ 	& $ 1.2 \pm 1.2 $ 	& $ 2.3 \pm 5.0 $ 	& $ 2.4 \pm 5.2 $ \\
$ 158.27444.129 $ 	& $ 48.7 \pm 3.0$ 	& $ 0.362 \pm 0.036 $ 	& $ 0.710 \pm 0.093 $ 	& $ 0.689 \pm 0.090 $ \\
$ 162.25865.442 $ 	& $ 61 \pm 10$  	& $ 0.75 \pm 0.21 $ 	& $ 1.01 \pm 0.50 $ 	& $ 0.95 \pm 0.47 $ \\
$ 162.25868.405 $ 	& $ 30.6 \pm 2.5$ 	& $ 0.386 \pm 0.057 $ 	& $ 1.00 \pm 0.19 $ 	& $ 1.04 \pm 0.20 $ \\
$ 176.18826.909 $ 	& $ 3.68 \pm 0.38$ 	& $ 7.27 \pm 0.70 $ 	& $ -- $   	& $ 1120 \pm 410 $ \\
$ 178.23531.931 $ 	& $ 13.1 \pm 2.9$ 	& $ 0.54 \pm 0.22 $ 	& $ 1.14 \pm 0.71 $ 	& $ 1.11 \pm 0.70 $ \\
$ 180.22240.202 $ 	& $ 454 \pm 20$ 	& $ 0.183 \pm 0.010 $ 	& $ 0.547 \pm 0.035 $ 	& $ 0.551 \pm 0.035 $ \\
$ 401.47991.1840 $ 	& $ 38 \pm 18$  	& $ 0.77 \pm 0.62 $ 	& $ 1.6 \pm 2.3 $ 	& $ 1.6 \pm 2.4 $ \\
$ 401.47994.1182 $ 	& $ 61 \pm 12$  	& $ 0.273 \pm 0.085 $ 	& $ -- $   	& $ 0.97 \pm 0.36 $ \\
$ 401.48052.861 $ 	& $ 83 \pm 15$  	& $ 0.304 \pm 0.091 $ 	& $ 0.74 \pm 0.28 $ 	& $ 0.70 \pm 0.27 $ \\
$ 401.48167.1934 $ 	& $ 48.5 \pm 8.2$  	& $ 2.83 \pm 0.62 $ 	& $ 37.1 \pm 24.0 $ 	& $ 39.8 \pm 26.0 $ \\
$ 401.48469.789 $ 	& $ 7.5 \pm 4.2$ 	& $ < 0.26 $ 	& $ 0.58 \pm 0.58 $ 	& $ 0.54 \pm 0.55 $ \\
$ 402.47678.1666 $ 	& $ 13.4 \pm 6.8$ 	& $ 0.45 \pm 0.43 $ 	& $ 1.0 \pm 1.4 $ 	& $ 1.1 \pm 1.5 $ \\
$ 402.47737.1590 $ 	& $ 11.0 \pm 3.5$ 	& $ 0.111 \pm 0.055 $ 	& $ 0.26 \pm 0.13 $ 	& $ 0.29 \pm 0.15 $ \\
$ 402.47742.3318 $ 	& $ 64.7 \pm 8.5$ 	& $ 0.065 \pm 0.020 $ 	& $ 0.72 \pm 0.14 $ 	& $ 0.61 \pm 0.12 $ \\
$ 402.47796.1893 $ 	& $ 21.8 \pm 1.4$ 	& $ < 0.03 $ 	& $ 0.801 \pm 0.081 $ 	& $ 0.832 \pm 0.078 $ \\
$ 402.47798.1259 $ 	& $ 97800 \pm 19500 $ & $ (6 \pm 20219)\times 10^{-7}$ 	& $ (2.59 \pm 0.52)\times 10^{-4} $ 	& $ (2.31 \pm 0.47)\times 10^{-4} $ \\
$ 402.47799.1736 $ 	& $ 80 \pm 44$  	& $ 0.84 \pm 0.70 $ 	& $ 2.5 \pm 4.0 $ 	& $ 2.7 \pm 4.4 $ \\
$ 402.47856.561 $ 	& $ 52000 \pm 16000$ 	& $ (2.0\pm 1.1)\times 10^{-5} $ & $ (3.8 \pm 1.2)\times 10^{-5} $ 	& $ (5.0 \pm 1.6)\times 10^{-5} $ \\
$ 402.48158.1296 $ 	& $ 20.7 \pm 4.9$ 	& $ 0.148 \pm 0.082 $
& $ 0.56 \pm 0.26 $ 	& $ 0.61 \pm 0.28 $ \\ 
$ 402.48280.502 $       & $ 119 \pm 28 $        & $ 0.035 \pm 0.011 $   & $ 0.064 \pm  0.018 $  & $ 0.084 \pm 0.024 $ \\              
$ 403.47491.770 $ 	& $ 45.3 \pm 5.3$ 	& $ 0.494 \pm 0.098 $ 	& $ 1.19 \pm 0.35 $ 	& $ 1.18 \pm 0.35 $ \\
$ 403.47550.807 $ 	& $ 10.9 \pm 1.9$ 	& $ 0.33 \pm 0.12 $ 	& $ 1.17 \pm 0.47 $ 	& $ 1.13 \pm 0.45 $ \\
$ 403.47610.576 $ 	& $ 14.5 \pm 1.2$ 	& $ 0.123 \pm 0.014 $ 	& $ 0.492 \pm 0.059 $ 	& $ 0.608 \pm 0.073 $ \\
$ 403.47845.495 $ 	& $ 13.9 \pm 2.7$ 	& $ 0.62 \pm 0.19 $ 	& $ 1.23 \pm 0.64 $ 	& $ 1.20 \pm 0.62 $ \\
\enddata
\end{deluxetable}

\begin{deluxetable}{lcc}
\tablecomments{For the first three events, blend fractions are taken
from Alcock et al.\ (2000c).
The solutions for 97-BLG-41 are designated as follows:\\
$^a$ rotating binary solution obtained by Albrow et al.\
(2000). Johnson's $I$ and $V$ results are reported in $f_{r_M}$ and
$f_{b_M}$ columns, respectively;\\
$^b$ binary with a planet obtained by Bennett et al.\ (1999).\\
For all cases, the statistical errors in blend fractions are $\lsim 0.05$.}
\tablecaption{Known blend fractions for clump events with binary lenses\label{tab:binaryblendfr}}
\tablewidth{0pt}
\tablehead{
\colhead{Event} & 
\colhead{$f_{r_M}$} & 
\colhead{$f_{b_M}$} 
}
\startdata
108.18951.593 (97-BLG-28) & 1.00 & 0.90 \\
118.18141.731 (94-BLG-4) & 1.06 & 1.04 \\
401.48408.649 (98-BLG-14) & 1.08 & 1.10\\
402.47862.1576 (97-BLG-41) & 0.95$^{a}$ & 0.92$^{a}$\\
& 0.86$^{b}$ & 0.84$^{b}$ \\
\enddata
\end{deluxetable}

\begin{deluxetable}{ccccccccc}
\tabletypesize{\small}
\tablewidth{0pt}
\tablecaption{Data on the 94 bulge fields.\label{tab:fielddat}}
\tablecomments{Efficiencies (columns 6 and 7) are averaged over $\amax$. }
\tablehead{\colhead{Field} & \colhead{$ l $}  & \colhead{$b$}  &
\colhead{$N_{\rm clumps}/100$} & \colhead{$N_{\rm exposures}$}  & 
\colhead{$\epsilon(\mathrm{50days})$}  & \colhead{$\epsilon(\mathrm{200days})$}}
\startdata
101  &   3.73   &   $-$3.02   & 629.5 &   804   &  0.41  &  0.61 \\ 
102  &   3.77   &   $-$4.11   & 444 &   421   &  0.24  &  0.46 \\ 
103  &   4.31   &   $-$4.62   & 368 &   331   &  0.22  &  0.37 \\ 
104  &   3.11   &   $-$3.01   & 652 &   1639  &  0.40  &  0.59 \\ 
105  &   3.23   &   $-$3.61   & 539 &   640   &  0.39  &  0.52 \\ 
106  &   3.59   &   $-$4.78   & 346 &   12    &  $ <0.003 $  &  $ <0.003 $ \\ 
107  &   4.00   &   $-$5.31   & 269 &   51    &  0.007  &  $ <0.003 $ \\ 
108  &   2.30   &   $-$2.65   & 790 &   1031  &  0.46  &  0.74 \\ 
109  &   2.45   &   $-$3.20   & 661 &   761   &  0.36  &  0.59 \\ 
110  &   2.81   &   $-$4.48   & 408 &   650   &  0.33  &  0.55 \\ 
111  &   2.99   &   $-$5.14   & 298 &   305   &  0.23  &  0.37 \\ 
112  &   3.40   &   $-$5.53   & 246.5 &   43    &  $ 0.005 $  &  $ <0.003 $ \\ 
113  &   1.63   &   $-$2.78   & 834 &   1127  &  0.49  &  0.73 \\ 
114  &   1.81   &   $-$3.50   & 617 &   776   &  0.37  &  0.59 \\ 
115  &   2.04   &   $-$4.85   & 325.5 &   357   &  0.24  &  0.35 \\ 
116  &   2.38   &   $-$5.44   & 268.5 &   314   &  0.21  &  0.38 \\ 
117  &   2.83   &   $-$6.00   & 198 &   39    &  0.006  &  $ <0.003 $ \\ 
118  &   0.83   &   $-$3.07   & 741.5 &   1053  &  0.44  &  0.74 \\ 
119  &   1.07   &   $-$3.83   & 542.5 &   1815  &  0.45  &  0.68 \\ 
120  &   1.64   &   $-$4.42   & 394 &   676   &  0.35  &  0.54 \\ 
121  &   1.20   &   $-$4.94   & 344 &   352   &  0.22  &  0.37 \\ 
122  &   1.57   &   $-$5.45   & 266 &   186   &  0.15  &  0.23 \\ 
123  &   1.95   &   $-$6.05   & 210.5 &   40    &  0.006  &  $ <0.003 $ \\ 
124  &   0.57   &   $-$5.28   & 265.5 &   337   &  0.22  &  0.39 \\ 
125  &   1.11   &   $-$5.93   & 189 &   287   &  0.21  &  0.37 \\ 
126  &   1.35   &   $-$6.40   & 182 &   36    &  $ <0.003 $  &  $ <0.003 $ \\ 
127  &   0.28   &   $-$5.91   & 192 &   178   &  0.13  &  0.24 \\ 
128  &   2.43   &   $-$4.03   & 516.5 &   711   &  0.35  &  0.55 \\ 
129  &   4.58   &   $-$5.93   & 234.5 &   31    &  0.006  &  $ <0.003 $ \\ 
130  &   5.11   &   $-$6.49   & 169.5 &   20    &  0.02  &  0.02 \\ 
131  &   4.98   &   $-$7.33   & 130 &   106   &  0.12  &  0.22 \\ 
132  &   5.44   &   $-$7.91   & 102 &   180   &  0.17  &  0.33 \\ 
133  &   6.05   &   $-$8.40   & 70.5 &   176   &  0.15  &  0.18 \\ 
134  &   6.34   &   $-$9.07   & 46.5 &   197   &  0.09  &  0.15 \\ 
135  &   3.89   &   $-$6.26   & 179 &   30    &  0.004  &  $ <0.003 $ \\ 
136  &   4.42   &   $-$6.82   & 137.5 &   207   &  0.17  &  0.26 \\ 
137  &   4.31   &   $-$7.60   & 101 &   193   &  0.15  &  0.36 \\ 
138  &   4.69   &   $-$8.20   & 95.5 &   201   &  0.16  &  0.21 \\ 
139  &   5.35   &   $-$8.65   & 81 &   191   &  0.11  &  0.11 \\ 
140  &   5.71   &   $-$9.20   & 64 &   209   &  0.13  &  0.25 \\ 
141  &   3.26   &   $-$6.59   & 159 &   28    &  0.03  &  0.03 \\ 
142  &   3.81   &   $-$7.08   & 126.5 &   218   &  0.15  &  0.36 \\ 
143  &   3.80   &   $-$8.00   & 97 &   210   &  0.17  &  0.25 \\ 
144  &   4.68   &   $-$9.02   & 50.5 &   210   &  0.10  &  0.18 \\ 
145  &   5.20   &   $-$9.50   & 49.5 &   210   &  0.12  &  0.23 \\ 
146  &   3.26   &   $-$7.54   & 106 &   207   &  0.19  &  0.33 \\ 
147  &   3.96   &   $-$8.81   & 66 &   208   &  0.19  &  0.20 \\ 
148  &   2.33   &   $-$6.71   & 161.5 &   229   &  0.18  &  0.30 \\ 
149  &   2.43   &   $-$7.43   & 112.5 &   236   &  0.19  &  0.34 \\ 
150  &   2.96   &   $-$8.01   & 98.5 &   229   &  0.18  &  0.28 \\ 
151  &   3.17   &   $-$8.89   & 85 &   194   &  0.14  &  0.23 \\ 
152  &   1.76   &   $-$7.07   & 124 &   235   &  0.24  &  0.32 \\ 
153  &   2.11   &   $-$7.87   & 85.5 &   231   &  0.17  &  0.35 \\ 
154  &   2.16   &   $-$8.51   & 82 &   247   &  0.19  &  0.20 \\ 
155  &   1.01   &   $-$7.44   & 95.5 &   247   &  0.17  &  0.36 \\ 
156  &   1.34   &   $-$8.12   & 88.5 &   249   &  0.14  &  0.33 \\ 
157  &   0.08   &   $-$7.76   & 105.5 &   258   &  0.20  &  0.37 \\ 
158  &   7.08   &   $-$4.44   & 246.5 &   260   &  0.20  &  0.29 \\ 
159  &   6.35   &   $-$4.40   & 258.5 &   464   &  0.27  &  0.47 \\ 
160  &   6.84   &   $-$5.04   & 248 &   208   &  0.15  &  0.26 \\ 
161  &   5.56   &   $-$4.01   & 366 &   467   &  0.32  &  0.45 \\ 
162  &   5.64   &   $-$4.62   & 301 &   382   &  0.22  &  0.38 \\ 
163  &   5.98   &   $-$5.22   & 259 &   251   &  0.19  &  0.36 \\ 
164  &   6.51   &   $-$5.90   & 167.5 &   20    &  0.002  &  $ <0.003 $ \\ 
165  &   7.01   &   $-$6.38   & 135 &   14    &  $ <0.003 $  &  $ <0.003 $ \\ 
166  &   7.10   &   $-$7.07   & 131.5 &   110   &  0.09  &  0.14 \\ 
167  &   4.88   &   $-$4.21   & 388.5 &   364   &  0.26  &  0.40 \\ 
168  &   5.01   &   $-$4.92   & 298 &   210   &  0.17  &  0.26 \\ 
169  &   5.40   &   $-$5.63   & 208 &   24    &  0.004  &  $ <0.003 $ \\ 
170  &   5.81   &   $-$6.20   & 170 &   17    &  $ <0.003 $  &  $ <0.003 $ \\ 
171  &   6.42   &   $-$6.65   & 120.5 &   125   &  0.15  &  0.19 \\ 
172  &   6.82   &   $-$7.61   & 89 &   132   &  0.13  &  0.18 \\ 
173  &   6.92   &   $-$8.41   & 87 &   146   &  0.06  &  0.09 \\ 
174  &   5.71   &   $-$6.92   & 143.5 &   144   &  0.18  &  0.22 \\ 
175  &   6.10   &   $-$7.55   & 111.5 &   97    &  0.06  &  0.10 \\ 
176  &   2.93   &   $-$2.30   & 814.5 &   423   &  0.25  &  0.43 \\ 
177  &   6.75   &   $-$3.82   & 319 &   416   &  0.30  &  0.43 \\ 
178  &   5.24   &   $-$3.42   & 477 &   376   &  0.27  &  0.41 \\ 
179  &   4.92   &   $-$2.83   & 510 &   349   &  0.20  &  0.39 \\ 
180  &   5.93   &   $-$2.69   & 478 &   343   &  0.21  &  0.38 \\ 
301  &   18.77   &   $-$2.05   & - &   925   &  0.31  &  0.46 \\ 
302  &   18.09   &   $-$2.24   & - &   365   &  0.28  &  0.37 \\ 
303  &   17.30   &   $-$2.33   & - &   369   &  0.23  &  0.41 \\ 
304  &   9.07   &   $-$2.70   & - &   458   &  0.30  &  0.50 \\ 
305  &   9.69   &   $-$2.36   & - &   411   &  0.29  &  0.46 \\ 
306  &   8.46   &   $-$3.03   & - &   423   &  0.28  &  0.41 \\ 
307  &   7.84   &   $-$3.37   & - &   435   &  0.27  &  0.43 \\ 
308  &   10.02  &   $-$2.98   & - &   360   &  0.26  &  0.40 \\ 
309  &   9.40   &   $-$3.31   & - &   347   &  0.23  &  0.40 \\ 
310  &   8.79   &   $-$3.64   & - &   387   &  0.26  &  0.43 \\ 
311  &   8.17   &   $-$3.97   & - &   396   &  0.28  &  0.40 \\ 
401  &   2.02   &   $-$1.93   & 964.5 &   429   &  0.25  &  0.45 \\ 
402  &   1.27   &   $-$2.09   & 1153 &   1256  &  0.26  &  0.45 \\ 
403  &   0.55   &   $-$2.32   & 1085 &   473   &  0.26  &  0.45 \\

\enddata
\end{deluxetable}

\begin{deluxetable}{lccccc}
\tablecaption{ Optical Depth by Field \label{tab:taubyfield} }
\tablewidth{0pt}
\tablecomments{Only fields with at least one clump giant event (and field 119) are shown.}
\tablehead{
\colhead{Field} & \colhead{$(l,b)$} & \colhead{$ N_{\rm stars}/100 $} &
\colhead{$N_{\rm clumps}/100$} & \colhead{ $N_{\rm events}$}  & 
\colhead{$\tau/10^{-6}$}
}
\startdata
101 & $(3.73,-3.02)$ &  6284  &  629.5   & 1  	& 1.62 $\pm$ 1.62 \\
102 & $(3.77,-4.11)$ &  6545  &  444\phm{.5}   & 1 	& 3.67 $\pm$ 3.67 \\
104 & $(3.11,-3.01)$ &  5813  &  652\phm{.5}   & 9 	& 8.76 $\pm$ 3.73 \\
105 & $(3.23,-3.61)$ &  6375  &  539\phm{.5}   & 1 	& 0.35 $\pm$ 0.35 \\
$108^*$ & $(2.30,-2.65)$ & 6498  &  790\phm{.5}   & 6        & 2.04 $\pm$ 0.92 \\
$109^*$ & $(2.45,-3.20)$ & 6926  &  661\phm{.5}   & 2        & 0.58 $\pm$ 0.41 \\
110  & $(2.81,-4.48)$ & 6649  &  408\phm{.5}   & 1        & 0.46 $\pm$ 0.46 \\
$113^*$ & $(1.63,-2.78)$ & 6252  &  834\phm{.5}   & 3        & 0.55 $\pm$ 0.35 \\
$114^*$ & $(1.81,-3.50)$ & 6665  &  617\phm{.5}   & 3        & 1.19 $\pm$ 0.74 \\
$118^*$ & $(0.83,-3.07)$ & 6347 &  741.5  & 7        & 2.85 $\pm$ 1.35 \\
$119^*$ & $(1.07,-3.83)$ &  7454 &  542.5   & 0        & ---  \\
121 & $(1.20,-4.94)$ & 5855   &  344\phm{.5}   & 1 	& 0.59 $\pm$ 0.59 \\
158 & $(7.08,-4.44)$ & 4703    &  246.5   & 1 	& 2.96 $\pm$ 2.96 \\
162 & $(5.64,-4.62)$ & 5914    &  301\phm{.5}   & 2 	& 3.11 $\pm$ 2.21 \\
176 & $(2.93,-2.30)$ & 7741   &  814.5   & 1 	& 0.37 $\pm$ 0.37 \\
178 & $(5.24,-3.42)$ & 8186   &  477\phm{.5}   & 1 	& 0.43 $\pm$ 0.43 \\
180 & $(5.93,-2.69)$ & 6388    &  478\phm{.5}   & 1 	& 4.62 $\pm$ 4.62 \\
$401^*$ &  $(2.02,-1.92)$ & 6630 &  964.5   & 7        & 5.13 $\pm$ 2.16 \\
$402^*$ &  $(1.27,-2.09)$ & 7098 &  1153\phm{1.5}  & 10\phm{0}      & 3.95 $\pm$ 1.50 \\
$403^*$ & $(0.55,-2.32)$ & 6798 &  1085\phm{1.5}  & 4        & 1.16 $\pm$ 0.66 \\
\hline\\
CGR\tablenotemark{a}& $(1.50,-2.68)$ & 60668\phm{0} &   7388.5\phm{7} & 42\phm{4}    & $2.17^{+0.47}_{-0.38}$ \\
\enddata
\tablenotetext{*}{Field is near the Galactic center and included in CGR.}
\tablenotetext{a} {Average towards the group of 9 fields designated as CGR.}
\end{deluxetable}

\begin{deluxetable}{ll|ll|ll|ll}

\tableheadfrac{0.05}
\tablecaption{One sigma upper limits on the optical depth for fields
with no clump events \label{tab:upperlim}}
\tablewidth{0pt}

\tablehead{ \colhead{field} & \colhead{$\tau^{\rm lim, 1\sigma}$} &
\colhead{Field} & \colhead{$\tau^{\rm lim, 1\sigma}$} &
\colhead{Field} & \colhead{$\tau^{\rm lim, 1\sigma}$} &
\colhead{Field} & \colhead{$\tau^{\rm lim, 1\sigma}$} }
\startdata
103 & $3.26 \times 10^{-6}$ & 128 & $1.45 \times 10^{-6}$ & 144 & $4.62 \times 10^{-5}$ & 161 & $2.52 \times 10^{-6}$ \\
106 & $7.90 \times 10^{-6}$ & 129 & $9.87 \times 10^{-5}$ & 145 & $4.10 \times 10^{-5}$ & 163 & $5.84 \times 10^{-6}$ \\
107 & $4.22 \times 10^{-5}$ & 130 & $1.13 \times 10^{-4}$ & 146 & $1.42 \times 10^{-5}$ & 164 & $3.68 \times 10^{-4}$ \\
111 & $4.56 \times 10^{-6}$ & 131 & $2.01 \times 10^{-5}$ & 147 & $3.10 \times 10^{-5}$ & 165 & $6.81 \times 10^{-4}$ \\
112 & $7.12 \times 10^{-5}$ & 132 & $2.04 \times 10^{-5}$ & 148 & $9.71 \times 10^{-6}$ & 166 & $2.52 \times 10^{-5}$ \\
115 & $3.58 \times 10^{-6}$ & 133 & $3.25 \times 10^{-5}$ & 149 & $1.39 \times 10^{-5}$ & 167 & $2.83 \times 10^{-6}$ \\ 
116 & $5.02 \times 10^{-6}$ & 134 & $5.45 \times 10^{-5}$ & 150 & $1.59 \times 10^{-5}$ & 168 & $6.30 \times 10^{-6}$ \\
117 & $8.37 \times 10^{-5}$ & 135 & $9.14 \times 10^{-5}$ & 151 & $2.67 \times 10^{-5}$ & 169 & $1.05 \times 10^{-4}$ \\
119 & $1.10 \times 10^{-6}$ & 136 & $1.23 \times 10^{-5}$ & 152 & $1.18 \times 10^{-5}$ & 170 & $4.54 \times 10^{-4}$ \\
120 & $1.91 \times 10^{-6}$ & 137 & $1.72 \times 10^{-5}$ & 153 & $1.83 \times 10^{-5}$ & 171 & $2.08 \times 10^{-5}$ \\
122 & $6.90 \times 10^{-6}$ & 138 & $1.97 \times 10^{-5}$ & 154 & $2.14 \times 10^{-5}$ & 172 & $2.91 \times 10^{-5}$ \\
123 & $7.70 \times 10^{-5}$ & 139 & $2.80 \times 10^{-5}$ & 155 & $1.65 \times 10^{-5}$ & 173 & $4.44 \times 10^{-5}$ \\
124 & $4.94 \times 10^{-6}$ & 140 & $3.42 \times 10^{-5}$ & 156 & $1.85 \times 10^{-5}$ & 174 & $1.36 \times 10^{-5}$ \\
125 & $7.95 \times 10^{-6}$ & 141 & $9.01 \times 10^{-5}$ & 157 & $1.39 \times 10^{-5}$ & 175 & $3.58 \times 10^{-5}$ \\
126 & $1.35 \times 10^{-4}$ & 142 & $1.28 \times 10^{-5}$ & 159 & $3.57 \times 10^{-6}$ & 177 & $3.09 \times 10^{-6}$ \\
127 & $1.02 \times 10^{-5}$ & 143 & $1.80 \times 10^{-5}$ & 160 & $8.12 \times 10^{-6}$ & 179 & $2.61 \times 10^{-6}$ \\
\enddata
\end{deluxetable}

\begin{deluxetable}{lr|lr}

\tableheadfrac{0.05}
\tablecomments{Events in CGR are displayed in the left column and
events in the remaining fields in the right column.}
\tablecaption{Events from the verification sample
\label{tab:verificationevents}}
\tablewidth{0pt}

\tablehead{ \colhead{Event} & \colhead{$\delta_{\rm blue}$} &
\colhead{Event} & \colhead{$\delta_{\rm blue}$} }
\startdata
108.18947.3618  & $ 0.04$ & 102.22466.140   & $ 0.55$ \\
108.18951.1221  & $ 0.31$ & 104.19992.858   & $ 0.20$ \\
108.18952.941   & $ 0.74$ & 104.20119.6312  & $-1.14$ \\
108.19334.1583  & $ 0.56$ & 104.20259.572   & $ 0.02$ \\
109.20640.360   & $ 0.25$ & 104.20515.498   & $ 0.66$ \\
113.19192.365   & $ 0.27$ & 104.20645.3129  & $ 1.08$ \\
114.19712.813   & $-0.20$ & 105.21813.2516  & $ 0.15$ \\
118.18014.320   & $ 0.40$ & 110.22455.842   & $ 0.70$ \\
118.18402.495   & $-1.35$ & 121.22032.133   & $ 0.26$ \\
118.18797.1397  & $ 1.28$ & 162.25865.442   & $-0.12$ \\
118.19182.891   & $ 1.11$ & 162.25868.405   & $ 0.21$ \\
401.47991.1840  & $ 0.27$ & 178.23531.931   & $ 0.16$ \\
401.47994.1182  & $-0.08$ & & \\
401.48052.861   & $-1.13$ & & \\
401.48167.1934  & $ 1.49$ & & \\
401.48469.789   & $-0.83$ & & \\
402.47678.1666  & $ 0.07$ & & \\
402.47799.1736  & $ 0.40$ & & \\
402.48158.1296  & $-1.39$ & & \\
403.47491.770   & $ 0.51$ & & \\
403.47550.807   & $ 0.28$ & & \\
403.47845.495   & $ 0.32$ & & \\
\enddata
\end{deluxetable}

\begin{deluxetable}{lcccc}
\tablecaption{Optical Depth by Field from Verification Sample \label{tab:vertau}}
\tablewidth{0pt}
\tablehead{
\colhead{Field} & \colhead{$N_{\rm events}^{\rm ver}$} &
\colhead{$\tau^{\rm ver}/10^{-6}$} & \colhead{$N_{\rm events}^{\rm orig}$} &
\colhead{$\tau/10^{-6}$}
}
\startdata
108 & 4 & $ 2.25 \pm 1.20 $ & 6 & $ 2.04 \pm 0.92 $\\
109 & 1 & $ 0.59 \pm 0.59 $ & 2 & $ 0.58 \pm 0.41 $\\
113 & 1 & $ 0.80 \pm 0.80 $ & 3 & $ 0.55 \pm 0.35 $\\
114 & 1 & $ 1.07 \pm 1.07 $ & 3 & $ 1.19 \pm 0.74 $\\
118 & 4 & $ 2.77 \pm 1.85 $ & 7 & $ 2.85 \pm 1.34 $\\
119 & 0 &  ---  & 0 &  --- \\
401 & 5 & $ 5.41 \pm 2.67 $ & 7 & $ 5.13 \pm 2.16 $\\
402 & 3 & $ 4.55 \pm 3.42 $ & 10\phm{1} & $ 3.95 \pm 1.50 $\\
403 & 3 & $ 1.29 \pm 0.85 $ & 4 & $ 1.16 \pm 0.66 $\\
& & & & \\
101 & 0 & --- & 1 & $ 1.62 \pm 1.62 $\\
102 & 1 & $ 3.67 \pm 3.67 $ & 1 & $ 3.67 \pm 3.67 $\\
104 & 5 & $ 6.07 \pm 3.18 $ & 9 & $ 8.76 \pm 3.73 $\\
105 & 1 & $ 0.35 \pm 0.35 $ & 1 & $ 0.35 \pm 0.35 $\\
110 & 1 & $ 0.46 \pm 0.46 $ & 1 & $ 0.46 \pm 0.46 $\\
121 & 1 & $ 0.59 \pm 0.59 $ & 1 & $ 0.59 \pm 0.59 $\\
158 & 0 & --- & 1 & $ 2.96 \pm 2.96 $\\
162 & 2 & $ 3.11 \pm 2.21 $ & 2 & $ 3.11 \pm 2.21 $\\
176 & 0 & --- & 1 & $ 0.37 \pm 0.37 $\\
178 & 1 & $ 0.43 \pm 0.43 $ & 1 & $ 0.43 \pm 0.43 $\\
180 & 0 & --- & 1 & $ 4.62 \pm 4.62 $\\
\enddata
\end{deluxetable}

\end{document}